\documentclass[useAMS,usenatbib]{mn2e}
\usepackage{graphicx}
\usepackage{ulem}
\title[3-mm spectral imaging of the Sagittarius B2 region]{Spectral imaging of
the Sagittarius B2 region in multiple 3-mm molecular lines with the Mopra
telescope}
\author[P. A. Jones et al.]{P. A. Jones$^{1}$
\thanks{E-mail:pjones@phys.unsw.edu.au (PAJ)}
M. G. Burton$^{1}$, M. R. Cunningham$^{1}$, K. M. Menten$^{2}$, P.
Schilke$^{2}$,
\newauthor{A. Belloche$^{2}$, S. Leurini$^{3}$, J. Ott$^{4}$,
A. J. Walsh$^{5}$}
\\
$^{1}$School of Physics, University of New South Wales, NSW 2052, Australia \\ 
$^{2}$Max-Planck-Institut f\"{u}r Radioastronomie, Auf dem H\"{u}gel 69, 53121 
Bonn, Germany \\
$^{3}$European Southern Observatory, Karl-Schwarzschild-Str. 2, 85748 
Garching, Germany \\
$^{4}$National Radio Astronomy Observatory, 520 Edgemont Road, 
Charlottesville, VA 22903, USA \\
$^{5}$School of Maths, Physics and IT, James Cook University, Qld 4814, 
Australia \\
}

\begin{document}

\date{Accepted . Received ; in original form 2007 XXX XX}

\pagerange{\pageref{firstpage}--\pageref{lastpage}} \pubyear{2007}

\maketitle

\label{firstpage}

\begin{abstract}

Using the Mopra telescope, we have undertaken a 3-mm spectral-line imaging 
survey of a 5 arcmin square area around Sgr~B2. We covered almost the 
complete spectral the range from 81.7 to 113.5 GHz, 
with 2.2 MHz wide spectral channels or $\sim 6$~km~s$^{-1}$\, and have 
observed 24 lines, with 0.033 MHz wide, or $\sim 0.1$~km~s$^{-1}$ channels.  
We discuss the distribution of around 50 lines, and present velocity-integrated emission images for 38 of the lines. In addition, we have detected around 120 
more lines, mostly concentrated at the particularly spectral line-rich 
Sgr~B2(N) source.

There are significant differences in molecular emission, pointing to both 
abundance and excitation differences throughout the region.
Seven distinct spatial locations are identified for the emitting species,
including peaks near the prominent star forming cores of Sgr B2(N), (M) and (S)
that are seen in IR-to-radio continuum images.  The other features are a 
'North Ridge' and a 'North Cloud' to the north of the Sgr B2 N-M-S cores, a 
'South-East Peak' and a 'West Ridge'.   

The column density, as evident through C$^{18}$O, peaks at
the Sgr B2(N) and (M) cores, where strong absorption is also evident in
otherwise generally-bright lines such as HCO$^{+}$, HCN and HNC.
Most molecules trace a ridge line to the west of the
Sgr B2 N-M-S cores, wrapping around the cores and extending NE to the
North Cloud.  This is most clearly evident in the species HC$_{3}$N, CH$_{3}$CN,
CH$_{3}$OH and OCS.  They are found to be closer in distribution to the cooler
dust traced by the sub-mm continuum than either the warmer dust seen in
the mid-IR or to the radio continuum.  The molecule CN, in contrast, is
reasonably uniform over the entire region mapped, aside from strong
absorption at the positions of the Sgr B2(N) and (M) cores.

\end{abstract}


\section{Introduction}

Sagittarius B2 (Sgr~B2) (G0.7-0.0) is a very massive and well-studied 
molecular cloud complex near the centre of the Galaxy. It contains multiple 
centres of (in many cases) spectacular star formation activity.
The name derives from low resolution radio observations where
Sagittarius A is the strong source at the Galactic Centre proper 
\citep{pidmin51} and B1, B2, C, D and E
refer to other radio and mid-IR features nearby
\citep{leq62, hoffreeme71}, albeit with some confusion in the literature
\citep{palgos96}. 

Sgr B2 is about 100 pc in projected distance
from the Galactic Centre and we assume its distance from the Sun to be 
identical to the latter's, $R_o$. 
An $R_o$ of $7.1 \pm 1.5$ kpc
was measured by \citet{rei+88} using a kinematic parallax method. 
A more precise distance is the `best-estimate' value of
$8.0 \pm 0.5$ kpc that \citet{rei93} derived by combining this with other data
for the Galactic Centre. The latter is corroborated by the value of 
$R_o = 7.9 \pm 0.4$ kpc that
has recently been determined from orbital solutions of a star moving around 
the super-massive central  
black hole, Sgr A$^*$ \citep{eis+03}. In the following we assume 
$R_o = 8$ kpc.

Sgr B2 presents itself as the strongest feature in images of emission in 
$^{12}$CO, $^{13}$CO \citep{oka+98} and CS \citep{tsu+99} that define the 
bar-shaped \citep{saw+04} Central Molecular Zone (CMZ), which stretches over 
the central few hundred pc of the Galaxy. The total mass of Sgr B2 is
$ > 5 \times 10^6$ M$_{\sun}$ and its peak H$_{2}$ column density 
$ \ge 10^{24}$ cm$^{-2}$ \citep{lisgol90}.

Recent star formation is indicated by a giant H~{\sc II} region \citep{meh+93},
with many compact and ultra-compact H~{\sc II} regions
\citep{gau+95}. There are multiple centres of maser emission from the water 
\citep{mcggosdep04}, hydroxyl \citep{gaucla90} and formaldehyde
\citep{mehgospal94} molecules, as well as class I and class II methanol masers
\citep{cas96, mehmen97}. The region's huge far-IR luminosity requires several 
young O-type stars as power sources, which are deeply
embedded in the molecular cores.

The star-forming centres are located in a
north-south line about 2 arcmin ($\sim$~5 pc) long, in components
labelled (north to south) Sgr~B2(N),
Sgr~B2(M) and Sgr~B2(S). All have prominent radio H~{\sc II} free-free,
millimetre and sub-millimetre \citep{gor+93, pie+00}
and infrared \citep{gol+92} emission. These cores have been extensively studied
with millimetre spectral-line surveys \citep{cum+86, tur89, num+98, num+00, 
bel+05, bel+07}.

Sgr~B2(N) is particularly rich in complex molecules:
it has been called the `Large Molecule Heimat' \citep{snykuamia94, mia+95}
or LMH. Sgr~B2(N) is considered to be in a more recent stage of star formation
than Sgr~B2(M) \citep{mia+95}, due to the presence of the complex molecules,
stronger H$_{2}$O masers, and the relatively large amount of dust.

The surrounding molecular cloud has complex kinematics. The densest
core emits around 60--65 km~s$^{-1}$, but there is a `hole' in the CO and 
CS emission around 40--50 km~s$^{-1}$
in this area \citep{sat+00}. This has been attributed \citep{has+94}
to a collision between the 40--50 km~s$^{-1}$ cloud and a cloud at 70--80
km~s$^{-1}$, triggering the star formation activity.
There is also a cloud 2 arcmin north of Sgr B2(M), and 1 arcmin north of
Sgr B2(N),
with chemical enhancement
in HNCO and HOCO$^+$ \citep{min+98}, which may be associated with the shock
from this collision.

We present here a multi-line spectral study in the 3-mm band,
of the central 12 pc of the Sgr~B2 complex, to probe the chemistry and
kinematics with a wide range of molecular tracers. The data were obtained with
a new 8 GHz wide spectrometer on the Mopra millimetre wave telescope in
Australia. These are the initial results of a project to map the CMZ in a 
variety of molecular species emitting in the 3-mm band.

\section[]{Observations and Data Reduction}

The observations were made with the 22-m Mopra radio telescope, in
on-the-fly mapping mode \citep{lad+05}.

During 2005 a new wide-bandwidth digital filterbank, MOPS, was installed.
This takes advantage of the wide bandwidth of the MMIC receiver, also
installed in 2005, which covers the range from 77 to 117 GHz and has a wide
front-end bandwidth. The MOPS can cover 8 GHz of bandwidth simultaneously,
in either a broad band mode covering the whole band in four 2.2 GHz wide 
spectra, or a zoom mode where several narrower spectral bands of 137 MHz can be
selected within the overall 8 GHz. In both the broad band and zoom modes, 
two polarisations are detected.

The observations made in 2006 June
had 1024 channels in each 2.2 GHz in the broad band mode giving channel
width 2.15 MHz or around 6.4 km~s$^{-1}$ (at 100 GHz). This is coarser velocity
sampling than desirable, but does allow the whole 8 GHz spectrum to be
covered for a single tuning, and most of the 3-mm band in 4 tunings. The narrow
band mode allowed a maximum of 8 zoom spectra of 137 MHz with 4096 channels
(0.033 MHz or 0.10 km~s$^{-1}$ at 100 GHz) to be observed (with a maximum of 
4 zooms in each 2.2 GHz section)
or 8 lines to be selected at high spectral resolution within the 8 GHz band
covered by a single tuning.

This period was while the MOPS was still being upgraded, and the performance
has since improved to allow up to 8192 channels of 0.27 MHz for each 2.2 GHz
window in the broad band mode and up to 16 zoom spectra simultaneously.
Further observations in this Mopra CMZ mapping project in 2007 and onwards
use this increased performance.\footnote{In 2007 we have observed the 
frequency range 85.3 to 93.3 GHz in broad band mode, over the area longitude
-0.2 to 0.9 deg., and latitude -0.20 to 0.12 deg., and will discuss these 
observations in a later paper.} 

The on-the-fly (OTF) observations covered an area $5 \times 5$ arcmin$^2$
centred on
($\alpha, \delta)_{\rm J2000} = 17^{\rm h}47^{\rm m}19\rlap.$\,$^{s}$\,$8, 
-28^{\circ}22'17''$, i.e.,  close to Sagittarius B2(N).
We observed this area in both the broad band and zoom modes, in several
tunings, as summarised in Table \ref{obs_log}. The broad band ranges are
calculated assuming an overall range of 8 GHz: the data cover a bit more
spectral
range with 2.2 GHz sub-band spectra separated by 2.0 GHz, but are poor at the
sub-band edges.

The OTF observations were made in a similar mode as for the 
Mopra G333-0.5/RCW106 survey
\citep{bai+06}. We used position switching for bandpass calibration with
an off-source reference position
(($\alpha, \delta)_{\rm J2000} = 17^{\rm h}51^{\rm m}03\rlap.$\,$^{s}$\,$6, 
-28^{\circ}22'47''$, or
$l = 1.093$ deg., $b = -0.735$ deg. observed before each 5 arcmin long source
scan. The spectra were read out with 2 seconds of integration time.
The scan lines were separated by about 10 arcsec, so around 30 scan lines were
needed, taking around an hour in total. Observations of SiO maser positions
were used to correct the telescope pointing, before every map, giving a 
pointing accuracy better than 10 arcsec. The system temperature was calibrated 
with a noise diode, and hot load (paddle).

\begin{table*}
\caption{Log of Mopra observations. The rms noise of the zoom bands is
given for the 9-point Hanning smoothed data with 0.13 MHz channels, whereas for 
the broad band data it is 2.15 MHz per channel.}
\begin{tabular}{cccccc}
\hline
Date     & Time & Mode & Central   & Broad band range and sub-band centres &
rms $T_{MB}$ \\
2006 Jun & UT   &      &   Freq.   &              &     \\
         & h m  &      &    GHz    &  GHz         &  K   \\
\hline
27 & 13 05 & broad & ~85.70  & ~81.70 -- 89.70 and 82.662, 84.734, 86.664,
88.736 & 0.12, 0.16, 0.11, 0.09 \\
27 & 15 20 & broad & ~94.13  & ~90.13 -- 98.13 and 91.093, 93.165,  95.095,
97.167 & 0.14, 0.16, 0.13, 0.17 \\
27 & 16 26 & broad & 102.78  & ~98.78 -- 106.78 and 99.742, 101.814, 103.744,
105.816 & 0.21, 0.33, 0.23, 0.18 \\
28 & 09 19 & broad & 109.48  & 105.48 -- 113.48 and 106.452, 108.524, 110.454,
112.526 & 0.15, 0.17, 0.24, 0.35 \\
   &       &       &         & \\
27 & 11 51 & zoom  & 85.70  & 85.146, 85.560, 86.112, 86.802, 87.354, 87.906,
88.596, 89.148 & 0.22 -- 0.29 \\
29 & 12 09 & zoom  & 94.13  & 90.678, 90.954, 91.920, 93.162, 94.404, 96.750,
97.302, 97.992 & 0.26 -- 0.38 \\
27 & 17 50 & zoom  & 102.78  & 99.328, 100.018, 100.570, 101.536, 102.088,
102.502, 103.054, 104.572 & 0.32 -- 0.42 \\
\hline
\end{tabular}
\label{obs_log}
\end{table*}

The OTF data were reduced into FITS data cubes with the {\sc livedata} and
{\sc gridzilla} 
packages\footnote{http://www.atnf.csiro.au/people/mcalabre/livedata.html}. 
{\sc livedata} is the processing software originally designed for the 
Parkes {\sc HI} multibeam survey and is used to apply system temperature 
calibration, bandpass calibration, heliocentric correction, spectral 
smoothing, and to write out the data in sdfits \citep{gar00} format.
{\sc gridzilla} is a re-gridding software package that is used to form three dimensional (RA-Dec-velocity) data cubes from bandpass-calibrated sdfits files 
(usually from {\sc livedata}).
The raw data files in 
rpfits\footnote{http://www.atnf.csiro.au/computing/software/rpfits.html} format, 
were corrected with {\sc livedata} for bandpass by the off-source spectra,
a robust second order polynomial fit to the baseline subtracted and output
as sdfits spectra. These were then regridded into data cubes using
{\sc gridzilla}, with a gaussian smoothing function for the interpolation.

The resolution of the Mopra beam varies between 36 arcsec at 86 GHz and
33 arcsec at 115 GHz \citep{lad+05}, so the resolution in the final data
varies between 39 and 36 arcsec after convolution with the 15 arcsec FWHM
gaussian in the {\sc gridzilla} interpolation. The main beam efficiency of Mopra
varies between 0.49 at 86 GHz, 0.44 at 100 GHz and 0.42 at 115 GHz
\citep{lad+05}. These parameters were measured, however by \citet{lad+05}
with a previous receiver and correlator.
Since we are more concerned in this paper with the spatial
and velocity structure, we have left the intensities throughout in this paper
in the T$_{A}^*$ scale,
without correction for the beam efficiency onto the T$_{MB}$ scale (except
for the rms noise in Table \ref{obs_log}).

The zoom mode data, with high resolution in velocity, were output as cubes
over the velocity range $-$30 to 170 km~s$^{-1}$, to reduce the file size, using
the appropriate rest frequency of the line targeted. The broad band
mode cubes were made with frequency as the third axis, over the whole 1024
channels of each sub-band. The pixel size was 12 arcsec.
The FITS cubes were then read into the {\sc MIRIAD} package for further
analysis.

The integrated spectra over the region were plotted for the broad
band mode data cubes, to identify the lines detected. Because of
ripples in the spectra, particularly at the bandpass
edges (mostly correlator ringing or 30 MHz ripples
due to standing waves between the main dish and the secondary), we did not 
always
reach the expected thermal noise sensitivity of around $T_{MB} = 0.1$~K (see
Table \ref{obs_log}). However, we identified
several dozen strong lines (see Table \ref{lines}) in the average
spectra over the broad-band data cubes. These lines were identified
using the NIST on-line database \citep{lov02}. For these lines we extracted
sub-regions of 100 channels ($\sim$ 650 km~s$^{-1}$) from the frequency cubes,
and relabelled the scale as velocity, by putting the appropriate rest frequency
into the file headers. These made low velocity resolution data cubes for
the broad band mode data.

Since the lines are broad compared to the 0.033 MHz frequency channels of the 
zoom mode data, we also made smoothed versions of the zoom mode cubes, with a 
9-point Hanning function, to improve the signal-to-noise of the spectra, and
using every fourth channel to reduce the file size, making 0.13 MHz channels.

For both the zoom mode and broad band mode data cubes, we then made integrated
emission images, by summing the data over velocity, using velocity range
over which the emission was well above the noise level. This velocity
range differed depending on whether the particular line had strong
line wings. These images are plotted and discussed in Section \ref{sec:results}.

In addition, we searched the broad band data cubes, visually plane-by-plane,
to identify line emission which was weak or not very extended, and so was
not obvious in the spectrum integrated over the whole area. There are around
120 of these additional lines, which are listed in Table \ref{extra_lines}.
Most of these are known lines, in the NIST on-line database \citep{lov02}.
The line around 107.63 GHz is attributed to multiple blended transitions of 
CH$_{3}$CH$_{2}$CN v = 1 (John Pearson, private communication).

The additional lines (Table \ref{extra_lines}) are discussed below, in 
Section \ref{sec:results}, but as the
line emission is weak and noisy, and mostly confined to a small area, the
images are not plotted here.

The mapped area of 5 arcmin corresponds to 12 pc, and the
resolution is 1.4 to 1.5 pc (using the Galactic Centre distance
$R_o = 8.0$~kpc).

\begin{table}
\caption{Summary of strong lines detected from the broad band mode
observations. The flag Z in the last column indicates lines for which there is
zoom mode data with higher velocity resolution. For most of these lines we
show integrated images in Figs. \ref{co_int}, \ref{cs_etc_int} to 
\ref{so_etc_int} and \ref{nh2cho_etc_int}.}
\begin{tabular}{ccccc}
\hline
Rough     & line ID          &                    & Exact     &     \\
Freq.     & molecule         & transition         & Rest Freq. &     \\
GHz       &                  &                    & GHz       &      \\
\hline
 81.88 &  HC$_{3}$N          & 9~--~8              &  81.881462  &   \\
 84.52 &  CH$_{3}$OH         & 5(-1,5) -- 4(0,4) E &  84.521206  &   \\
 85.14 &  OCS                & 7~--~6              &  85.139104  & Z \\
 85.27 &  CH$_{3}$CH$_{2}$OH & 6(0,6) -- 5(1,5)    &  85.265507  &   \\
 85.34 &  c-C$_{3}$H$_{2}$   & 2(1,2) -- 1(0,1)    &  85.338906  &   \\
 85.46 &  CH$_{3}$CCH        & 5(3) -- 4(3)          &  85.442600  &   \\
       &                     & 5(2) -- 4(2)          &  85.450765  &   \\
       &                     & 5(1) -- 4(1)          &  85.455665  &   \\
       &                     & 5(0) -- 4(0)          &  85.457299  &   \\
 85.53 &  HOCO$^{+}$         & 4(0,4) -- 3(0,3)    &  85.531480  & Z \\
 86.09 &  SO                 & 2(2) -- 1(1)        &  86.093983  & Z \\
 86.34 &  H$^{13}$CN         & 1~--~0 F=1-1        &  86.338735  &   \\
       &                     & 1~--~0 F=2-1        &  86.340167  &   \\
       &                     & 1~--~0 F=0-1        &  86.342256  &   \\
 86.75 &  H$^{13}$CO$^{+}$   & 1~--~0              &  86.754330  &   \\
 86.85 &  SiO                & 2~--~1 v=0          &  86.847010  & Z \\
 87.09 &  HN$^{13}$C         & 1~--~0 F=0-1        &  87.090735  &   \\
       &                     & 1~--~0 F=2-1        &  87.090859  &   \\
       &                     & 1~--~0 F=1-1        &  87.090942  &   \\
 87.32 &  C$_{2}$H           & 1~--~0 3/2-1/2 F=2-1 & 87.316925  & Z \\
       &                     & 1~--~0 3/2-1/2 F=1-0 & 87.328624  &   \\
 87.40 &  C$_{2}$H           & 1~--~0 1/2-1/2 F=1-1 & 87.402004  & Z \\
       &                     & 1~--~0 1/2-1/2 F=0-1 & 87.407165  &   \\
 87.93 &  HNCO               & 4(0,4) -- 3(0,3)    &  87.925238  & Z \\
 88.63 &  HCN                & 1~--~0 F=1-1        &  88.6304157 & Z \\
       &                     & 1~--~0 F=2-1        &  88.6318473 &   \\
       &                     & 1~--~0 F=0-1        &  88.6339360 &   \\
 89.19 &  HCO$^{+}$          & 1~--~0              &  89.188526  & Z \\
 90.66 &  HNC                & 1~--~0 F=0-1        &  90.663450  & Z \\
       &                     & 1~--~0 F=2-1        &  90.663574  &   \\
       &                     & 1~--~0 F=1-1        &  90.663656  &   \\  	
 90.98 &  HC$_{3}$N          & 10~--~9             &  90.978989  & Z \\
 91.99 &  CH$_{3}$CN         & 5(3)~--~4(3) F=6-5  &  91.971310  & Z \\
       &                     & 5(3)~--~4(3) F=4-3  &  91.971465  &   \\
       &                     & 5(2)~--~4(2) F=6-5  &  91.980089  &   \\
       &                     & 5(1)~--~4(1)        &  91.985316  &   \\
       &                     & 5(0)~--~4(0)        &  91.987089  &   \\
 92.49 &  $^{13}$CS          & 2~--~1              &  92.494303  &   \\
 93.17 &  N$_{2}$H$^{+}$     & 1~--~0 F$_1$=1-1 F=0-1   &  93.171621  & Z \\
       &                     & 1~--~0 F$_1$=1-1 F=2-2   &  93.171917  &   \\
       &                     & 1~--~0 F$_1$=1-1 F=1-0   &  93.172053  &   \\
       &                     & 1~--~0 F$_1$=2-1 F=2-1   &  93.173480  &   \\
       &                     & 1~--~0 F$_1$=2-1 F=3-2   &  93.173777  &   \\
       &                     & 1~--~0 F$_1$=2-1 F=1-1   &  93.173967  &   \\
       &                     & 1~--~0 F$_1$=0-1 F=1-2   &  93.176265  &   \\
\hline
\end{tabular}
\label{lines}
\end{table}

\begin{table}
\begin{tabular}{ccccc}
\multicolumn{2} {l} {{\bf Table \ref{lines} } continued.} &  &  & \\
\hline
Rough     & line ID          &                    & Exact     &     \\
Freq.     & molecule         & transition         & Rest Freq. &     \\
GHz       &                  &                    & GHz       &      \\
\hline
 94.41 &  $^{13}$CH$_{3}$OH  & 2(-1,2) -- 1(-1,1) E  &  94.405223  & Z \\
       &                     & 2(0,2) -- 1(0,1) A+   &  94.407129  &   \\
       &                     & 2(0,2) -- 1(0,1) E    &  94.410895  &   \\
       &                     & 2(1,1) -- 1(1,0) E    &  94.420439  &   \\
 95.17 &  CH$_{3}$OH         & 8(0,8) -- 7(1,7) A+   &  95.169516  &   \\
 95.91 &  CH$_{3}$OH         & 2(1,2) -- 1(1,1) A+   &  95.914310  &   \\
 96.41 &  C$^{34}$S          & 2~--~1                &  96.412961  &   \\
 96.74 &  CH$_{3}$OH         & 2(-1,2) -- 1(-1,1) E  &  96.739393  & Z \\
       &                     & 2(0,2) -- 1(0,1) A+   &  96.741377  &   \\
       &                     & 2(0,2) -- 1(0,1) E    &  96.744549  &   \\
       &                     & 2(1,1) -- 1(1,0) E    &  96.755507  &   \\
 97.30 &  OCS                & 8~--~7                &  97.301209  & Z \\
 97.58 &  CH$_{3}$OH         & 2(1,1) -- 1(1,0) A-   &  97.582808  &   \\
 97.98 &  CS                 & 2~--~1                 &  97.980953  & Z \\
 99.30 &  SO                 & 3(2) -- 2(1)        &  99.299905  & Z \\
100.08 &  HC$_{3}$N          & 11~--~10            & 100.076385  & Z \\
100.63 &  NH$_{2}$CN         & 5(1,4) -- 4(1,3)    & 100.62950~  & Z \\
101.48 &  H$_{2}$CS          & 3(1,3) -- 2(1,2)    & 101.477764  &   \\
102.07 &  NH$_{2}$CHO        & 5(1,5) -- 4(1,4)    & 102.064263  & Z \\
       &  H$_{2}$COH$^{+}$   & 4(0,4) -- 3(1,3)    & 102.065856  &   \\
102.55 &  CH$_{3}$CCH        & 6(3) -- 5(3)        & 102.530346  & Z \\
       &                     & 6(2) -- 5(2)        & 102.540143  &   \\
       &                     & 6(1) -- 5(1)        & 102.546023  &   \\
       &                     & 6(0) -- 5(0)        & 102.547983  &   \\
103.04 &  H$_{2}$CS          & 3(0,3) -- 2(0,2)    & 103.040416  & Z \\
104.03 &  SO$_{2}$           & 3(1,3) -- 2(0,2)    & 104.029410  &   \\
104.62 &  H$_{2}$CS          & 3(1,2) -- 2(1,1)    & 104.616988  & Z \\
105.79 &  CH$_{2}$NH         & 4(0,4) -- 3(1,3)    & 105.794057  &   \\
106.91 &  HOCO$^{+}$         & 5(0,5) -- 4(0,4)    & 106.913524  &   \\
108.89 &  CH$_{3}$OH         & 0(0,0) -- 1(-1,1) E & 108.893929  &   \\
109.17 &  HC$_{3}$N          & 12~--~11            & 109.173638  &   \\
109.25 &  SO                 & 2(3) -- 1(2)        & 109.252212  &   \\
109.46 &  OCS                & 9~--~8              & 109.463063  &   \\
109.78 &  C$^{18}$O          & 1~--~0              & 109.782173  &   \\
109.91 &  HNCO               & 5(0,5) -- 4(0,4)    & 109.905753  &   \\
110.20 &  $^{13}$CO          & 1~--~0              & 110.201353  &   \\
110.38 &  CH$_{3}$CN         & 6(3)~--~5(3) F=7-6  & 110.364469  &   \\	    	
       &                     & 6(3)~--~5(3) F=5-4  & 110.364524  &   \\    	
       &                     & 6(2)~--~5(2) F=7-6  & 110.375052  &   \\    	
       &                     & 6(1)~--~5(1) F=7-6  & 110.381404  &   \\
       &                     & 6(0)~--~5(0) F=7-6  & 110.383522  &   \\
112.36 &  C$^{17}$O          & 1~--~0              & 112.358988  &   \\
113.17 &  CN                 & 1--0 1/2-1/2 F=1/2-3/2 & 113.144192 &  \\
       &                     & 1--0 1/2-1/2 F=3/2-1/2 & 113.170528 &  \\
       &                     & 1--0 1/2-1/2 F=3/2-3/2 & 113.191317 &  \\
113.49 &  CN                 & 1--0 3/2-1/2 F=3/2-1/2 & 113.488140 &  \\
       &                     & 1--0 3/2-1/2 F=5/2-3/2 & 113.490982 &  \\
       &                     & 1--0 3/2-1/2 F=1/2-1/2 & 113.499639 &  \\
       &                     & 1--0 3/2-1/2 F=3/2-3/2 & 113.508944 &  \\
\hline
\end{tabular}
\end{table}

\begin{table}
\caption{Summary of weaker lines detected from the broad band mode
observations. We do not show the integrated images for these lines here.
The flags in the last column indicate the spatial distribution
of the line: N = peak at Sgr B2(N); M = peak at Sgr B2(M); B = peaks at both
Sgr B2(N) and Sgr B2(M); E = extended. Lines marked as `unidentified' in 
column 2, have been noted in previous surveys and included in the NIST database
\citep{lov02} with the rest frequency in column 4. Lines marked as `U' in 
column 2 are not in the NIST database. These rest frequencies are quoted to the 
nearest MHz assuming radial velocity around 61 km~s$^{-1}$ appropriate for
Sgr B2(N) and Sgr B2(M), Section \ref{sec:discuss}.}
\begin{tabular}{ccccc}
\hline
Rough     & line ID          &                    & Exact     &     \\
Freq.     & molecule         & transition         & Rest Freq. &     \\
GHz       &                  &                    & GHz       &     \\
\hline
 82.46 &    CH$_{3}$OCH$_{3}$      &  11(1,10)--11(0,11)  &    82.456986 & N \\
       &                           &   AE+EA              &              & \\
       &    CH$_{3}$CH$_{2}$CN     &  9(1,8)--8(1,7)      &    82.458611 & \\
       &    CH$_{3}$OCH$_{3}$      &  11(1,10)--11(0,11)  &    82.458660 & \\
       &                           &   EE                 &              & \\
       &    CH$_{3}$OCH$_{3}$      &  11(1,10)--11(0,11)  &    82.460334 & \\
       &                           &   AA                 &              & \\
 83.69 &    SO$_{2}$               &  8(1,7)--8(0,8)      &    83.688086 & M \\
 85.09 &    NH$_{2}$CHO            &  4(2,2)--3(2,1)      &    85.093268 & N \\
 85.69 &     U                     &                      &    85.686~~~ & B \\
 87.85 &    NH$_{2}$CHO            &  4(1,3)--3(1,2)      &    87.848871 & E \\
 88.17 &    H$^{13}$CCCN           &        10--9         &    88.166808 & N \\
 88.24 &    HNCO                   &  4(1,3)--3(1,2)      &    88.239027 & N \\
 89.32 &    CH$_{3}$OCHO           &  8(1,8)--7(1,7) E    &    89.314589 & N \\
       &    CH$_{3}$OCHO           &  8(1,8)--7(1,7) A    &    89.316668 & \\
 89.57 &    CH$_{3}$CH$_{2}$CN     &  10(6)--9(6)         &    89.562318 & N \\
       &    CH$_{3}$CH$_{2}$CN     &  10(7)--9(7)         &    89.565034 & \\
       &    CH$_{3}$CH$_{2}$CN     &  10(5)--9(5)         &    89.568100 & \\
       &    CH$_{3}$CH$_{2}$CN     &  10(8)--9(8)         &    89.573057 & \\
 89.59 &    CH$_{3}$CH$_{2}$CN     &  10(4,7)--9(4,6)     &    89.590033 & N \\
       &    CH$_{3}$CH$_{2}$CN     &  10(4,6)--9(4,5)     &    89.591017 & \\
 90.45 &    CH$_{3}$CH$_{2}$CN     &  10(2,8)--9(2,7)     &    90.453354 & N \\
 90.60 &    HC$^{13}$CCN           &        10--9         &    90.593059 & E \\
       &    HCC$^{13}$CN           &        10--9         &    90.601791 & \\
 91.20 &    HC$_{3}$N            & 10--9 v$_{6}$=1 $l$=1 f &   91.199796 & N \\
       &    HC$_{3}$N            & 10--9 v$_{7}$=l $l$=1 e &   91.202607 & \\
 91.33 &    HC$_{3}$N            & 10--9 v$_{7}$=1 $l$=1 f &   91.333308 & N \\
 91.55 &    CH$_{3}$CH$_{2}$CN     &  10(1,9)--9(1,8)     &    91.549117 & N \\
       &    SO$_{2}$               &  18(5,13)--19(4,16)  &    91.550442 & \\
 91.60 &    unidentified           &                      &    91.603~~~ & N \\
 91.84 &    unidentified           &                      &    91.848~~~ & N \\
 92.04 &     U                     &                      &    92.035~~~ & M \\
 92.26 &    CH$_{3}$CN        & 5(0)--4(0) v$_{8}$=1 $l$=1 &   92.261440 & N \\
       &    CH$_{3}$CN        & 5(2)--4(2) v$_{8}$=1 $l$=1 &   92.263992 & \\
 92.43 &    CH$_{2}$CHCN           &  10(1,10)--9(1,9)    &    92.426260 & N \\
 93.60 &    CH$_{3}$CHO            &  5(-1,5)--4(-1,4) E  &    93.595238 & E \\
 93.87 &    CCS                    &  8(7)--7(6)          &    93.870098 & E \\
       &    NH$_{2}$CHO            &  3(2,2)--4(1,3)      &    93.871700 & \\
 94.28 &    CH$_{2}$CHCN           &  10(0,10)--9(0,9)    &    94.276640 & N \\
 94.54 &    CH$_{3}$OH             &  8(3,5)--9(2,7) E    &    94.541806 & N \\
 94.76 &     U                     &                      &    94.759~~~ & N \\
 94.91 &    CH$_{2}$CHCN           &  10(4,7)--9(4,6)     &    94.913139 & N \\
       &    CH$_{2}$CHCN           &  10(4,6)--9(4,5)     &    94.913250 & \\
 94.92 &     U                     &                      &    94.924~~~ & N \\
 94.94 &     U                     &                      &    94.940~~~ & N \\
 95.15 &    unidentified           &                      &    95.145~~~ & E \\
 95.33 &    CH$_{2}$CHCN           &  10(2,8)--9(2,7)     &    95.325490 & N \\
 95.44 &    CH$_{3}$CH$_{2}$CN     &  11(1,11)--10(1,10)  &    95.442479 & N \\
       &    t-CH$_{3}$CH$_{2}$OH   &  16(2,14)--16(1,13)  &    95.444067 & \\
 95.95 &    CH$_{3}$CHO            &  5(0,5)--4(0,4) E    &    95.947439 & E \\
 95.96 &    CH$_{3}$CHO            &  5(0,5)--4(0,4) A++  &    95.963465 & E \\
\hline
\end{tabular}
\label{extra_lines}
\end{table}

\begin{table}
\begin{tabular}{ccccc}
\multicolumn{2} {l} {{\bf Table \ref{extra_lines} } continued.} &  &  & \\
\hline
Rough     & line ID          &                    & Exact     &     \\
Freq.     & molecule         & transition         & Rest Freq. &     \\
GHz       &                  &                    & GHz       &     \\
\hline
 96.49 &    CH$_{3}$OH             &  2(1,2)--1(1,1) E    &    96.492164 & N \\
       &                           &  v$_{t}$=1           &              & \\
       &    CH$_{3}$OH             &  2(0,2)--1(0,1) E    &    96.493553 & \\
       &                           &  v$_{t}$=1           &              & \\
 96.98 &    O$^{13}$CS             &        8--7          &    96.988123 & E \\
 97.70 &    SO$_{2}$               &  7(3,5)--8(2,6)      &    97.702340 & M \\
 97.72 &    $^{34}$SO              &  3(2)--2(1)          &    97.715401 & M \\
 98.18 &    CH$_{3}$CH$_{2}$CN     &  11(2,10)--10(2,9)   &    98.177578 & N \\
       &    CH$_{3}$OCHO           &  8(7,1)--7(7,0) E    &    98.182199 & \\
 98.90 &    CH$_{3}$CHO            &  5(1,4)--4(1,3) A--  &    98.900951 & E \\
 99.02 &     U                     &                      &    99.021~~~ & M \\
 99.65 &    HC$^{13}$CCN           &        11--10        &    99.651863 & N \\
       &    HCC$^{13}$CN           &        11--10        &    99.661471 & \\
 99.68 &    CH$_{3}$CH$_{2}$CN     &  11(2,9)--10(2,8)    &    99.681511 & N \\
100.03 &    SO                     &  4(5)--4(4)          &   100.029565 & B \\
100.32 &    HC$_{3}$N          & 11--10 v$_{7}$=1 $l$=1 e &   100.322349 & N \\
100.41 &     U                     &                      &   100.406~~~ & M \\
100.46 &    CH$_{3}$OCH$_{3}$      &  6(2,5)--6(1,6)      &   100.460412 & N \\
       &                           &  EA+AE               &              & \\
       &    CH$_{3}$OCH$_{3}$      &  6(2,5)--6(1,6) EE   &   100.463066 & \\
       &    CH$_{3}$OCH$_{3}$      &  6(2,5)--6(1,6) AA   &   100.465708 & \\
100.61 &    CH$_{3}$CH$_{2}$CN     &  11(1,10)--10(1,9)   &   100.614291 & N \\
100.71 &    HC$_{3}$N           & 11--10 v$_{7}$=2 $l$=0 &  100.708837 & N \\
       &    HC$_{3}$N          & 11--10 v$_{7}$=2 $l$=2 e & 100.710972 & \\
       &    HC$_{3}$N          & 11--10 v$_{7}$=2 $l$=2 f & 100.714306 & \\
100.88 &    SO$_{2}$              &  2(2,0)--3(1,3)       &   100.878105 & M \\
101.03 &    CH$_{2}$CO            &  5(2,4)--4(2,3)       &   101.024438 & N \\
       &    CH$_{3}$SH            &  4(-1)--3(-1) E       &   101.029750 & \\
101.14 &    CH$_{3}$SH            &  4(0)--3(0) A         &   101.139160 & E \\
       &    CH$_{3}$SH            &  4(0)--3(0) E         &   101.139650 & \\
101.33 &    H$_{2}$CO             &  6(1,5)--6(1,6)       &   101.332987 & N \\
101.98 &    CH$_{2}$CO            &  5(1,4)--4(1,3)       &   101.981426 & E \\
103.57 &    CH$_{2}$CHCN          &  11(0,11)--10(0,10)   &   103.575401 & N \\
104.05 &  CH$_{3}$CH$_{2}$CN      &  12(1,12)--11(1,11)   &   104.051278 & N \\
104.21 &    CH$_{2}$CHCN          &  11(2,10)--10(2,9)    &   104.212655 & N \\
104.24 &    SO$_{2}$              &  10(1,9)--10(0,10)    &   104.239293 & B \\
104.30 &    CH$_{3}$OH            &  11(-1,11)--10(-2,9)  &   104.300396 & N \\
       &                          &   E                   &              & \\
104.35 &    CH$_{3}$OH            &  10(4,7)--11(3,8)     &   104.354861 & N \\
       &                          &   A-                  &              & \\
104.41 &    CH$_{2}$CHCN          &  11(5,*)--10(5,*)     &   104.408903 & N \\
       &    CH$_{3}$OH            &  10(4,6)--11(3,9)     &   104.410489 & \\
       &                          &   A+                  &              & \\
       &    CH$_{2}$CHCN          &  11(4,8)--10(4,7)     &   104.411262 & \\
       &    CH$_{2}$CHCN          &  11(4,7)--10(4,6)     &   104.411485 & \\
104.49 &   t-CH$_{3}$CH$_{2}$OH   &  7(0,7)--6(1,6)       &   104.487254 & E \\
104.80 &   t-CH$_{3}$CH$_{2}$OH   &  5(1,5)--4(0,4)       &   104.808618 & E \\
104.96 &    CH$_{2}$CHCN          &  11(2,9)--10(2,8)     &   104.960550 & N \\
105.06 &    CH$_{3}$OH            &  13(1,13)--12(2,10)   &   105.063761 & N \\
       &                          &   A+  	          &              & \\
105.30 &     U                    &                       &   105.299~~~ & M \\
105.46 &    NH$_{2}$CHO           &  5(0,5)--4(0,4)       &   105.464216 & E \\
       &    CH$_{3}$CH$_{2}$CN    &  12(0,12)--11(0,11)   &   105.469300 & \\
105.54 &     U                    &                       &   105.537~~~ & N \\
105.57 &    CH$_{3}$OH            &  14(-2,13)--14(1,13)  &   105.576385 & N \\
       &                          &   E                   &              & \\
105.77 &    CH$_{3}$OCH$_{3}$     &  13(1,12)--13(0,13)   &   105.768276 & N \\
       &                          &   EA+AE               &              & \\
       &    CH$_{3}$OCH$_{3}$     &  13(1,12)--13(0,13)   &   105.770340 & \\
       &                          &   EE                  &              & \\
       &    CH$_{3}$OCH$_{3}$     &  13(1,12)--13(0,13)   &   105.772403 & \\
       &                          &   AA                  &              & \\
\hline
\end{tabular}
\end{table}

\begin{table}
\begin{tabular}{ccccc}
\multicolumn{2} {l} {{\bf Table \ref{extra_lines} } continued.} &  &  & \\
\hline
Rough     & line ID          &                    & Exact     &     \\
Freq.     & molecule         & transition         & Rest Freq. &     \\
GHz       &                  &                    & GHz       &     \\
\hline
105.97 &    NH$_{2}$CHO           &  5(2,4)--4(2,3)       &   105.972593 & N \\
106.11 &     U                    &                       &   106.107~~~ & N \\
106.13 &    NH$_{2}$CHO           &  5(3,3)--4(3,2)       &   106.134418 & B \\
106.35 &    CCS                   &  9(8)--8(7)           &   106.347740 & E \\
106.54 &    NH$_{2}$CHO           &  5(2,3)--4(2,2)       &   106.541674 & N \\
106.64 &    CH$_{2}$CHCN          &  11(1,10)--10(1,9)    &   106.641394 & N \\
106.74 &    $^{34}$SO             &  2(3)--1(2)           &   106.743374 & M \\
107.01 &    CH$_{3}$OH            &  3(1,3)--4(0,4) A+    &   107.013770 & B \\
107.04 &     U                    &                       &   107.042~~~ & N \\
107.06 &    SO$_{2}$              &  27(3,25)--26(4,22)   &   107.060225 & M \\
107.10 &    unidentified          &                       &   107.1032~~ & E \\
107.16 &    CH$_{3}$OH            &  15(-2,14)--15(1,14)  &   107.159915 & N \\
       &                          &   E                   &              & \\
107.19 &    $^{13}$CH$_{3}$CN     &  6(1)--5(1)           &   107.194547 & N \\
       &    $^{13}$CH$_{3}$CN     &  6(0)--5(0)           &   107.196564 & \\
107.48 &    CH$_{3}$CH$_{2}$CN    &  17(2,16)--17(1,17)   &   107.481465 & N \\
       &    CH$_{3}$CH$_{2}$CN    &  12(7,*)--11(7,*)     &   107.485181 & \\
       &    CH$_{3}$CH$_{2}$CN    &  12(6,*)--11(6,*)     &   107.486962 & \\
       &    CH$_{3}$CH$_{2}$CN    &  12(8,*)--11(8,*)     &   107.491579 & \\
107.50 &    CH$_{3}$CH$_{2}$CN    &  12(5,8)--11(5,7)     &   107.502426 & N \\
       &    CH$_{3}$CH$_{2}$CN    &  12(5,7)--11(5,6)     &   107.502473 & \\
107.54 &    CH$_{3}$CH$_{2}$CN    &  12(11,*)--11(11,*)   &   107.539857 & N \\
       &    CH$_{3}$OCHO          &  9(2,8)--8(2,7) A     &   107.543746 & \\
       &    CH$_{3}$CH$_{2}$CN    &  12(4,9)--11(4,8)     &   107.543924 & \\
       &    CH$_{3}$CH$_{2}$CN    &  12(4,8)--11(4,7)     &   107.547599 & \\
107.59 &    CH$_{3}$CH$_{2}$CN    &  12(3,10)--11(3,9)    &   107.594046 & N \\
107.63 &    CH$_{3}$CH$_{2}$CN    &  v = 1, multiple      &   107.636~~~ & N \\
107.73 &    CH$_{3}$CH$_{2}$CN    &  12(3,9)--11(3,8)     &   107.734738 & N \\
107.84 &    SO$_{2}$              &  12(4,8)--13(3,11)    &   107.843478 & M \\
108.65 &    $^{13}$CN             &  1/2--1/2 F=2-1,      &   108.651297 & E \\
       &                          & F$_{1}$=0, F$_{2}$=1-0 &            & \\
       &    $^{13}$CN             &  1/2--1/2 F=2-2,      &   108.657646 & \\
       &                          & F$_{1}$=0, F$_{2}$=1-1 &            & \\
       &    $^{13}$CN             &  1/2--1/2 F=1-2,      &   108.658948 & \\
       &                          & F$_{1}$=1, F$_{2}$=1-1 &            & \\
108.71 &    HC$^{13}$CCN          &        12--11         &   108.710523 & N \\
       &    HCC$^{13}$CN          &        12--11         &   108.721008 & \\
108.78 &    $^{13}$CN             &  3/2--1/2 F=3-2,      &   108.780201 & E \\
       &                          &  F$_{1}$=1,F$_{2}$=2-1 &            & \\
       &    $^{13}$CN             &  3/2--1/2 F=2-1       &   108.782374 & \\
       &                          &  F$_{1}$=1,F$_{2}$=2-1 &            & \\
       &    $^{13}$CN             &  3/2--1/2 F=1-0       &   108.786982 & \\
       &                          &  F$_{1}$=1,F$_{2}$=2-1 &            & \\
\hline
\end{tabular}
\end{table}

\begin{table}
\begin{tabular}{ccccc}
\multicolumn{2} {l} {{\bf Table \ref{extra_lines} } continued.} &  &  & \\
\hline
Rough     & line ID          &                    & Exact     &     \\
Freq.     & molecule         & transition         & Rest Freq. &     \\
GHz       &                  &                    & GHz       &     \\
\hline
108.94 &    CH$_{3}$CH$_{2}$CN    &  12(2,10)--11(2,9)    &   108.940596 & N \\
109.14 &    CH$_{3}$OH            &  26(0,26)--26(-1,26)  &   109.137570 & N \\
       &                          &   E                   &              & \\ 
109.15 &    CH$_{3}$OH            &  16(-2,15)--16(1,15)  &   109.153210 & N \\
       &                          &   E                   &              & \\
109.44 &    HC$_{3}$N          & 12--11 v$_{6}$=1 $l$=1 f & 109.438572 & N \\
       &    HC$_{3}$N          & 12--11 v$_{7}$=1 $l$=1 e & 109.441944 & \\
109.49 &    HNCO                &  5(1,5)--4(1,4)         &   109.496007 & E \\
109.60 &    HC$_{3}$N          & 12--11 v$_{7}$=1 $l$=1 f & 109.598751 & B \\
109.65 &   CH$_{3}$CH$_{2}$CN   &  12(1,11)--11(1,10)     &   109.650301 & N \\
109.75 &    NH$_{2}$CHO         &  5(1,4)--4(1,3)         &   109.753499 & E \\
       &    SO$_{2}$            &  17(5,13)--18(4,14)     &   109.757587 & \\
109.87 &    HC$_{3}$N          & 12--11 v$_{7}$=2 $l$=2 f & 109.870188 & B \\
       &    HNCO               & 5(1,5)--4(1,4) v$_{6}$=1 & 109.870278 & \\
       &    HNCO                &  5(2,4)--4(2,3)         &   109.872366 & \\
       &    HNCO                &  5(2,3)--4(2,2)         &   109.872773 & \\
110.29 &    HNCO                &  5(1,4)--4(1,3)         &   110.298098 & E \\
110.33 &    CH$_{3}^{~13}$CN    &  6(2)--5(2)             &   110.320438 & N \\
       &    CH$_{3}^{~13}$CN    &  6(1)--5(1)             &   110.326795 & \\
       &    CH$_{3}^{~13}$CN    &  6(0)--5(0)             &   110.328914 & \\
       &    CH$_{3}$CN          &  6(5)--5(5) F=7-6       &   110.330627 & \\
       &    CH$_{3}$CN          &  6(5)--5(5) F=5-4       &   110.330872 & \\
110.35 &    CH$_{3}$CN          &  6(4)--5(4) F=7-6       &   110.349659 & E \\
       &    CH$_{3}$CN          &  6(4)--5(4) F=5-4       &   110.349797 & \\
110.69 &    CH$_{3}$CN          &  6(2)--5(2) v$_{8}$=1 &   110.695506 & N \\
       &                        &   $l$=-1              &              & \\
       &    CH$_{3}$CN          &  6(4)--5(4) v$_{8}$=1 &   110.698701 & \\
       &                        &   $l$=1               &              & \\
110.71 &    CH$_{3}$CN          &  6(1)--5(1) v$_{8}$=1 &   110.706251 & N \\
       &                        &   $l$=-1              &              & \\
       &    CH$_{3}$CN          &  6(3)--5(3) v$_{8}$=1 &   110.709313 & \\
       &                        &   $l$=+1              &              & \\
       &    CH$_{3}$CN          &  6(0)--5(0) v$_{8}$=1 &   110.712166 & \\
       &                        &   $l$=1               &              & \\
       &    CH$_{3}$CN          &  6(2)--5(2) v$_{8}$=1 &   110.716212 & \\
       &                        &   $l$=1               &              & \\
111.29 &    CH$_{3}$OH          &  7(2,5)--8(1,8) A+    &   111.289601 & N \\
112.64 &    CH$_{3}$CH$_{2}$CN  &  13(1,13)--12(1,12)   &   112.646233 & N \\
112.84 &     U                  &                       &   112.839~~~ & N \\
113.12 &    CN                  & 1--0 J=1/2-1/2        &   113.123337 & E \\
       &                        & F=1/2-1/2             &              & \\
\hline
\end{tabular}
\end{table}

\section[]{Results}
\label{sec:results}

We present here the integrated emission images, analysis of these images
and the data cubes. An area of $5.2 \times 5.2$ arcmin$^{2}$ is plotted for each
image, generally using the same rifgt acsension and declination scale for the 
axes, to allow easy comparison. Images from the
broad-band 109-GHz tuning cover a region with slight offset in right
ascension to the other images, as these data had a small ($\sim$ 24 arcsec) 
systematic shift in 
position, which has been corrected. (The origin of this offset is not clear, 
but is probably due to a poor pointing correction made just before these data 
were collected). Integrated emission images
use the mean of broad band and zoom mode, if the zoom mode data were available
(or we use the better image if one of the broad or zoom data had problems).
The OTF scanning direction was in right ascension, and some of the images
show stripe artifacts in this direction.

We plot positions
of radio sources with crosses, to make the alignment of different features more
obvious. The radio positions are taken from the 9.1 GHz continuum peaks
of \citet{hun+99} obtained with the Australia Telescope Compact Array (ATCA),
supplemented by a few positions of peaks from 20-cm Very Large Array (VLA)
data for sources outside the area of \citet{hun+99}. Note in particular
that the peak near the centre is Sgr~B2(N) at J2000 17 47 20.4, -28 22 12,
with Sgr~B2(M) at 17 47 20.5, -28 23 05 and Sgr~B2(S) at 17 47 20.5, -28 23 44
in a line almost exactly to the south (labelled in Fig. \ref{co_int}).
We also plot with open squares, some
mid-infrared sources with positions fitted from the 21\,$\mu$m (band E)
Midcourse Space Experiment (MSX) data \citep{pri+01}. Note that the four mid-IR
peaks all correspond to radio sources, including Sgr~B2(M) and Sgr~B2(S) but
that Sgr~B2(N) does not have strong emission at 21\,$\mu$m. See Section \ref{sec:discuss} for plots of the radio and mid~IR
continuum, and discussion of the alignment of the different molecular lines
with the radio and mid-IR continuum features.

In the figure captions we give peak integrated brightness and contour level
steps, in K km~s$^{-1}$, on the T$_{A}^*$ scale, that is not corrected for beam
efficiency. The contours are in equal linear steps. In most cases the lowest
contour level is the same as the step size, but this is not the case for some
of the strongest lines (such as $^{12}$CO) where the whole 5 arcmin square
area is filled with emission well above the zero level.

In this section we present maps for many of the lines measured.  We
summarise these line maps in Table \ref{lines}, 
whereas in Table \ref{extra_lines} we list all the
other (weaker) lines detected, for which we do not present maps.
We also discuss the velocities and line widths at the emission peaks for the
various maps presented.  These are summarised in Table \ref{peaks_list}.
We use the rough frequency in GHz,
rounded to two decimal places, in the figures, Tables
\ref{lines} and \ref{extra_lines} and text below, as a convenient shorthand to
refer to the lines.

\subsection{$^{13}$CO, C$^{18}$O and C$^{17}$O}

\begin{figure*}
\includegraphics[width = 5.2 cm,angle=-90]{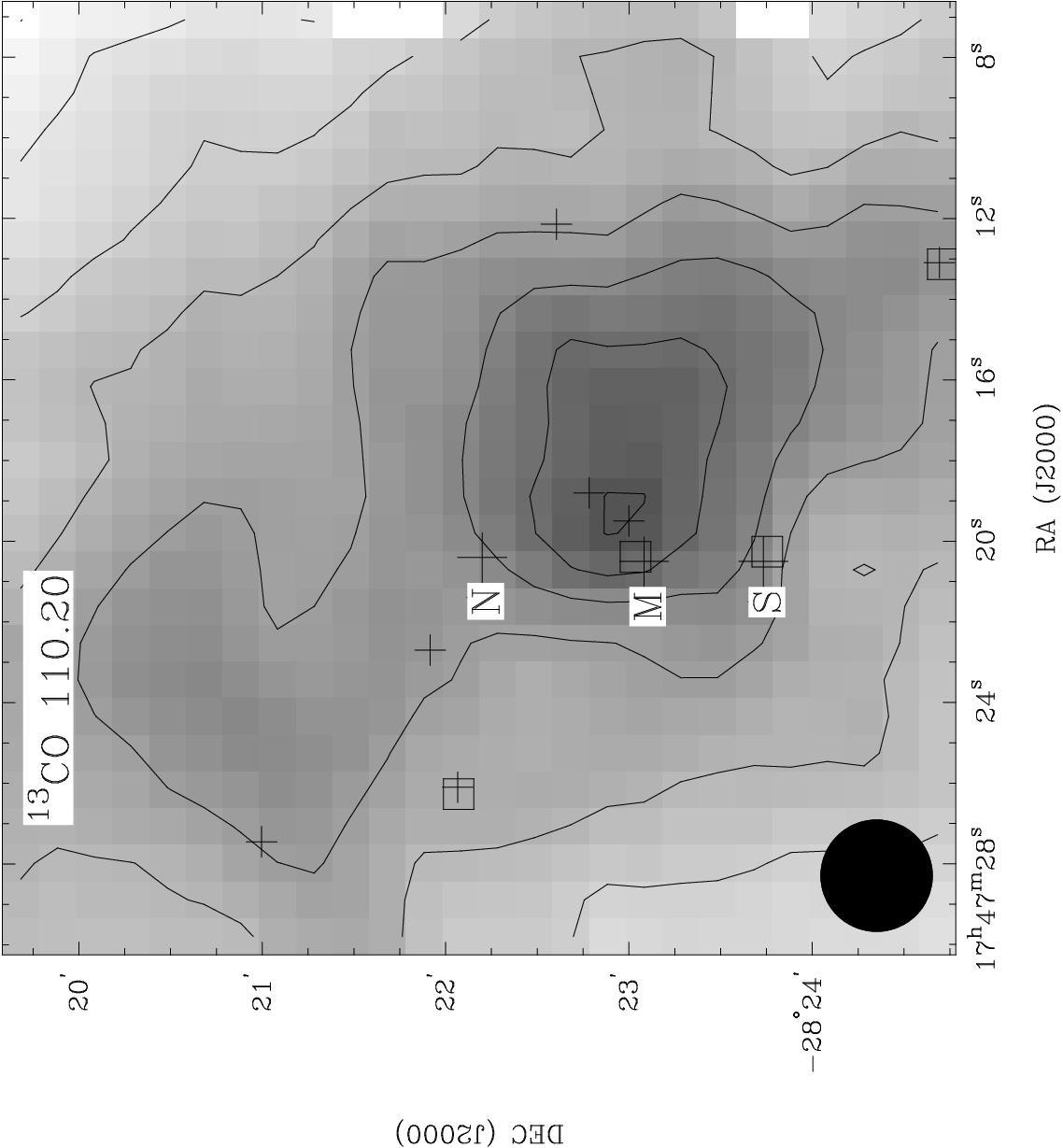}
\includegraphics[width = 5.2 cm,angle=-90]{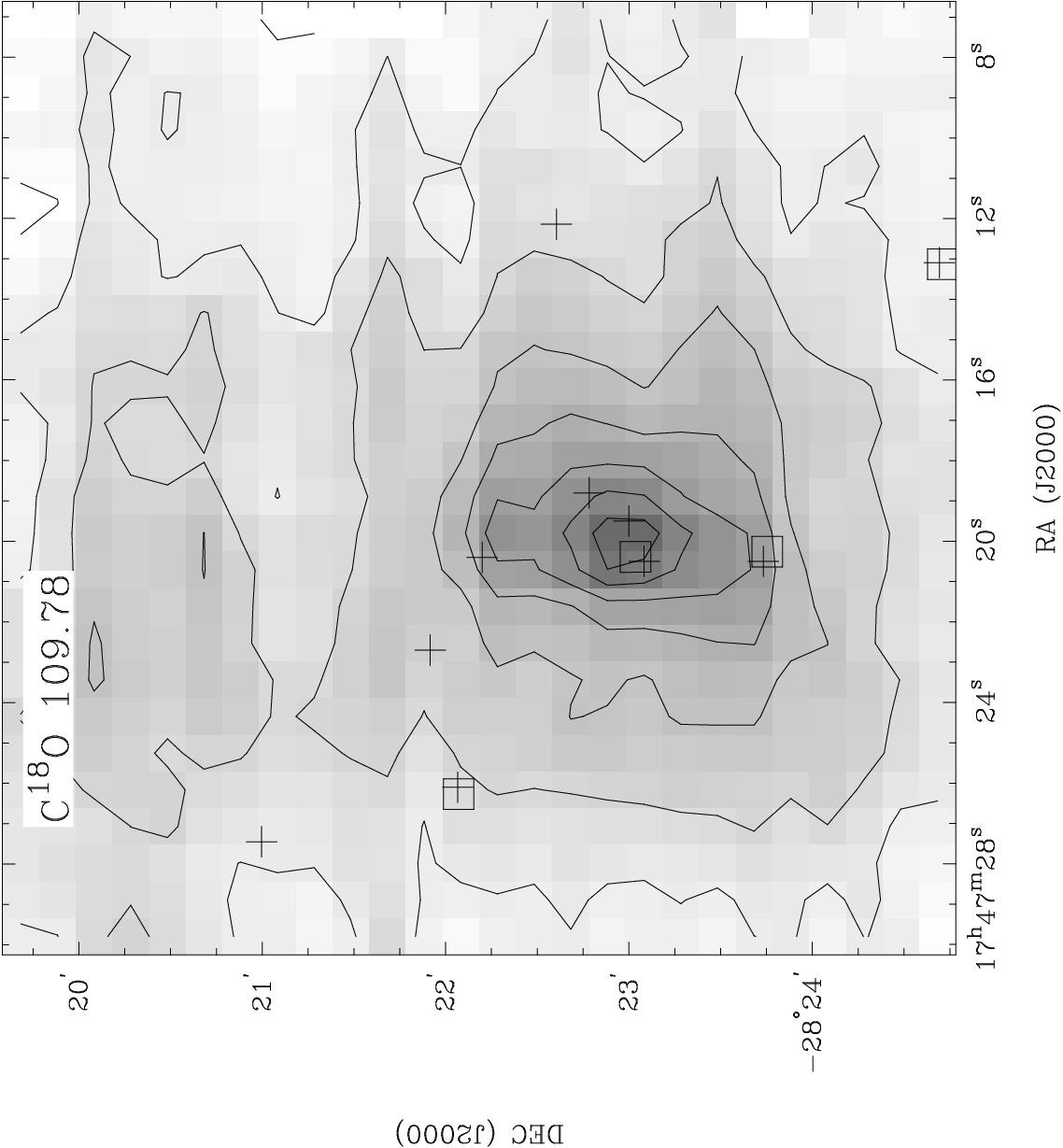}
\caption{Integrated emission for $^{13}$CO and C$^{18}$O. In this, and
subsequent images, the crosses indicate positions of radio peaks, as described 
in Section \ref{sec:results}, including in particular the positions of 
Sgr~B2(N),
(M) and (S). The squares show mid-IR sources. The optically thin C$^{18}$O
peaks near Sgr~B2(M), while the $^{13}$CO shows the widespread diffuse
emission. The peak brightness and contour steps are 240 K km~s$^{-1}$ and 20 
K km~s$^{-1}$
for $^{13}$CO, and 48 K km~s$^{-1}$ and 5 K km~s$^{-1}$ for C$^{18}$O. The 
beam size is shown
in the bottom left corner of the $^{13}$CO image.}
\label{co_int}
\end{figure*}

The isotopic carbon monoxide
$^{13}$CO 1~--~0 (110.20 GHz) and C$^{18}$O 1~--~0 (109.78 GHz)
integrated emission is shown in Fig. \ref{co_int}.
The $^{13}$CO emission is optically thick in the densest regions, with the ratio
of the peak integrated emission of $^{13}$CO/C$^{18}$O of around 5, rather
than $\sim$~9
for optically thin emission near the Galactic Centre
\citep{lisgol89}. The dense peaks are therefore
better traced by C$^{18}$O, which shows two peaks associated with Sgr~B2(M)
and Sgr~B2(N), with fitted positions (J2000) 17 47 20.3, -28 23 06 and
17 47 19.5, -28 22 15,
LSR velocities 63 and 68 km~s$^{-1}$ and full width at half maximum 21 and 22 
km~s$^{-1}$
respectively.

\begin{figure*}
\includegraphics[width = 17 cm]{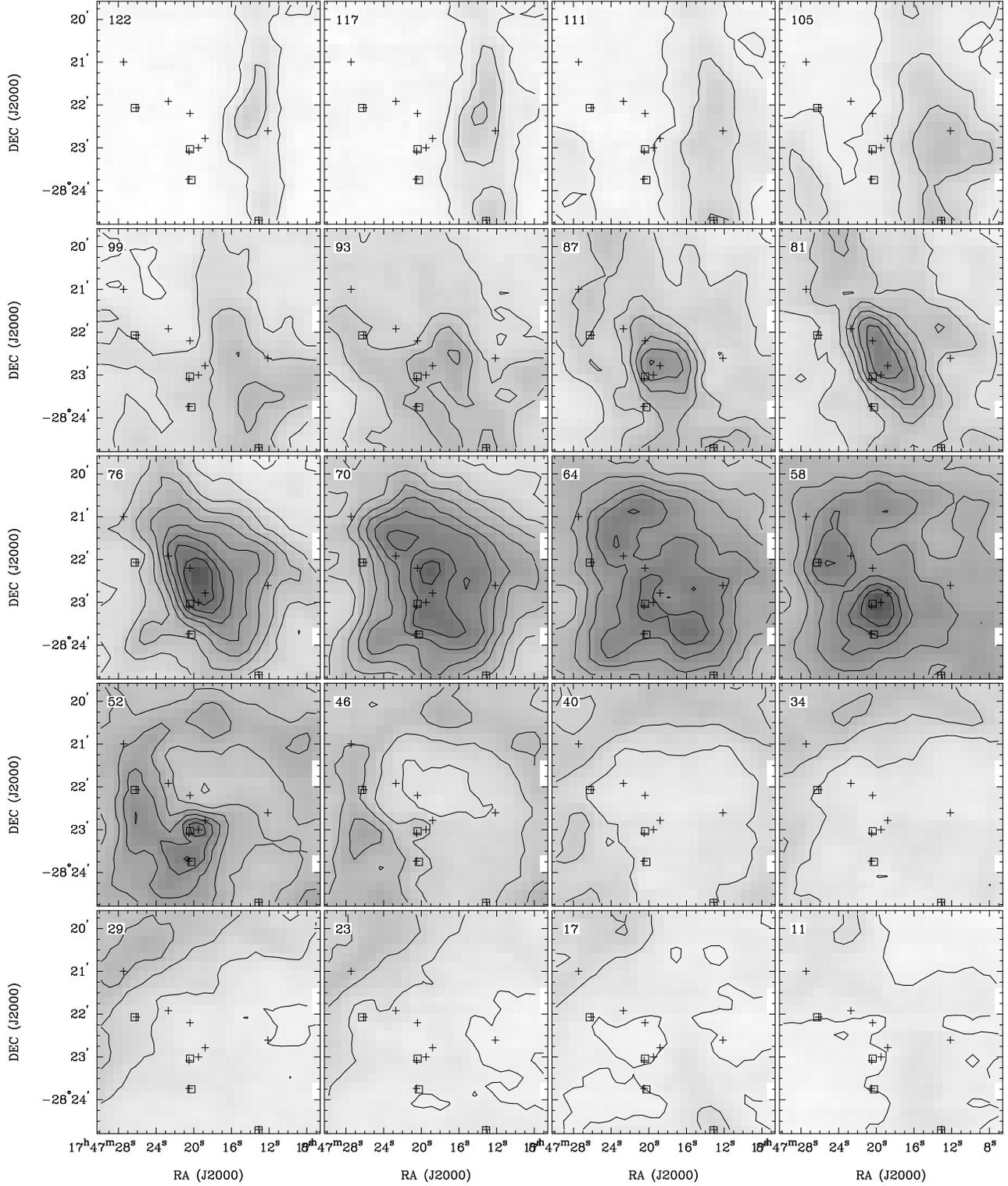}
\caption{Velocity channel images of $^{13}$CO, separated by the channel spacing
of 6 km~s$^{-1}$. Note the west ridge peaking at
117 km~s$^{-1}$, the north ridge at 64 km~s$^{-1}$, and how the hole at 
40 -- 50 km~s$^{-1}$ matches
the clump at 70 -- 80 km~s$^{-1}$. The contours are steps of 0.5 K, and the peak
is 5.45 K. The crosses and squares are the same as for Figure \ref{co_int}.}
\label{13co_planes}
\end{figure*}

The $^{13}$CO data cube, with
intensity as a function of velocity (Fig. \ref{13co_planes}), agrees well
with the results of \citet{sat+00} and \citet{has+07}, showing the
low velocity `hole' at
40--50 km~s$^{-1}$ and the high velocity `clump' at 70--80 km~s$^{-1}$. 
However, the broad
band data here are with poorer velocity and spatial resolution than that of
\citet{sat+00} or \citet{has+07}, so we do not resolve
details in the spatial and velocity structure that they attribute to their
cloud-cloud collision model \citep{has+94}. The integrated $^{13}$CO and
C$^{18}$O images (Fig. \ref{co_int}) also show the northern emission ridge
or `Edge' \citep{has+94} with peak at 17 47 24.2, -28 20 49 (in $^{13}$CO) 
with central velocity 65 km~s$^{-1}$ (width 42 km~s$^{-1}$) from $^{13}$CO and 
63 km~s$^{-1}$ (width 36 km~s$^{-1}$)
from C$^{18}$O. There is also the higher-velocity
ridge to the west in $^{13}$CO (Fig. \ref{13co_planes})
with peak at 17 47 14.0 -28 22 14,
velocity 109 km~s$^{-1}$ (width 32 km~s$^{-1}$).

We have also imaged the weaker C$^{17}$O 1~--~0 (112.36 GHz)
data, which shows the densest CO peak near Sgr~B2(M) at around 64 km~s$^{-1}$.
However, the C$^{17}$O data are affected by the bandpass ripples,
so we do not show the integrated image here, or consider further quantitative
analysis (such as line ratios).

\subsection{CS, $^{13}$CS and C$^{34}$S}

\begin{figure*}
\includegraphics[width = 5.2 cm,angle=-90]{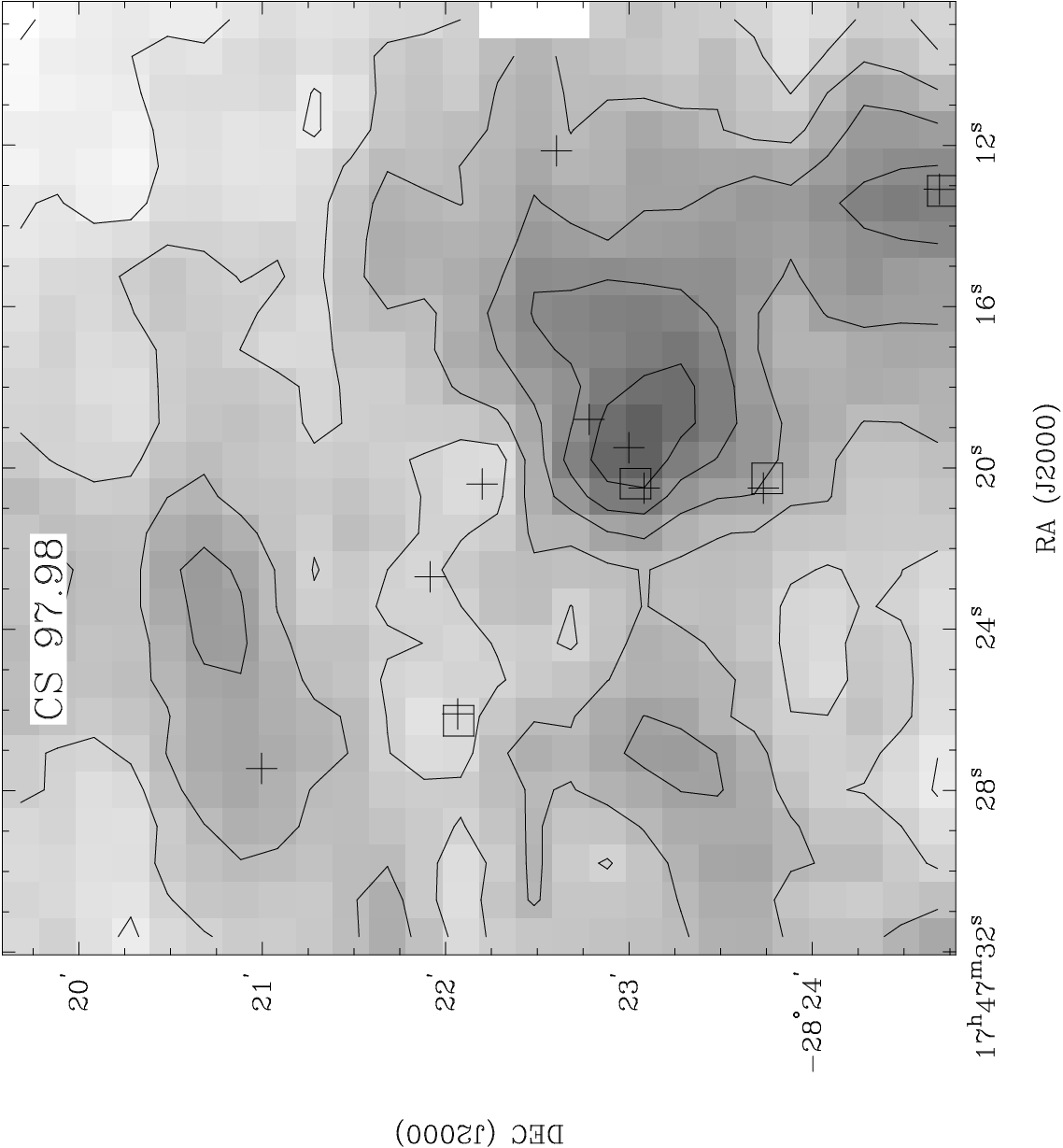}
\includegraphics[width = 5.2 cm,angle=-90]{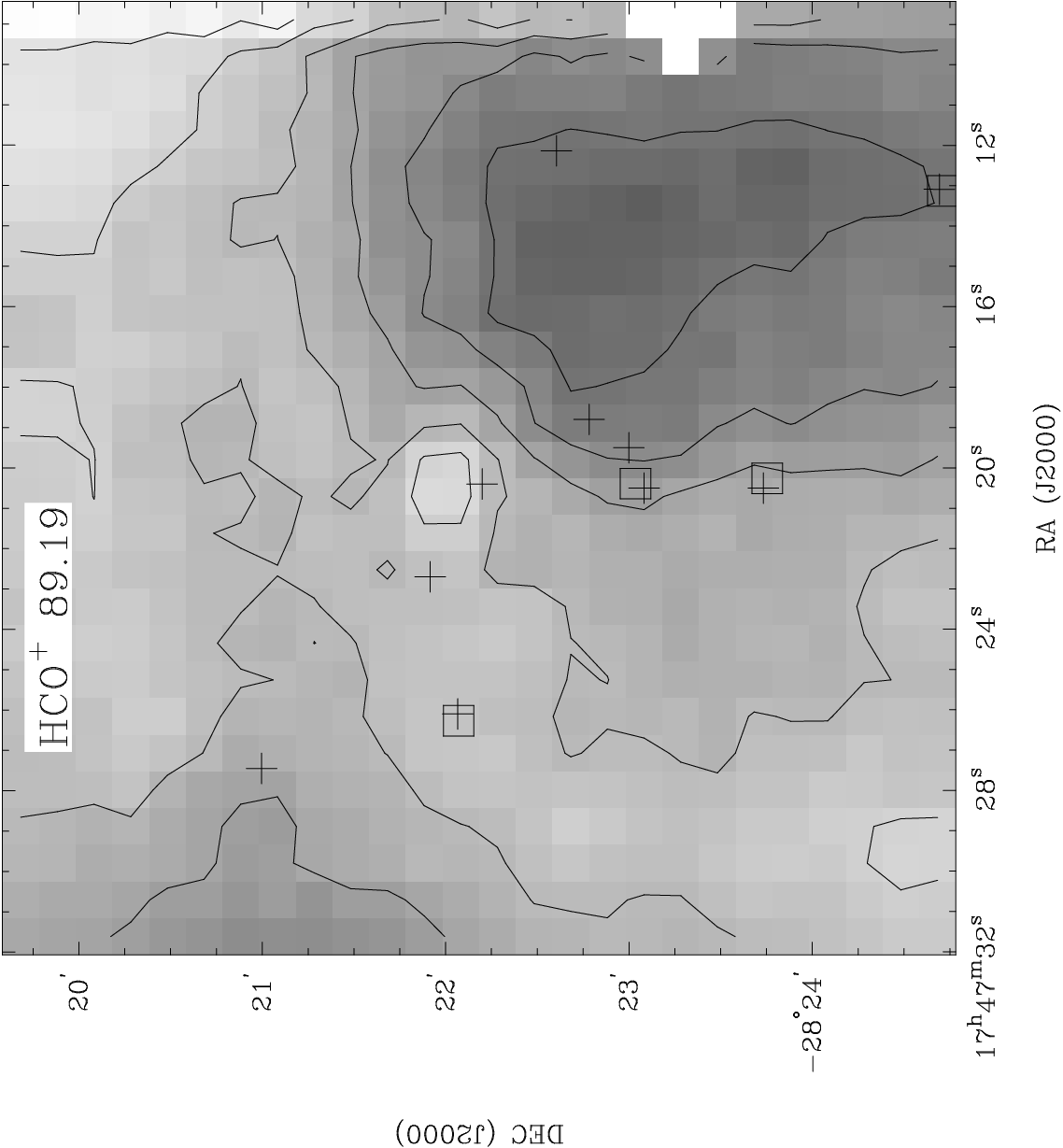}
\includegraphics[width = 5.2 cm,angle=-90]{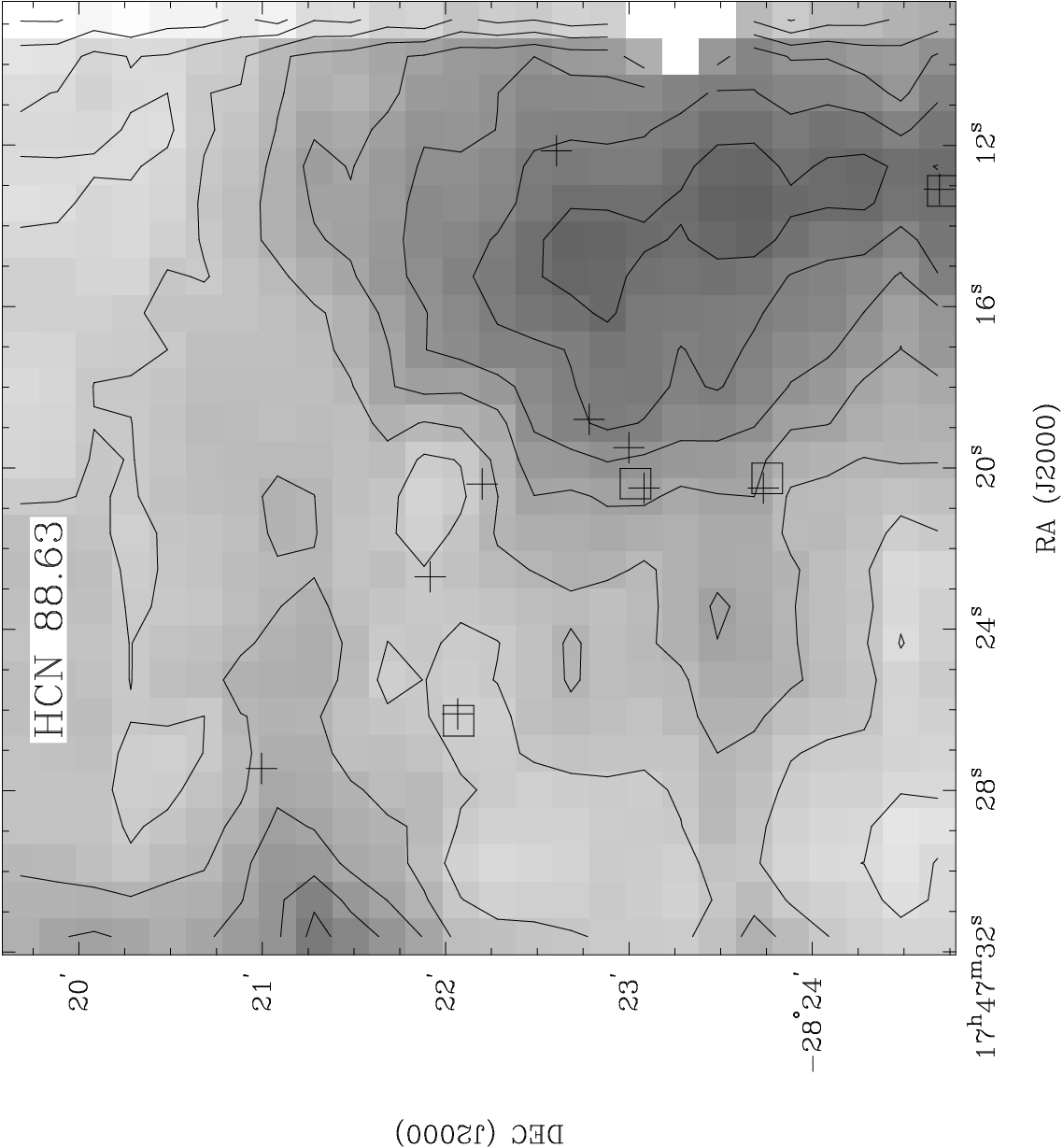}
\includegraphics[width = 5.2 cm,angle=-90]{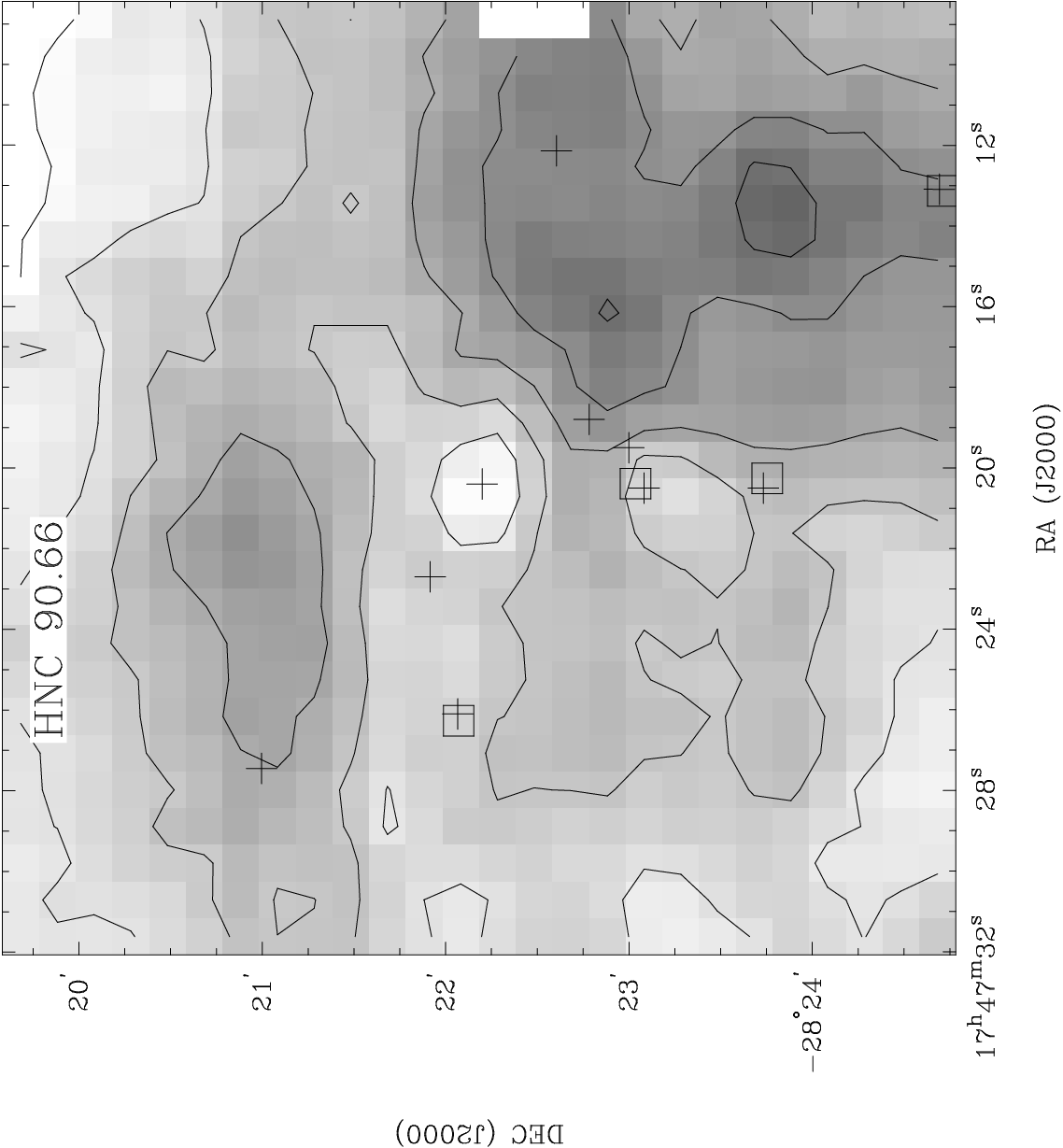}
\includegraphics[width = 5.2 cm,angle=-90]{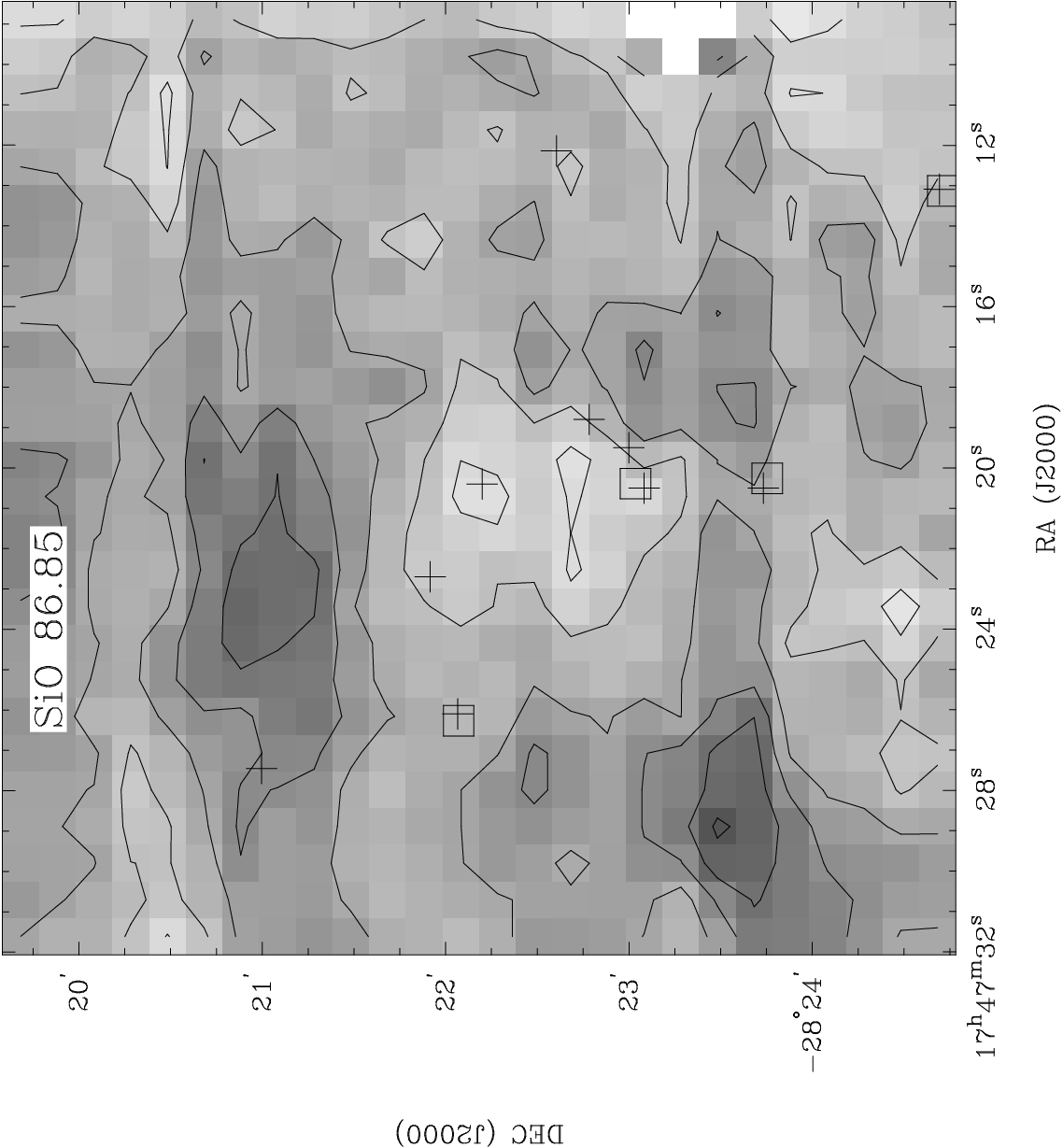}
\includegraphics[width = 5.2 cm,angle=-90]{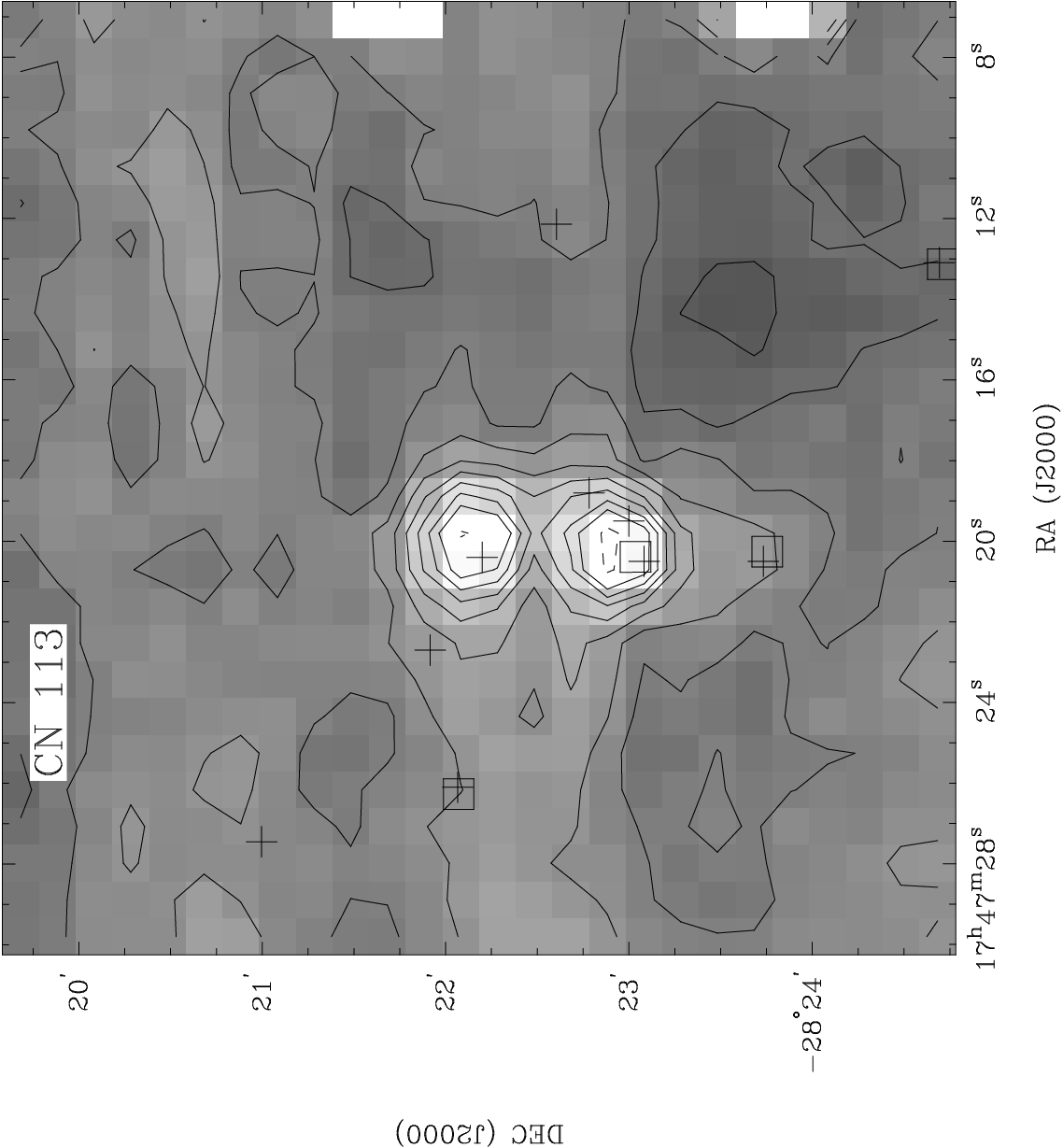}
\caption{Integrated emission for CS (contour step 5 K km~s$^{-1}$, peak 
74 K km~s$^{-1}$),
HCO$^{+}$ (step 5 K km~s$^{-1}$, peak 65 K km~s$^{-1}$), HCN (step 5 K 
km~s$^{-1}$, peak 94 K km~s$^{-1}$), HNC (step 5 K km~s$^{-1}$, peak 57 
K km~s$^{-1}$), SiO (step 2 K km~s$^{-1}$, peak 18.4 K km~s$^{-1}$) and CN 
(step 10 K km~s$^{-1}$, peak 93 K km~s$^{-1}$). Note that
the grey-scale is darker for stronger emission, so the lighter shades near
Sgr~B2(N) (and Sgr~B2(M) for SiO, HNC and CN) indicate lower integrated emission
due to absorption.}
\label{cs_etc_int}
\end{figure*}

The carbon monosulphide
CS 2~--~1 (97.98 GHz) integrated emission is shown in Fig. \ref{cs_etc_int}.
The CS data cube (not shown here) shows that the main peak near Sgr~B2(M) has 
a minimum
around velocity 62 km~s$^{-1}$, due to self-absorption at the position
and velocity where the brightest CO is found. The CS also traces the
low velocity
`hole' at $\sim 35$ km~s$^{-1}$ similar to the results of \citet{sat+00} and
\citet{tsu+99} using the CS 1~--~0 line at 48.99 GHz.

The CS 2~--~1 emission near Sgr~B2(M) shows a velocity gradient, with the 
emission
wings on either side of the 62 km~s$^{-1}$ self-absorption offset: the peak 
around
85 km~s$^{-1}$ is at 17 47 19.8, -28 22 56 and the peak around 50 km~s$^{-1}$ 
is at
17 47 19.5, -28 23 06. This is shown at higher resolution in BIMA observations
of \citet{meh95} who attribute this to an outflow.
The blue-shifted wing is stronger, so that the integrated CS emission peaks
at around 17 47 19.2, -28 23 03 to the south-west of Sgr~B2(M).
There is very little CS emission from Sgr~B2(N) indicating that it is
underabundant in CS, relative to Sgr~B2(M).

The CS data cube also shows: the `south-east CS peak' noted by \citet{yus+96},
centred at 17 47 27.1, -28 23 13, at 41 km~s$^{-1}$, width 20 km~s$^{-1}$; the 
north
ridge with peak at 17 47 22.3, -28 20 49, at 61 km~s$^{-1}$, width 
57 km~s$^{-1}$;
and the west ridge with peak at 17 47 14.9, -28 22 37, at
119 km~s$^{-1}$, width 14 km~s$^{-1}$ \citep{sat+00}.

We also have data (not plotted here)
from $^{13}$CS 2~--~1 (92.49 GHz) and C$^{34}$S 2~--~1 (96.41 GHz)
transitions, which are much weaker, but are optically thin and do not
suffer as much from the self-absorption. These confirm the lower CS emission
from Sgr~B2(N) than from Sgr~B2(M), and show that the peak near Sgr~B2(M) is 
at 17 47 18.7, -28 23 11 with velocity around 54 km~s$^{-1}$, width 
15 km~s$^{-1}$.

\subsection{HCO$^{+}$, HCN, HNC, H$^{13}$CO$^{+}$, H$^{13}$CN and HN$^{13}$C}

The integrated emission distributions of formylium (HCO$^{+}$) 1~--~0 
(89.19 GHz),
hydrogen cyanide HCN 1~--~0 (88.63 GHz)
and hydrogen isocyanide HNC 1~--~0 (90.66 GHz) are shown in Fig.
\ref{cs_etc_int}.
The distributions are qualitatively similar, but require careful interpretation
as they are strongly affected by self-absorption. In particular, the
low level of integrated emission in the centre, near Sgr~B2(N) and Sgr~B2(M),
is due to absorption, as is shown, for example, in the spectra and
integrated images of \citet{jac+99}.

The whole area is filled with emission over a wide velocity range. Fitting
spectra at the east edge of the imaged area, away from the strong
absorption in the centre, we find a peak velocity of 70 km~s$^{-1}$, width 
71 km~s$^{-1}$ for HCO$^{+}$,
velocity 69 km~s$^{-1}$, width 77 km~s$^{-1}$ for HCN and velocity 55 
km~s$^{-1}$,
width 67 km~s$^{-1}$ for HNC. This component is enhanced in the area of the 
north ridge in HNC, with peak position 17 47 22.0, -28 20 55, and single
component fit velocity 59 km~s$^{-1}$, width 54 km~s$^{-1}$ (but there is some 
self-absorption
at this position, making the single component gaussian not a very good fit).
The west ridge adds to this wide component, in the integrated images (Fig. 
\ref{cs_etc_int}), but is not well separated in velocity. Multi-component
fits to the spectra show that it peaks at 17 47 14.8 -28 22 36 in HCO$^{+}$ 
(we cannot get a good fit to this component in velocity),
peak 17 47 14.7 -28 22 34, velocity 119 km~s$^{-1}$,
width 24 km~s$^{-1}$ in HCN and peak 17 47 14.9 -28 22 34,
velocity 112 km~s$^{-1}$, width 21 km~s$^{-1}$ in HNC.

The HCO$^{+}$, HCN and HNC data cubes show a peak near Sgr~B2(M), much like
that in CS, with self-absorption around 65 km~s$^{-1}$. The spectra from this 
area show two components around 46 km~s$^{-1}$ and 90 km~s$^{-1}$, which are 
interpreted as a single
component with an absorption dip. There is also absorption of the Sgr~B2
continuum emission, by gas along the line of sight, giving a broad negative
feature to the spectra between velocities -120 and 20 km~s$^{-1}$. Quantitative
analysis of the peak near Sgr~B2(M) is affected by the absorption.

The fitted peak positions (at around 90 km~s$^{-1}$)
are 17 47 20.1, -28 22 34 in HCO$^{+}$,
17 47 20.1, -28 22 32 in HCN  and 17 47 19.8, -28 22 56 in HNC.
This is near the CS peak and the C$^{18}$O peak, but as for CS, there may
be a gradient of position with velocity.
We also have data (not plotted here) of the corresponding weaker isotopologue
lines H$^{13}$CO$^{+}$ 1~--~0 (86.75 GHz), H$^{13}$CN 1~--~0 (86.34 GHz) and
HN$^{13}$C 1~--~0 (87.09 GHz) which also show some self-absorption, but are less
affected, and hence better for the velocity fits. The velocities are
peak 50 km~s$^{-1}$, width 8 km~s$^{-1}$ for H$^{13}$CO$^{+}$, peak 47 
km~s$^{-1}$, width 12 km~s$^{-1}$
for H$^{13}$CN and peak 52 km~s$^{-1}$, width 16 km~s$^{-1}$ for HN$^{13}$C.
The HN$^{13}$C fit is in good agreement with the fit to the CS
peak but the other two are a bit
lower in velocity and narrower, presumably due to the effect of the absorption.

Note that the HCN, HNC, H$^{13}$CN and HN$^{13}$C lines are triplets
with hyperfine splitting, but that the spread of frequency for HNC and
HN$^{13}$C is only 0.21 MHz, so this will have negligible effect
on the fitted velocity widths. For HCN and H$^{13}$CN however,
the frequency range
is 3.5 MHz, corresponding to velocity range 12 km~s$^{-1}$, so that the blending
of the hyperfine components would contribute to increasing the fitted velocity
width.

\subsection{SiO}

The integrated silicon monoxide SiO 2~--~1 (86.85 GHz) emission is also 
shown in Fig. \ref{cs_etc_int}.

The SiO data cube and integrated image shows 
similar features to the CS 2~--~1 data, but the SiO line is weaker
and, thus, has lower lower signal to noise ratio. 
The integrated emission peak near Sgr~B2(M)
shows absorption at around 65 km~s$^{-1}$. The emission
peak is at 17 47 18.9, -28 22 49,
velocity 50 km~s$^{-1}$, width 11 km~s$^{-1}$, but this is affected by the 
absorption
with a second velocity component to the fit at 87 km~s$^{-1}$, width 31 
km~s$^{-1}$,
on the redshifted side of the absorption. There are also
the north ridge, peak 17 47 22.5,  -28 21 06, velocity 58 km~s$^{-1}$, width 
41 km~s$^{-1}$,
and the south-east `CS peak' at 17 47 27.1, -28 23 12, velocity 45 km~s$^{-1}$,
width 29 km~s$^{-1}$. These three peaks in integrated SiO, and the absorption
near Sgr~B2(M) and Sgr~B2(N) are also seen in the integrated SiO image
of \citet{mar+97}. Higher resolution BIMA data of the peak near Sgr~B2(M)
are interpreted by \citet{liu+98} as an outflow (like the  CS data.)

\subsection{CN}

Emission from the cyanide radical is observed in two 
groups of blended hyperfine components at
CN 1~--~0  J=1/2--1/2 (113.17 GHz) and
1~--~0 J=3/2--1/2 (113.49 GHz), each of which consists of several
components. The distribution of the integrated emission
from the two sets of lines is very similar, so the
sum of the two sets is plotted here, in 
Fig. \ref{cs_etc_int}. The most striking feature of the data
is the strong absorption associated with Sgr~B2(M) and Sgr~B2(N)
giving a deficit in the integrated emission in Fig. \ref{cs_etc_int}.
This absorption is due to spiral clouds along the line of sight 
\citep{grewil94} against the strong continuum of the Sgr~B2(M) and Sgr~B2(N)
cores, rather than absorption in the Sgr~B2 complex itself.

Because of the multiple components, the data cubes are rather
complicated with overlapping velocity and frequency structure. The J=3/2--1/2
(113.49 GHz) data cube shows the peak near Sgr~B2(M) with position
17 47 20.1, -28 22 50, velocity 94 km~s$^{-1}$ (from the strongest
component), width 19 km~s$^{-1}$. There is also
absorption over a wide velocity range down to -100 km~s$^{-1}$, 
at the continuum peaks. We therefore
interpret the velocity of the peak fit as being biased high due to the
absorption. This is confirmed by the optically thin $^{13}$CN lines having
velocity around 52 km~s$^{-1}$ \citep{ger+84}. There is widespread CN emission
over the whole area, with velocity 52 km~s$^{-1}$, fitted at the north ridge 
position,
with broad lines (but the fitted velocity width of 113 km~s$^{-1}$ includes the
confusion of the multiple components). The CN emission is widespread
compared to the distribution of other molecules studied here.
The J=1/2--1/2 (113.17 GHz) data cube shows deep absorption features at
Sgr~B2(M) and Sgr~B2(N) but is too complicated to do much more interpretation,
with the multiple components blended.

We also detect (Table \ref{extra_lines}) the weak lines of the $^{13}$CN
isotopologue
J=1/2--1/2 (108.65 GHz) and J=3/2--1/2 (108.78 GHz) in extended emission and
absorption at Sgr~B2(N) and  Sgr~B2(M).

\subsection{HC$_{3}$N}

\begin{figure*}
\includegraphics[width = 5.2 cm,angle=-90]{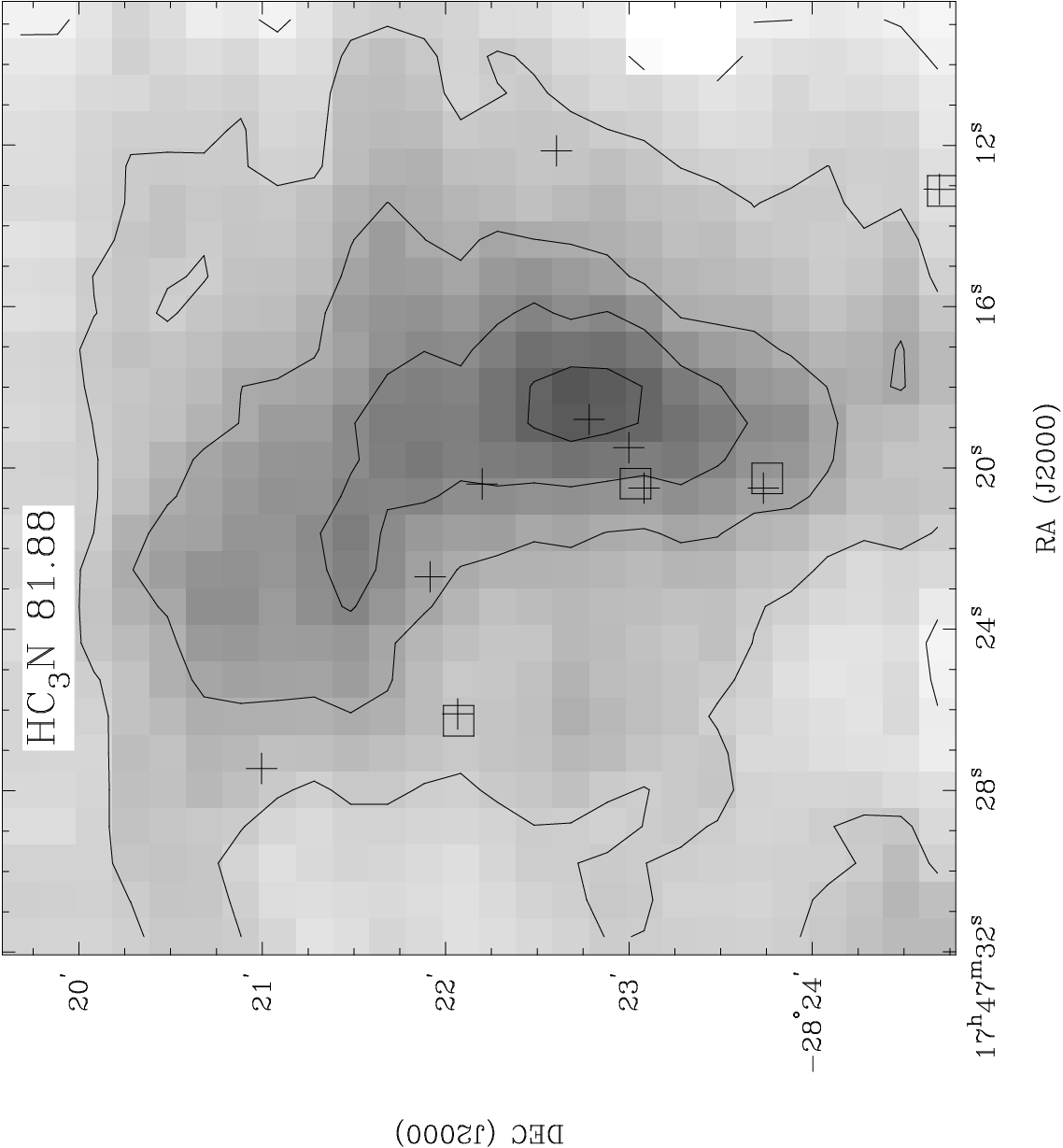}
\includegraphics[width = 5.2 cm,angle=-90]{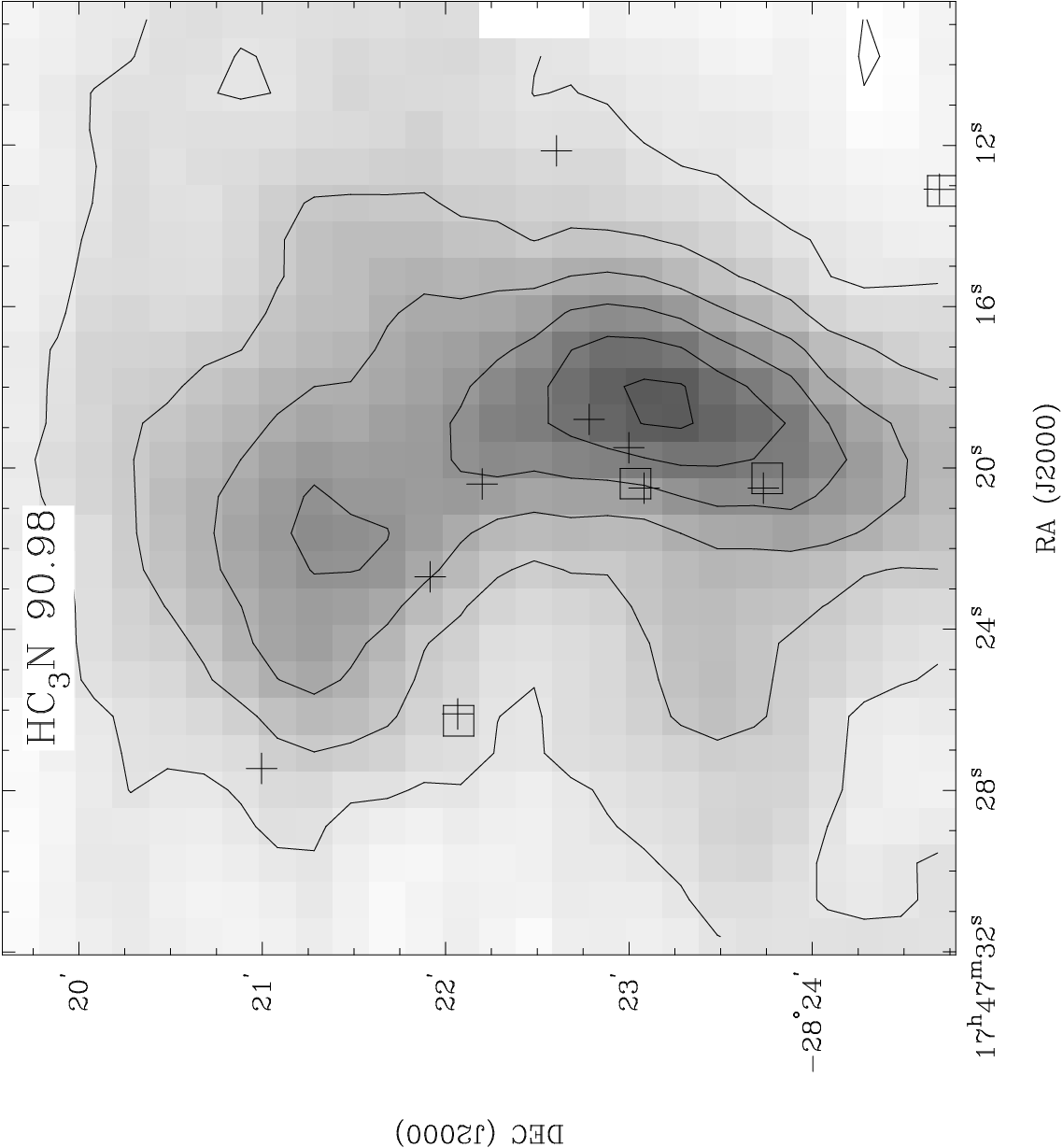}
\includegraphics[width = 5.2 cm,angle=-90]{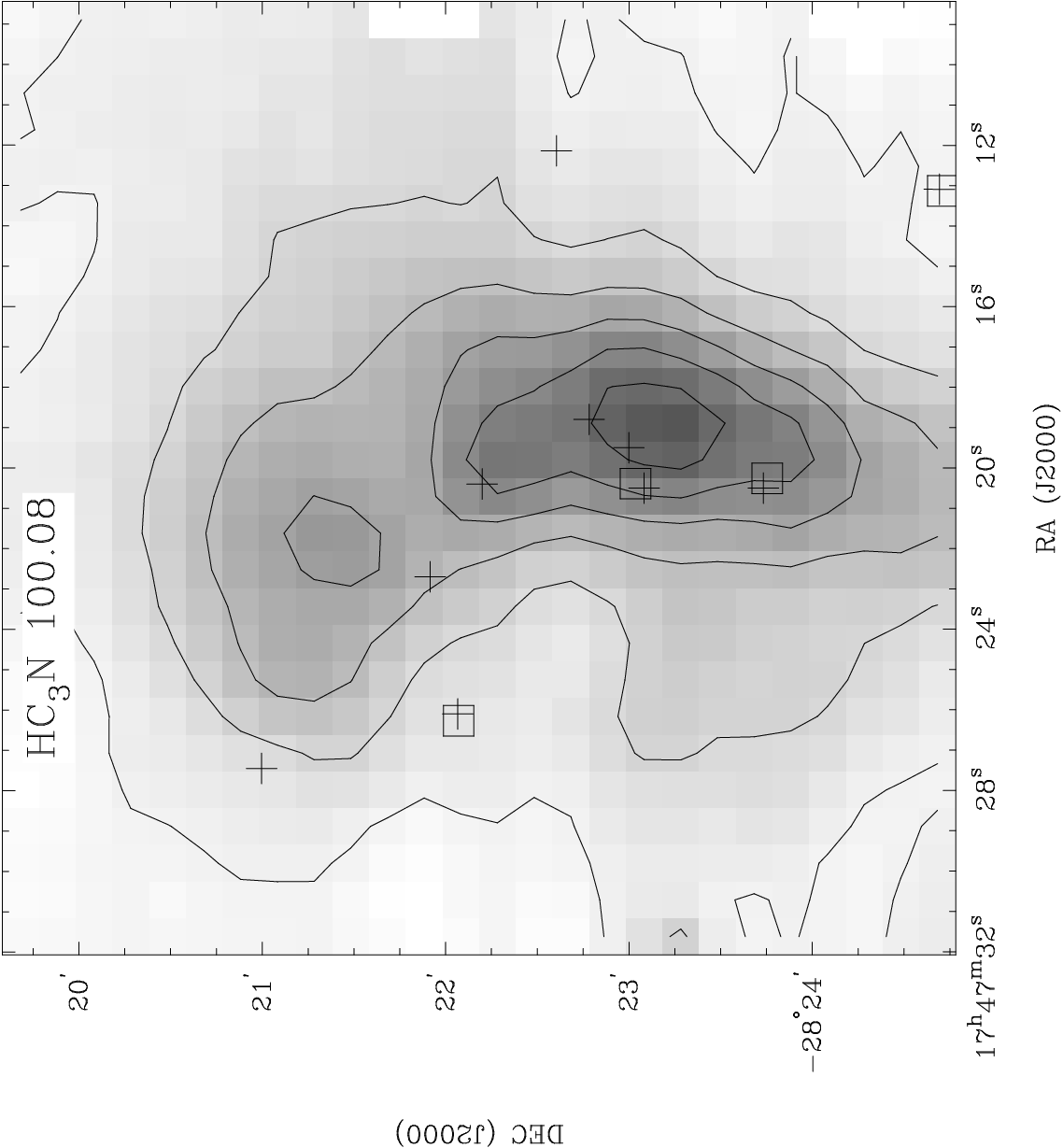}
\includegraphics[width = 5.2 cm,angle=-90]{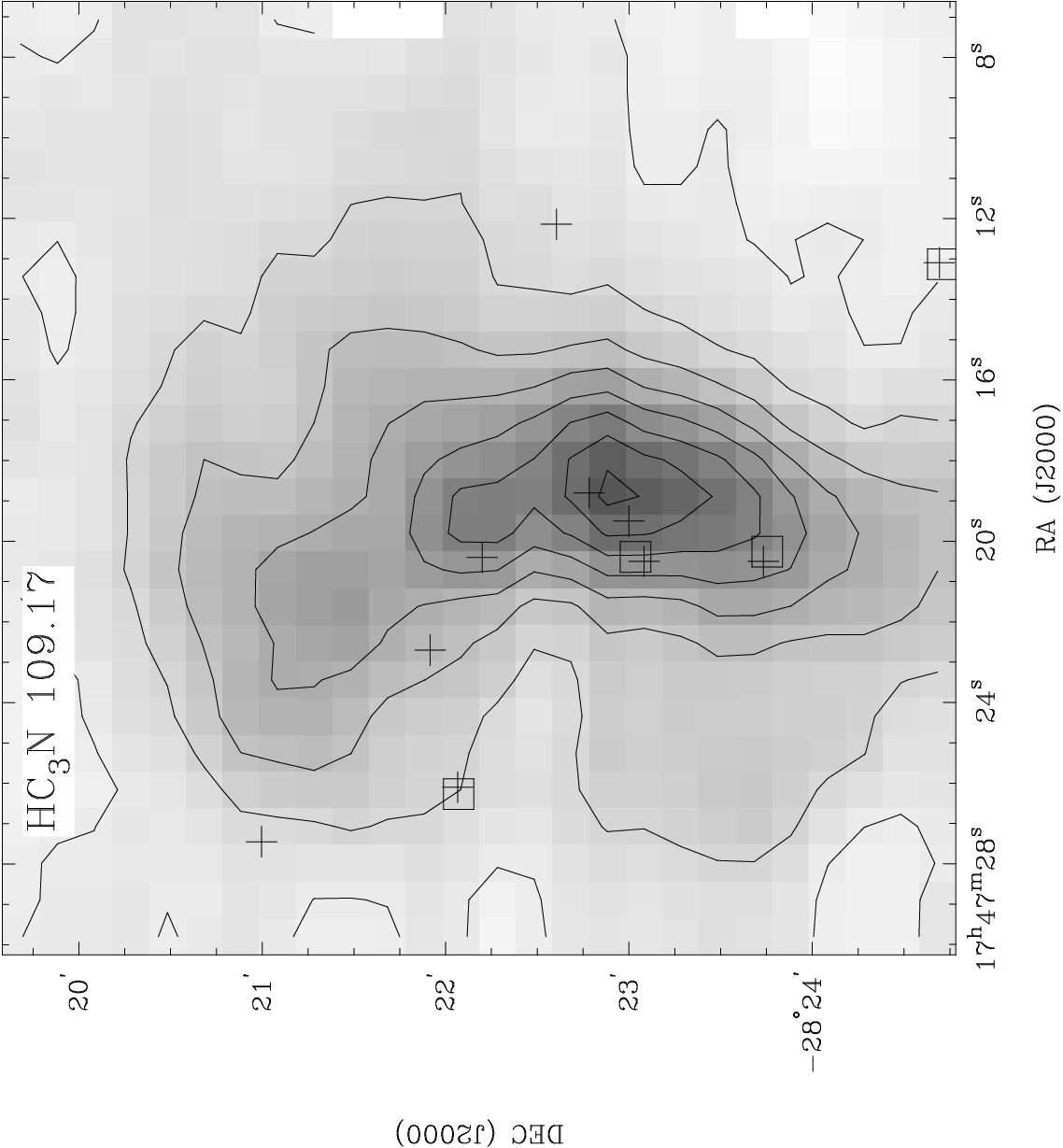}
\includegraphics[width = 5.2 cm,angle=-90]{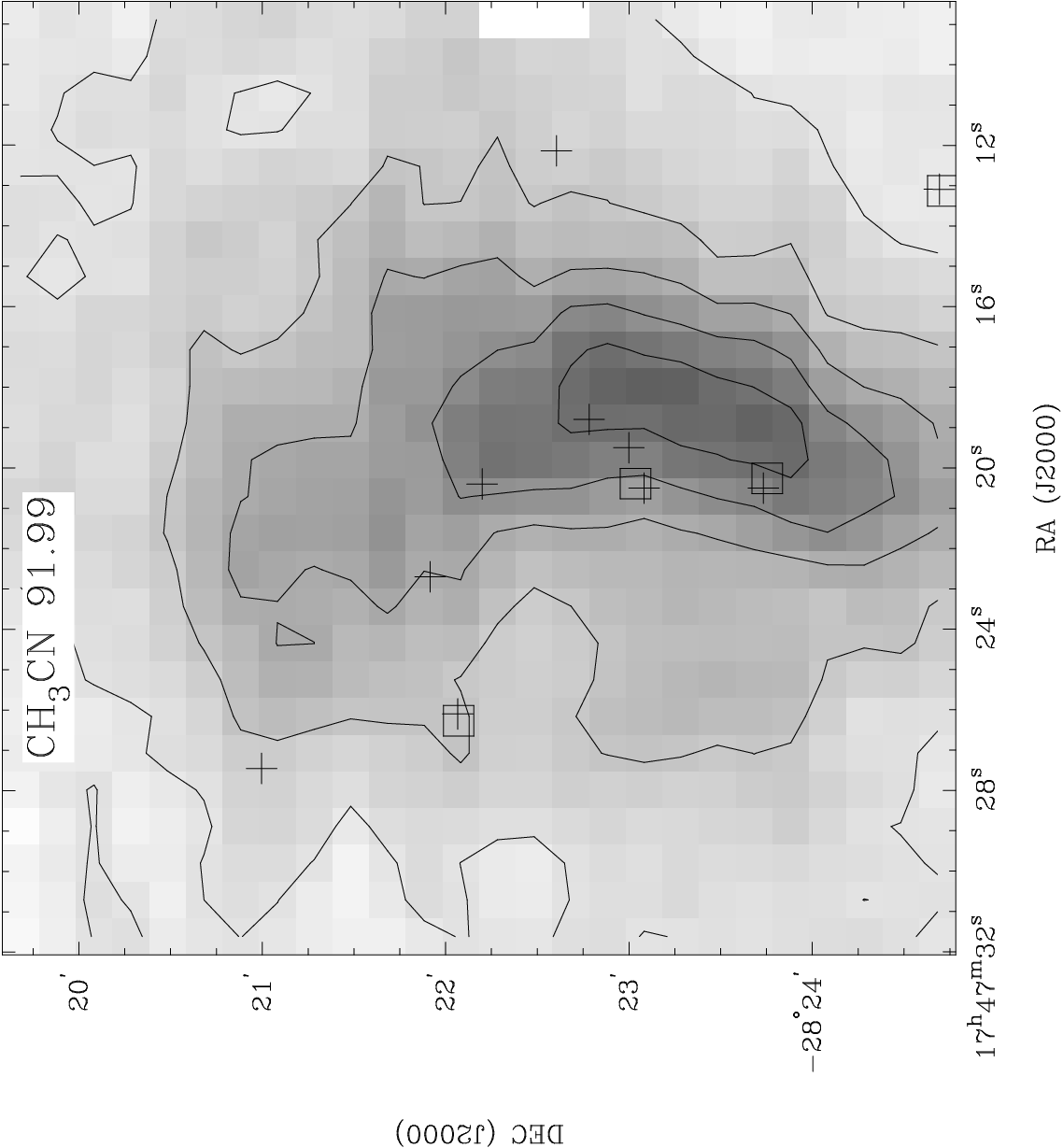}
\includegraphics[width = 5.2 cm,angle=-90]{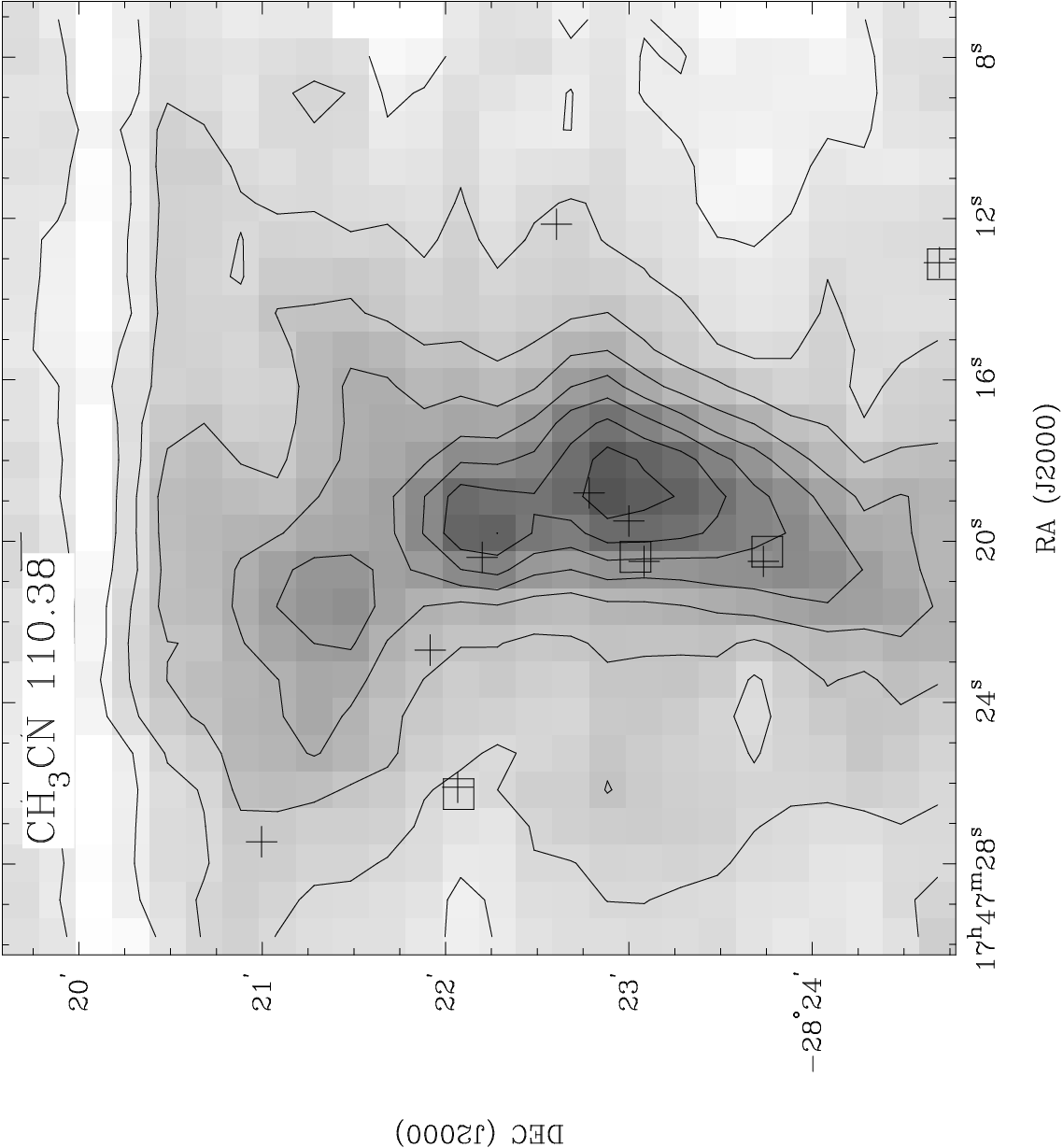}
\caption{Integrated emission for HC$_{3}$N (contour step 10 K km~s$^{-1}$; 
81.88 GHz,
peak 55 K km~s$^{-1}$; 90.98 GHz, peak 72 K km~s$^{-1}$; 100.08 GHz, peak 
68 K km~s$^{-1}$;
109.17 GHz, peak 83 K km~s$^{-1}$) and CH$_{3}$CN (contour step 5 
K km~s$^{-1}$; 91.99 GHz,
peak 28 K km~s$^{-1}$; 110.38 GHz, peak 43 K km~s$^{-1}$). These two molecules
trace an arc from the north cloud, then west of the radio and mid-IR continuum
peaks.}
\label{hc3n_ch3cn_int}
\end{figure*}

The integrated emission from cyanoacetylene
HC$_{3}$N 9~--~8 (81.88 GHz), 10~--~9 (90.98 GHz),
11~--~10 (100.08 GHz) and 12~--~11 (109.17 GHz) is shown in Fig.
\ref{hc3n_ch3cn_int}. All four transitions show similar structure, which is a
ridge of emission to the west of radio continuum peaks
Sgr~B2(N), Sgr~B2(M) and Sgr~B2(S),
looping to the east, north of Sgr~B2(N).

This is similar to the single-dish results
of \citet{lisgol91} for the 12~--~11 transition, \citet{chu+94} for the 10~--~9
and 12~--~11 transitions, and \citet{dev+97} for the
11~--~10 transition. Higher resolution interferometer
images of the HC$^{13}$CCN 9~--~8
transition at 81.53 GHz are given by \citet{kuasny96}
and multiple transitions are given by \citet{dev+00}. The high resolution
interferometer observations of the HC$_{3}$N 1~--~0 transition at 9.10 GHz
by \citet{hun+99} show weak maser emission, and so preferential emission
at the radio continuum peaks. While interesting in its own right, this does
not trace the molecular distribution well.

The data cubes show that the emission has several peaks with different
velocities, which are merged together in the integrated emission images.
We fit four peaks, from north to south, with a systematic velocity
gradient: (a)
the north cloud at 17 47 21.4, -28 21 29, north of Sgr~B2(N), velocity 68 
km~s$^{-1}$,
width 23 km~s$^{-1}$; (b)
peak at 17 47 18.7, -28 22 12, near Sgr~B2(N), velocity 67 km~s$^{-1}$, width 
23 km~s$^{-1}$;
(c) peak at 17 47 18.6, -28 23 04, near Sgr~B2(M), velocity 60 km~s$^{-1}$,
width 22 km~s$^{-1}$;
and (d) peak at 17 47 19.9, -28 23 55, near Sgr~B2(S), velocity 58 km~s$^{-1}$,
width 20 km~s$^{-1}$.
In addition to these four peaks we fit the north ridge at peak
17 47 21.0, -28 20 54, velocity 62 km~s$^{-1}$, width 27 km~s$^{-1}$, and the 
south-east peak at
17 47 26.3, -28 23 04, velocity 55 km~s$^{-1}$, width 23 km~s$^{-1}$.

We can calculate column densities of molecules in the upper level $N_u$ from
the intensities of the transitions, using the simple
assumption that lines are optically thin and in local thermodynamic
equilibrium (LTE) by
\begin{equation}
N_u = (8 \pi \nu^2 k/h c^3 A_{ul}) \int T_B dv 
\end{equation}
where $A_{ul}$ is the Einstein coefficient, and $\int T dv$ is the
integral over velocity of the brightness temperature $T_B$ of the emission line.
Using the multiple HC$_{3}$N transitions we can,
in principle, plot an excitation diagram of column density in that level
(expressed as $\ln(N_{u}/g_{u})$) versus the energy of the level
(expressed as $E_{u}/k$), to determine the
total column density $N$ and
excitation temperature $T_{ex}$, using the equation
\begin{equation}
(N_u/g_u) = (N/Q_T) \exp(-E_{u}/kT_{ex}) 
\end{equation}
where $Q_T$ is the partition function at excitation temperature $T_{ex}$,
and $g_u$ is the statistical weight of the upper level.
In practice, for the lines here in the 3-mm band,
we do not have enough range in the energy levels for this to be very reliable
($E_{u}/k = 20$ to 34 K for these lines). However,
we can determine that there are spatial variations in the excitation
temperature, between the peaks, with the cloud north of Sgr~B2(N) giving
$T_{ex} = 28$~K (20 -- 46 K in the 1$\sigma$ range) and the others hotter
with limits $> 43$~K, $> 76$~K and $> 41$~K (at the 1$\sigma$ level)
for the peaks near
Sgr~B2(N), Sgr~B2(M) and Sgr~B2(S) respectively. This is confirmed by
considering the  spatial variation in ratios of the different transitions,
and is consistent with the results of \citet{chu+94} and the
higher kinetic temperature in these hot dense cores \citep{dev+97}.
This analysis is complicated towards Sgr~B2(N), as the IRAM 30-m survey
of \citet{bel+05, bel+07} shows that the HC$_{3}$N is somewhat optically 
thick there.

We also detect (Table \ref{extra_lines}) seven vibrationally
excited lines of HC$_{3}$N
at 91.20, 91.33, 100.32, 100.71, 109.44, 109.60 and 109.87 GHz, concentrated
at Sgr~B2(N), as the higher upper energy transitions are excited in this hot
region. We detect weak lines of the isotopologues H$^{13}$CCCN, HC$^{13}$CCN and
HCC$^{13}$CN at 88.17, 90.60, 99.65 and 108.71 GHz, many of which appear
to peak at Sgr~B2(N), but as the lines are weak the spatial distribution
is not clear.

\subsection{CH$_{3}$CN}

The integrated emission from methyl cyanide CH$_{3}$CN 5~--~4 (91.99 GHz) and
6~--~5 (110.38 GHz) is shown in Fig. \ref{hc3n_ch3cn_int}.
There are multiple components for each of these transitions.
The integrated emission for the two sets of lines is similar, and similar
to that of the four HC$_{3}$N lines. It is also
similar to the single-dish results
of \citet{dev+97} for the 5~--~4 transition. We fit five components similar
to that in HC$_{3}$N above:
(a) the north cloud at 17 47 21.3, -28 21 28, velocity 68 km~s$^{-1}$, width 
35 km~s$^{-1}$;
(b) peak near Sgr~B2(N) at 17 47 19.1, -28 22 12, velocity 66 km~s$^{-1}$,
width 30 km~s$^{-1}$;
(c) peak near Sgr~B2(M) at 17 47 18.8, -28 23 11, velocity 61 km~s$^{-1}$,
width 33 km~s$^{-1}$;
(d) peak near Sgr~B2(S) at 17 47 19.9, -28 23 54, velocity 59 km~s$^{-1}$,
width 32 km~s$^{-1}$;
and (e) the north ridge at 17 47 23.5, -28 21 01, velocity 64 km~s$^{-1}$,
width 41 km~s$^{-1}$
The ratio of integrated emission of the two lines indicates that the
peaks near Sgr~B2(N) and Sgr~B2(M) have higher excitation temperature than
the surrounding area, but the overlapping components and low
signal to noise make more quantitative analysis difficult.

We also detect weak emission (Table \ref{extra_lines}) from several more
transitions of CH$_{3}$CN at 92.26, 110.33, 110.35, 110.69 and 110.71 GHz,
and the isotopologues $^{13}$CH$_{3}$CN at 107.19 GHz and possibly
CH$_{3}^{~13}$CN at
110.33 GHz (as a blend). These are concentrated at the position of Sgr~B2(N).

\subsection{CH$_{3}$OH and $^{13}$CH$_{3}$OH}

\begin{figure*}
\includegraphics[width = 5.2 cm,angle=-90]{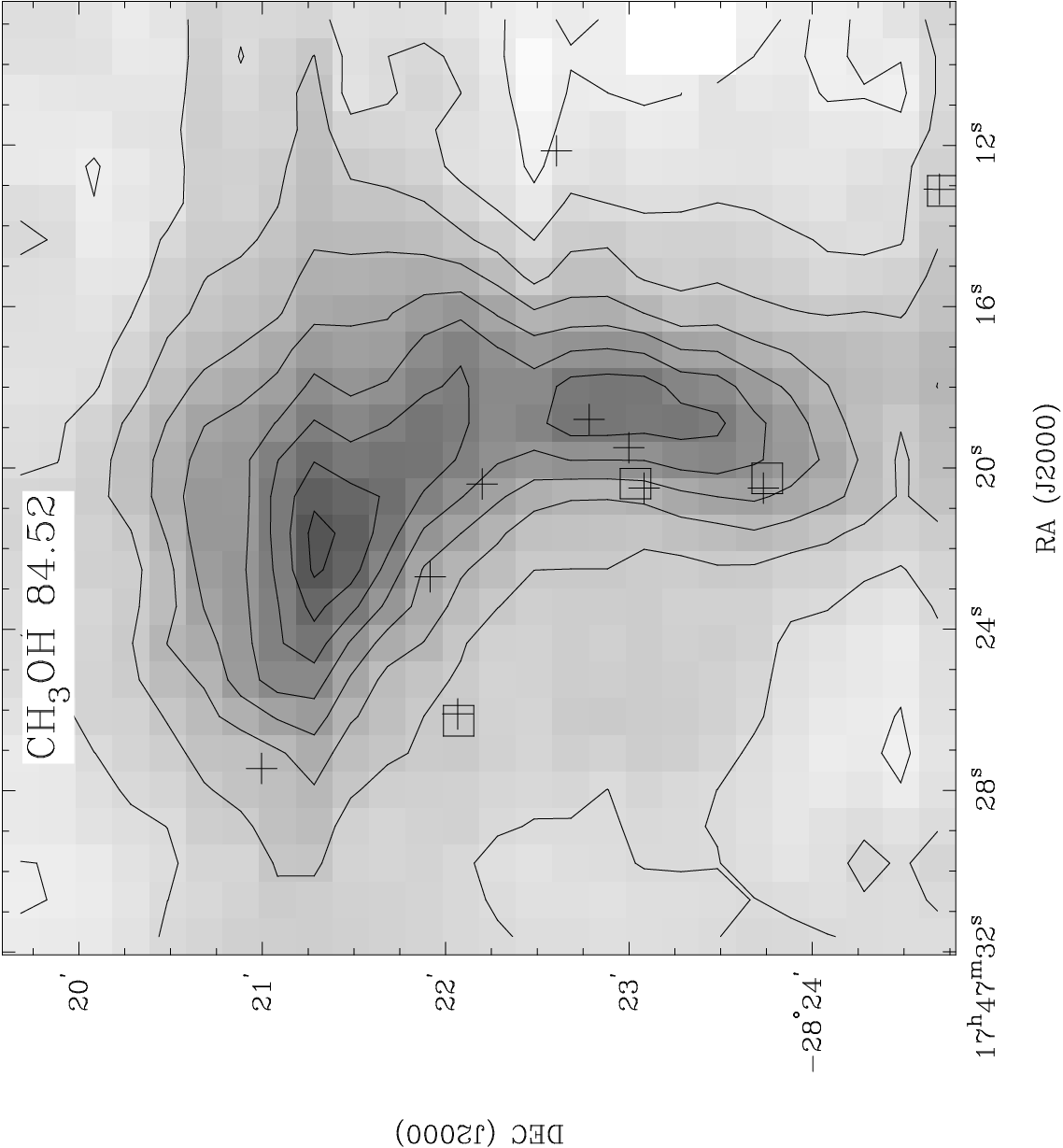}
\includegraphics[width = 5.2 cm,angle=-90]{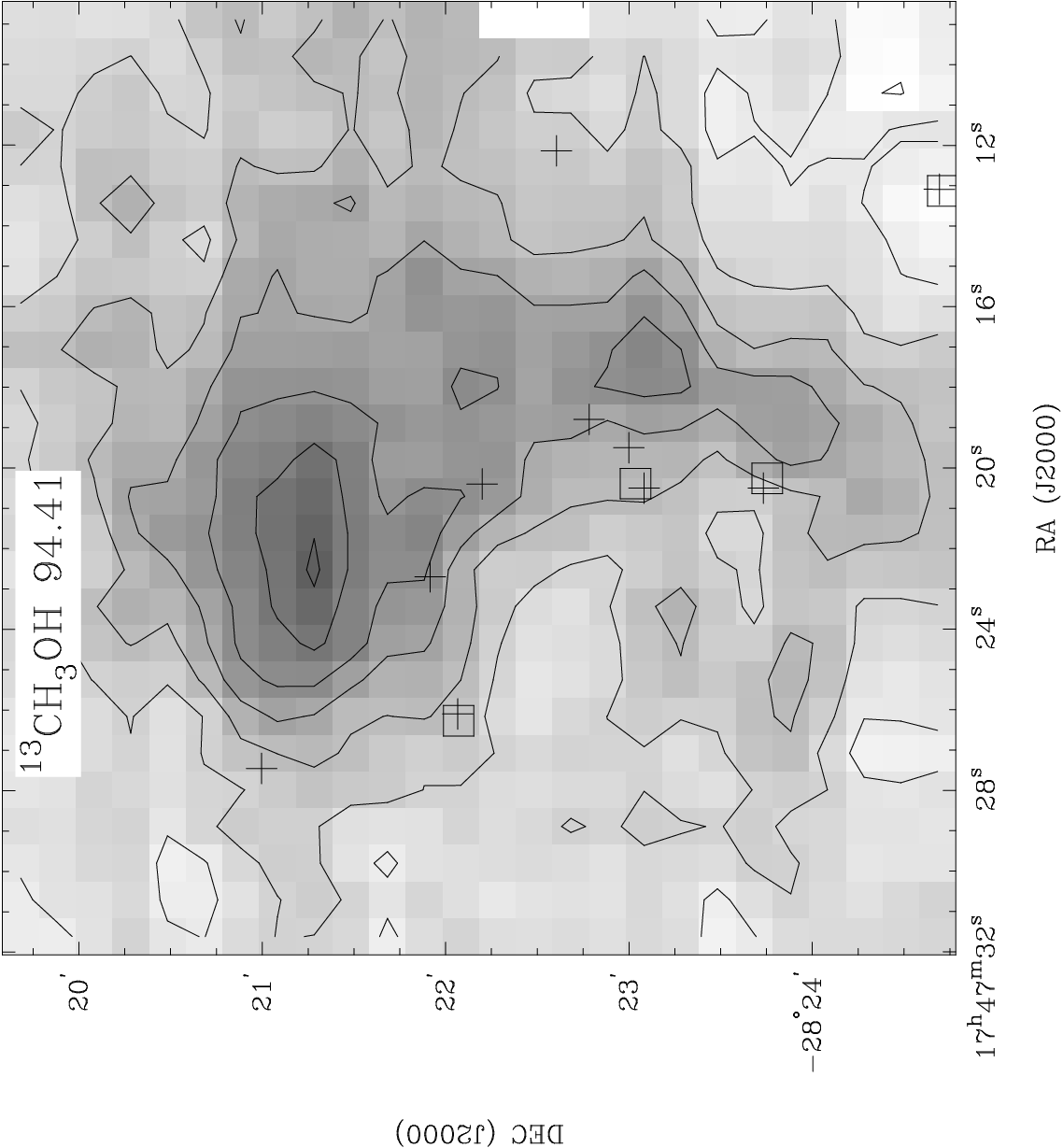}
\includegraphics[width = 5.2 cm,angle=-90]{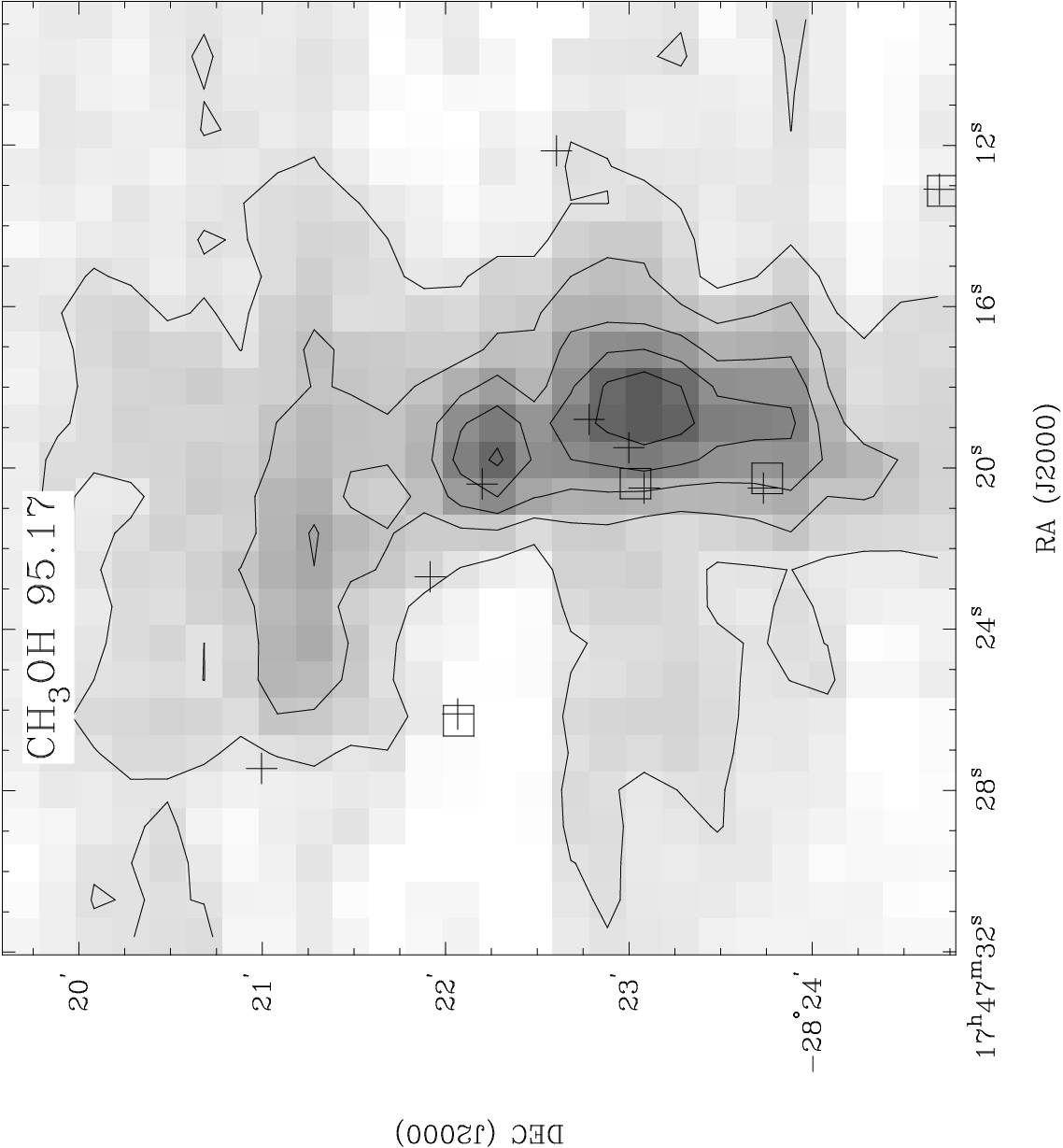}
\includegraphics[width = 5.2 cm,angle=-90]{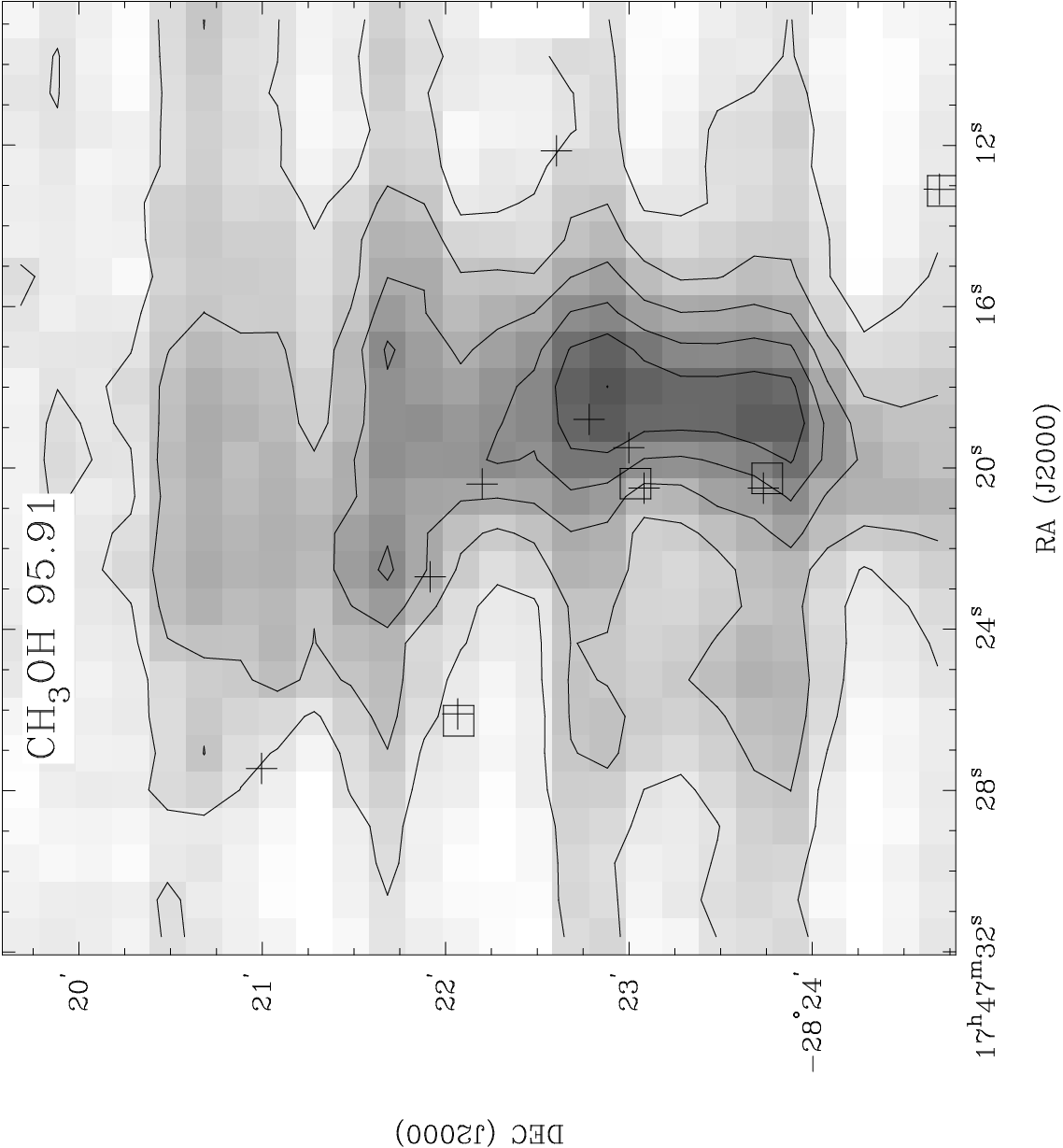}
\includegraphics[width = 5.2 cm,angle=-90]{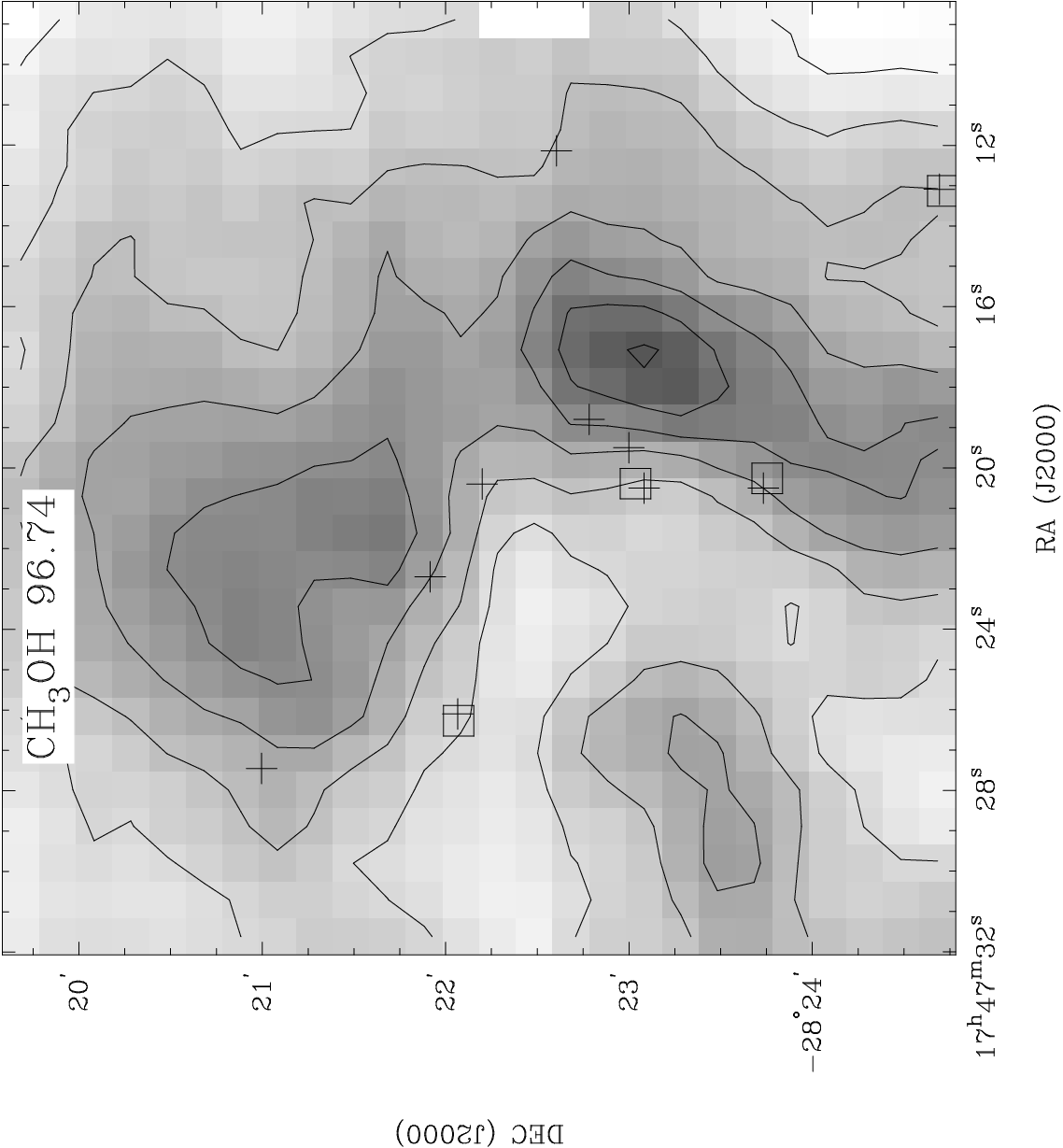}
\includegraphics[width = 5.2 cm,angle=-90]{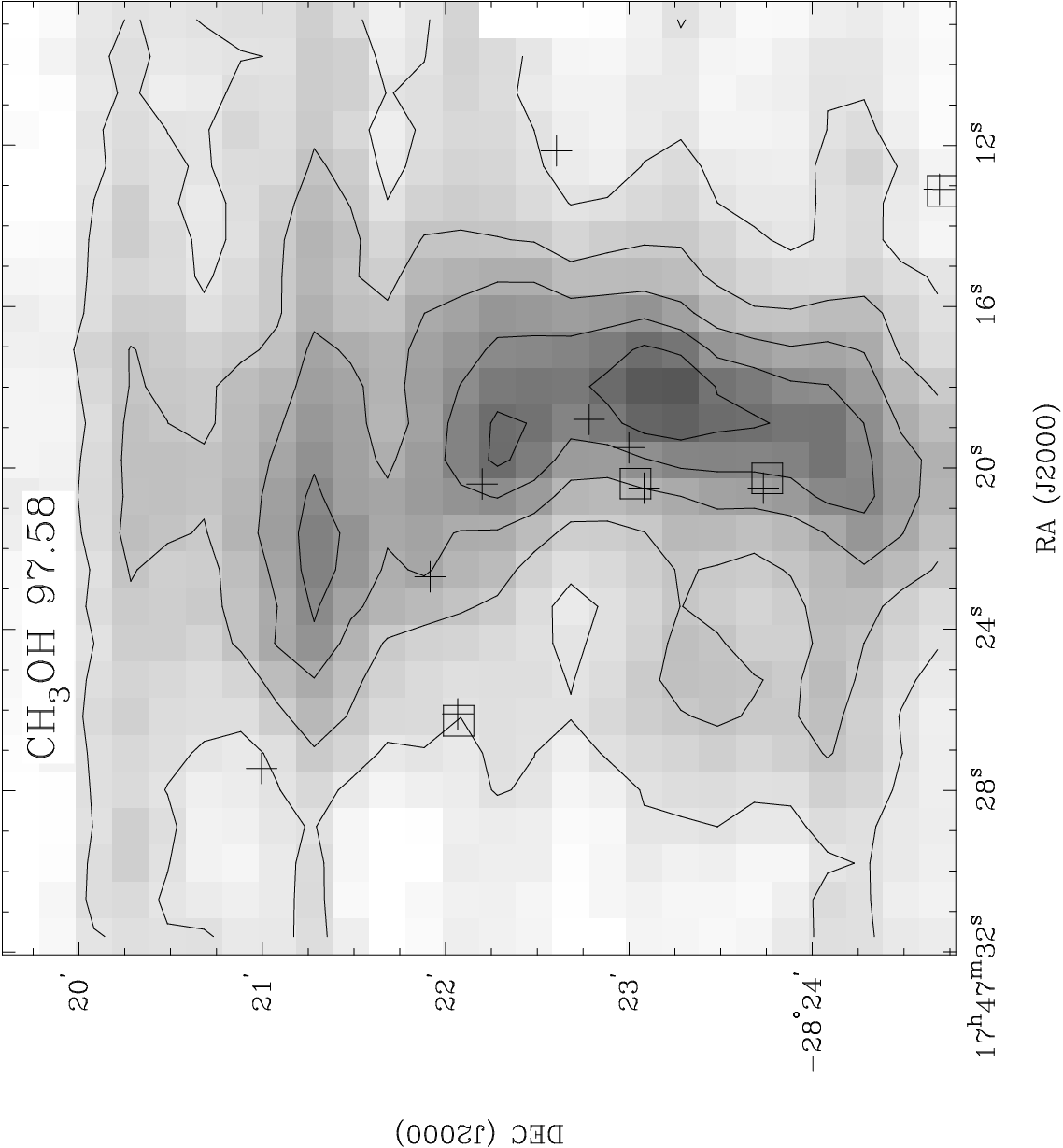}
\caption{Integrated emission for CH$_{3}$OH (84.52 GHz, contour step 10 K 
km~s$^{-1}$,
peak 93 K km~s$^{-1}$; 95.17 GHz, step 4 K km~s$^{-1}$, peak 23 K km~s$^{-1}$; 
95.91 GHz, step
5 K km~s$^{-1}$, peak 30 K km~s$^{-1}$; 96.74 GHz, step 10 K km~s$^{-1}$, 
peak 131 K km~s$^{-1}$; 97.58 GHz, step 5 K km~s$^{-1}$, peak 30 K km~s$^{-1}$)
and $^{13}$CH$_{3}$OH (94.41 GHz, step 2 K km~s$^{-1}$, peak 16.4 K 
km~s$^{-1}$). These lines
show the arc from the north cloud, west of the radio and mid-IR continuum peaks,
with differences in the relative intensities of the peaks related to the
complicated excitation of the different levels.}
\label{ch3oh_int}
\end{figure*}

In Fig. \ref{ch3oh_int} we show the integrated emission of five transitions of
methanol CH$_{3}$OH: 5(-1,5) -- 4(0,4) E (84.52 GHz), 8(0,8) -- 7(1,7) A+
(95.17 GHz), 2(1,2) -- 1(1,1) A+ (95.91 GHz), 2(0,2) -- 1(0,1) A+ blend (96.74 
GHz) and 2(1,1) -- 1(1,0) A- (97.58 GHz).
In addition, we show the integrated emission of the isotopologue
$^{13}$CH$_{3}$OH 2(0,2) -- 1(0,1) A+ blend (94.41 GHz), and we have data, not
plotted here for the CH$_{3}$OH 0(0,0) -- 1(-1,1) E (108.89 GHz) transition.
The distribution of integrated emission is quite different for the different
transitions.

Methanol is a very useful tracer of physical conditions, described as
`the Swiss army knife of star formation' \citep{leu+05}, particularly when
using simultaneous fits to multiple lines \citep{leu+04}. However, the excitation
conditions of methanol can be very complicated, with collisional and
radiative excitation.
For example, both the 84.52 GHz and 95.17 GHz transitions here can be masers
\citep{cra+92}.
Also the A- and E-types can be considered separate species, which have different
abundances. We do not attempt to model the different CH$_{3}$OH lines here, but
restrict ourselves to describing their overall features.

The different lines mostly trace the same spatial and velocity structure,
despite the different relative intensities of the features. These are: (a) the
north cloud at 17 47 21.4, -28 21 20, velocity 68 km~s$^{-1}$, width 
25 km~s$^{-1}$;
(b) the peak near Sgr B2(N) at 17 47 18.8, -28 22 14, velocity 67 km~s$^{-1}$,
width 19 km~s$^{-1}$;
(c) the peak near Sgr B2(M) at 17 47 18.2, -28 23 11, velocity 61 km~s$^{-1}$,
width 22 km~s$^{-1}$;
and (d) the peak near Sgr B2(S) at 17 47 19.9, -28 23 57, velocity 59 km~s$^{-1}$,
width 20 km~s$^{-1}$.
The 96.74 GHz CH$_{3}$OH line and the 94.41 GHz $^{13}$CH$_{3}$OH line are
blends of multiple transitions, so the velocity structure is confused. The
96.74 GHz line also shows absorption at Sgr B2(N) and Sgr B2(M). Because
it is the strongest line, however, it shows features not seen in the other
weaker lines: the south-east peak (seen in CS) at 17 47 26.7, -28 23 07,
velocity 56 km~s$^{-1}$, width 34 km~s$^{-1}$; the western ridge at 17 47 15.0, 
-28 22 44, velocity 120 km~s$^{-1}$, width 21 km~s$^{-1}$;
and a peak to the north-west of the
main ridge-line at 17 47 14.5, -28 21 41, velocity 70 km~s$^{-1}$ (and width 
unclear due to blending with other features).

There are thirteen more weak CH$_{3}$OH lines detected here
(Table \ref{extra_lines})
concentrated at the position of Sgr~B2(N), that are higher upper energy lines
excited in the hot core.

\subsection{CH$_{3}$CH$_{2}$OH}

We have also detected and imaged the ethanol CH$_{3}$CH$_{2}$OH 6(0,6) -- 5(1,5)
(85.27 GHz) transition, but as the line is weak, and the data are affected
by scanning stripes, the integrated emission is not shown here. The emission
is centred on the north cloud, and the line fit gives velocity 68 km~s$^{-1}$, 
width
21 km~s$^{-1}$. We expect from \citet{req+06} that the ethanol CH$_{3}$CH$_{2}$OH
column density follows that of methanol CH$_{3}$OH, but the distributions
of line emission here differ due to excitation differences.

The CH$_{3}$CH$_{2}$OH 7(0,7) -- 6(1,6) and 5(1,5) -- 4(0,4) (104.49 and
104.80 GHz) transitions show weak extended emission (Table \ref{extra_lines}).

\subsection{HNCO}

\begin{figure*}
\includegraphics[width = 5.2 cm,angle=-90]{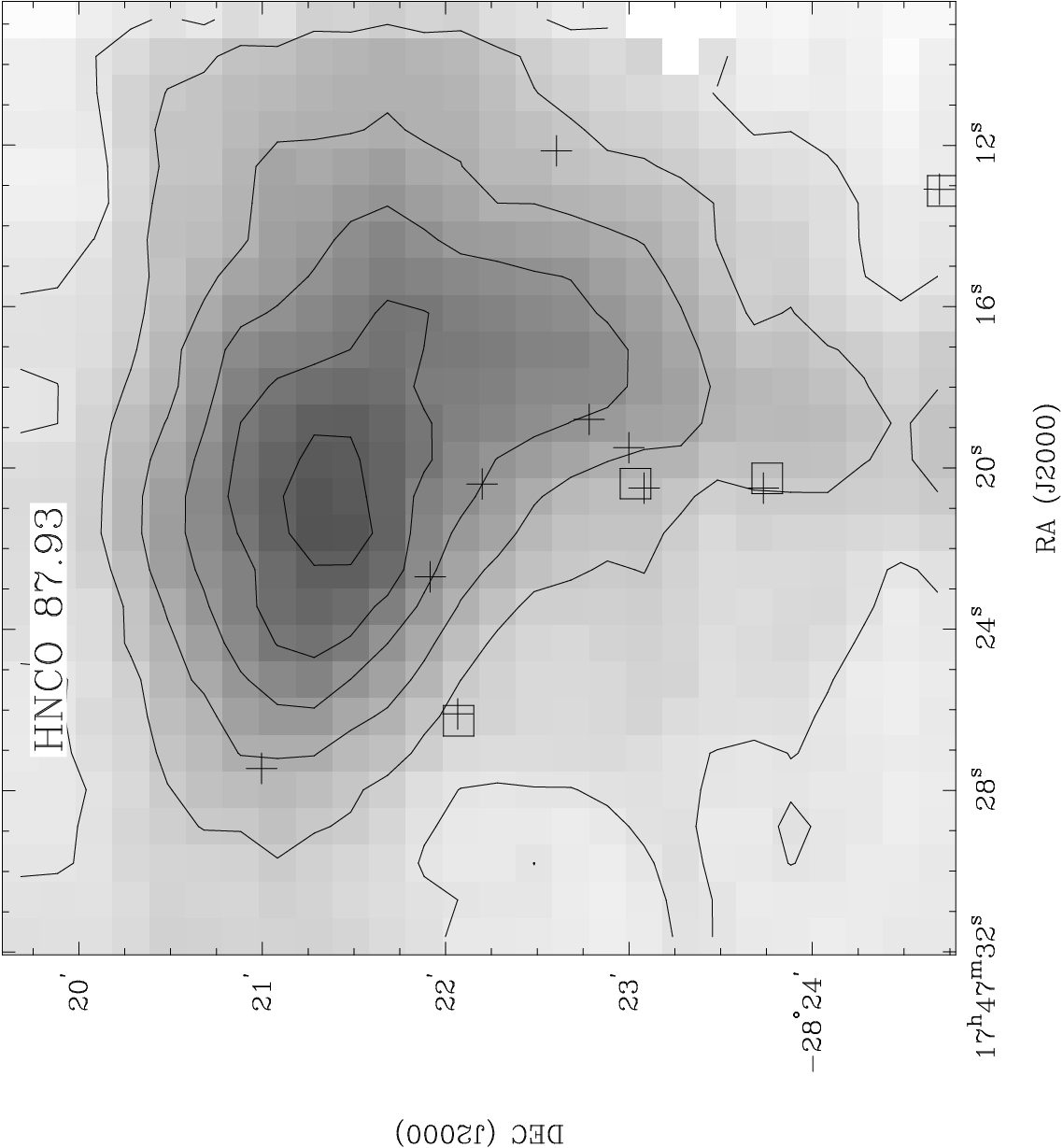}
\includegraphics[width = 5.2 cm,angle=-90]{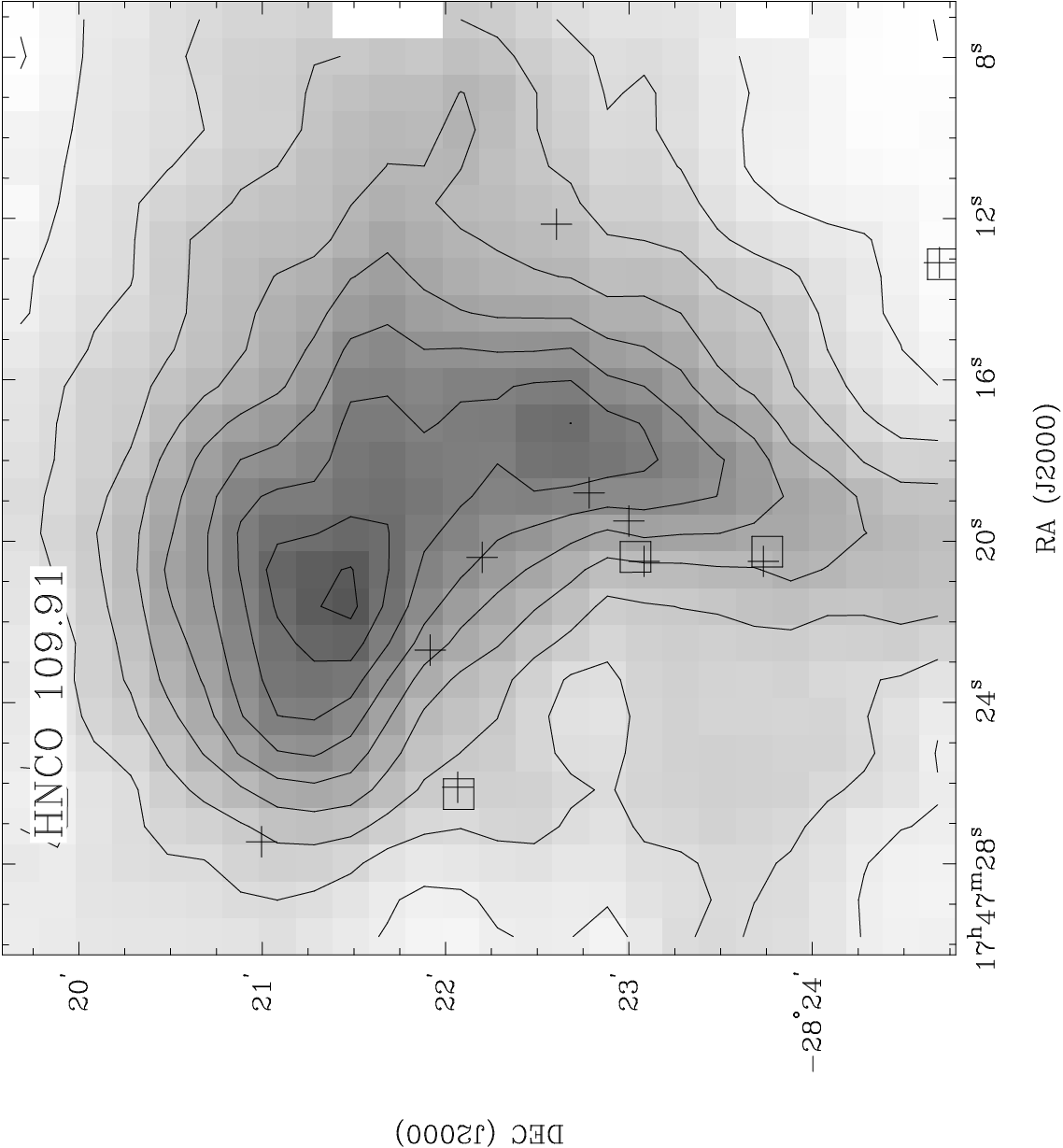}
\includegraphics[width = 5.2 cm,angle=-90]{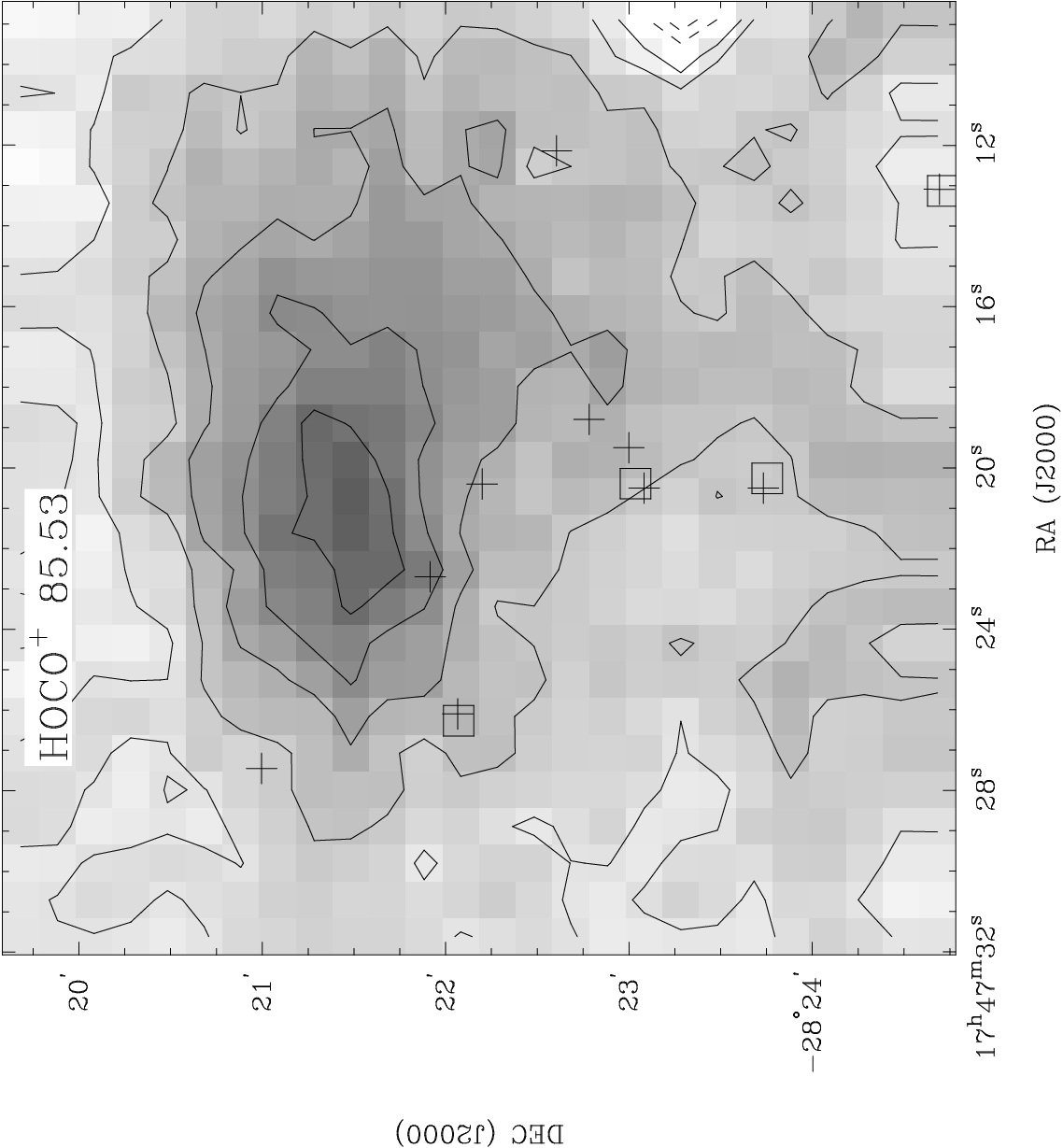}
\includegraphics[width = 5.2 cm,angle=-90]{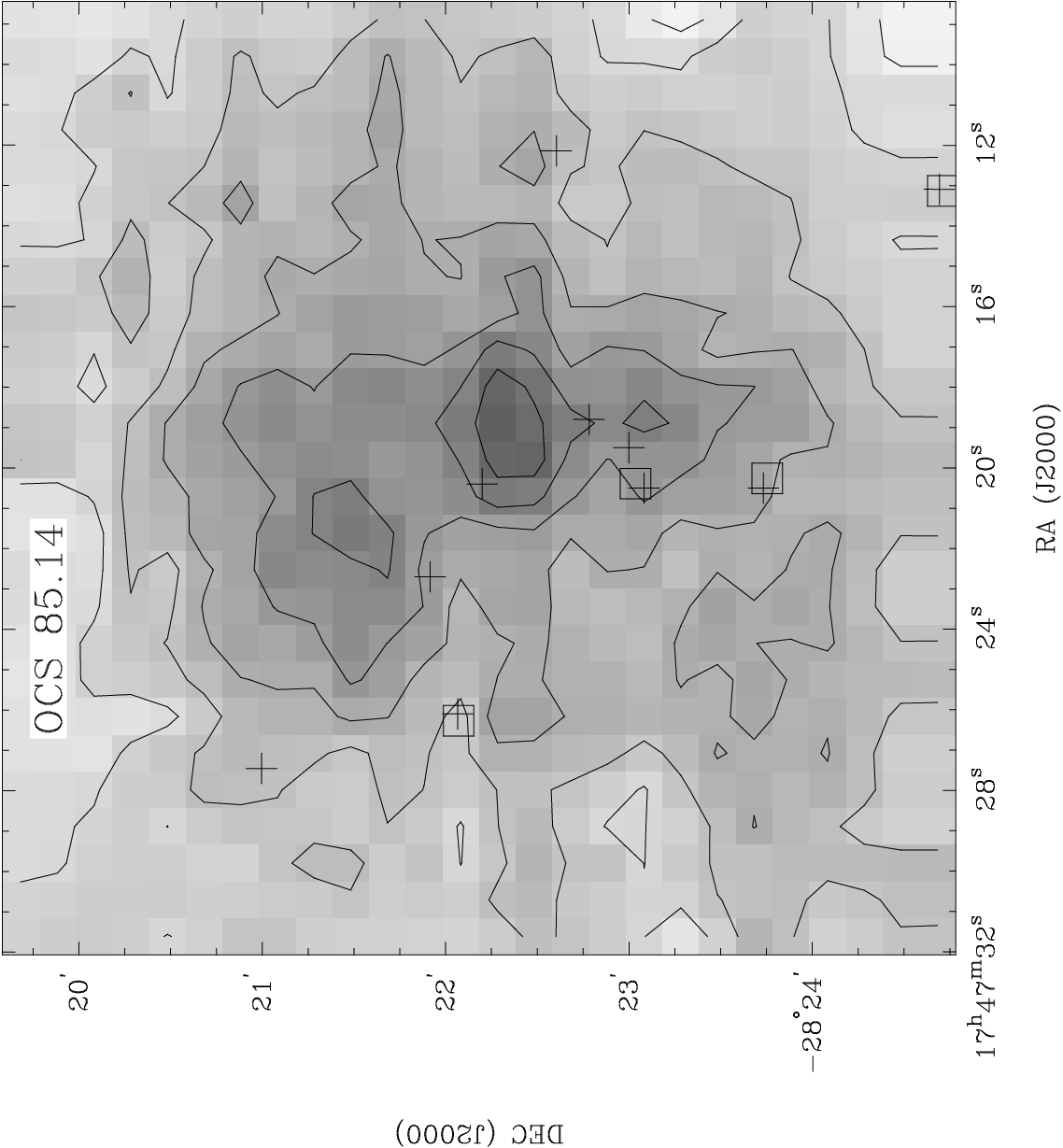}
\includegraphics[width = 5.2 cm,angle=-90]{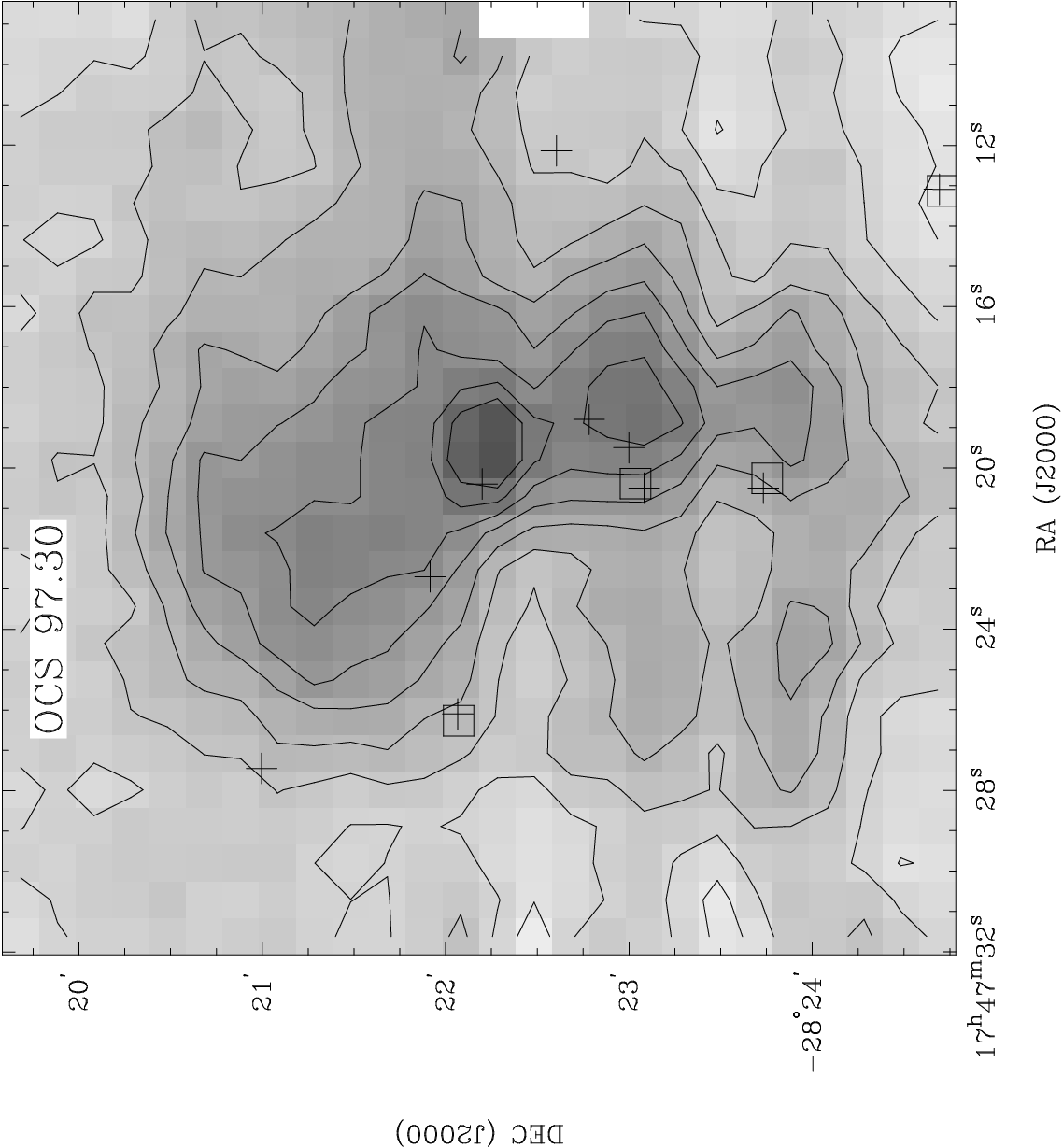}
\includegraphics[width = 5.2 cm,angle=-90]{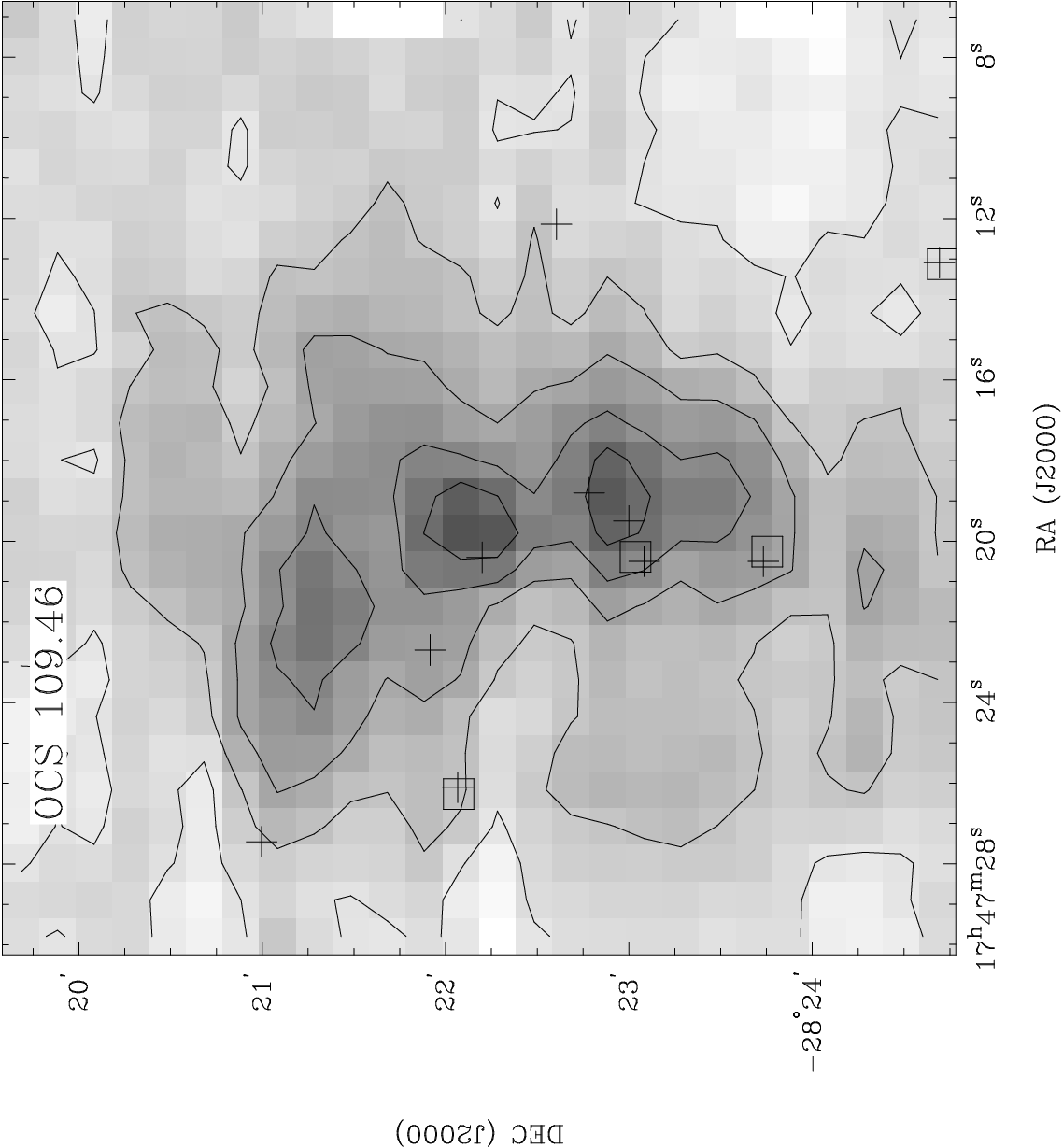}
\caption{Integrated emission for HNCO (contour step 10 K km~s$^{-1}$; 87.93 
GHz, peak
84 K km~s$^{-1}$; 109.91 GHz, peak 113 K km~s$^{-1}$), HOCO$^{+}$ (85.53 GHz, 
step 2 K km~s$^{-1}$, peak 13.4 K km/s) and OCS (85.14 GHz, step 2 
K km~s$^{-1}$, peak 15.6 K km~s$^{-1}$; 97.30
GHz, step 2 K km~s$^{-1}$, peak 24 K km~s$^{-1}$; 109.46 GHz, step 4 K 
km~s$^{-1}$, peak 24 K km~s$^{-1}$).}
\label{hnco_etc_int}
\end{figure*}

The integrated emission from isocyanic acid HNCO 4(0,4)~--~3(0,3) (87.93 GHz)
and 5(0,5)~--~4(0,4) (109.91 GHz)
is shown in Fig. \ref{hnco_etc_int}. The cloud 2 arcmin
north of Sgr B2(M) is particularly prominent in HNCO, as pointed out by
\citet{wil+96} from observations of the 21.98 GHz 1~--~0 line, and as discussed
in \citet{min+98} including observations, as here, of the 4(0,4)~--~3(0,3)
and 5(0,5)~--~4(0,4) lines. We find similar integrated emission in the
4(0,4)~--~3(0,3) to \citet{min+98}, and the velocity gradient in the data cubes,
which they attribute to collapse. The 5(0,5)~--~4(0,4) line at 109.91 GHz here
also shows the ridge west of the Sgr~B2(N), Sgr~B2(M) and Sgr~B2(S) radio
and infrared continuum peaks, but the
ridge is less clearly broken into clumps than in other molecules, such as
HC$_{3}$N. We fit the north cloud at peak 17 47 21.6, -28 21 20,
velocity 65 km~s$^{-1}$, width 25 km~s$^{-1}$
and the peak near Sgr~B2(M) at 17 47 18.2, -28 23 01,
velocity 66 km~s$^{-1}$, width 29 km~s$^{-1}$. From the ratio of the two 
lines, the
peak near Sgr~B2(M) has a higher excitation temperature, but the difference
in energy of the upper levels is too small ($E_{u}/k = 10.5$ to 15.8 K)
to get reliable excitation temperatures from this comparison.

We detect four more weak lines of HNCO at 88.24,
109.49, 109.87 and 110.29 GHz (Table \ref{extra_lines}) concentrated
at Sgr~B2(N) and Sgr~B2(M), which are higher upper energy transitions
excited in the hot cores.

\subsection{HOCO$^{+}$}

The integrated emission from protonated CO$_2$ HOCO$^{+}$ 4(0,4)~--~3(0,3)
(85.53 GHz) is shown in Fig. \ref{hnco_etc_int}. The distribution is
similar to that of HNCO, with the north cloud prominent, as also shown in
\citet{minirvzui88} and \citet{min+98}. We also have data for the
5(0,5)~--~4(0,4) (106.91 GHz) transition, but this is not shown here, as the
image is qualitatively similar, but poorer quality, being
affected by scanning stripes. There is a ridge west of
Sgr~B2(N), Sgr~B2(M) and Sgr~B2(S), similar to that for HNCO. We fit the
north cloud as
peak 17 47 21.1, -28 21 29, velocity 67 km~s$^{-1}$, width 23 km~s$^{-1}$, and
the peak near Sgr~B2(M) at 17 47 18.4, -28 23 21, velocity 63 km~s$^{-1}$,
width 22 km~s$^{-1}$. We also find the peak near Sgr~B2(M) has a higher 
excitation temperature, from the ratio of the peaks in the two lines. The
difference
in energy of the upper levels is however too small ($E_{u}/k = 10.3$ to 15.4 K)
for the calculated excitation temperatures to be very reliable, but the data
do suggest higher excitation temperatures than the 13 K of \citet{minirvzui88}.

\subsection{OCS}

The integrated emission from carbonyl sulphide OCS 7~--~6 (85.14 GHz),
8~--~7 (97.30 GHz) and 9~--~8 (109.46 GHz) is shown in Fig. \ref{hnco_etc_int}.
The emission traces the north cloud, and ridge line, with the peaks
near Sgr~B2(N), Sgr~B2(M) and Sgr~B2(S) quite compact and hence distinct
in the integrated emission. This is unlike the more continuous ridge line
seen in HC$_{3}$N (Fig. \ref{hc3n_ch3cn_int}), as shown by the
higher resolution data from  \citet{gol+87} for the OCS 9~--~8 and HC$_{3}$N
12~--~11 transitions. We fit (a) the north cloud at peak 17 47 21.3,
-28 21 18, velocity 65 km~s$^{-1}$, width 23 km~s$^{-1}$;
(b) the peak near Sgr~B2(N) at 17 47 19.8, -28 22 12, velocity 66 km~s$^{-1}$,
width 21 km~s$^{-1}$;
(c) the peak near Sgr~B2(M) at 17 47 18.6, -28 23 08, velocity 62 km~s$^{-1}$,
width 21 km~s$^{-1}$;
and (d) the peak near Sgr~B2(S) at 17 47 19.5, -28 23 53, velocity 58 
km~s$^{-1}$, width 19 km~s$^{-1}$.
Despite having three transitions, we cannot get reliable excitation temperatures
due to the small range of upper energy levels ($E_{u}/k = 16.3$ to 26.3 K)
and the low signal to noise, but we do note that the peaks near Sgr B2(N) and
Sgr B2(M) have higher excitation temperatures than the north cloud.

We also detect (Table \ref{extra_lines}) the O$^{13}$CS 9~--~8 line at 96.98 GHz,
which has a similar extended distribution, with the strongest peak near
Sgr~B2(N), although the IRAM 30-m survey of \citet{bel+05, bel+07} indicates
this line is blended with several other lines at Sgr~B2(N) and Sgr~B2(M).

\subsection{SO}

The integrated emission from sulphur monoxide SO 2(2) -- 1(1) (86.09 GHz),
3(2) -- 2(1) (99.30 GHz) and  2(3) -- 1(2) (109.25 GHz)
is shown in Fig. \ref{so_etc_int}. The distribution of the 86.09 GHz
and 109.25 GHz transitions is similar, with compact peaks near Sgr B2(N)
and Sgr B2(M), as shown by \citet{gol+87} at higher resolution for the 109.25
GHz transition. The 99.30 GHz transition, however, shows quite a different
distribution tracing the north cloud and ridge-line to the west, and with 
absorption in the data cube at Sgr B2(N) and Sgr B2(M). This is presumably 
because the 86.09 GHz and 109.25 GHz transitions trace the more excited gas
($E_{u}/k = 19.3$ and 21.0 K) than the 99.30 GHz transition ($E_{u}/k = 9.2$ K).
We fit: (a) the north cloud (99.30 GHz) at peak 17 47 21.3, -28 21 20,
velocity 66 km~s$^{-1}$, width 25 km~s$^{-1}$; (b) peak near Sgr B2(N) (86.09 
and 109.25 GHz)
at 17 47 19.3, -28 22 08, velocity 66 km~s$^{-1}$, width 27 km~s$^{-1}$; and 
(c) peak near
Sgr B2(M) at 17 47 19.8, -28 22 56, velocity 61 km~s$^{-1}$, width 20 
km~s$^{-1}$.

We also detect (Table \ref{extra_lines}) the SO 4(5)~--~4(4) line at 100.03 GHz,
concentrated at Sgr~B2(M) and Sgr~B2(N), and the isotopologue $^{34}$SO
3(2)~--~2(1) and 2(3)~--~1(2) lines at 97.72 and 106.74 GHz, at Sgr~B2(M).

\subsection{SO$_{2}$}

\begin{figure*}
\includegraphics[width = 5.2 cm,angle=-90]{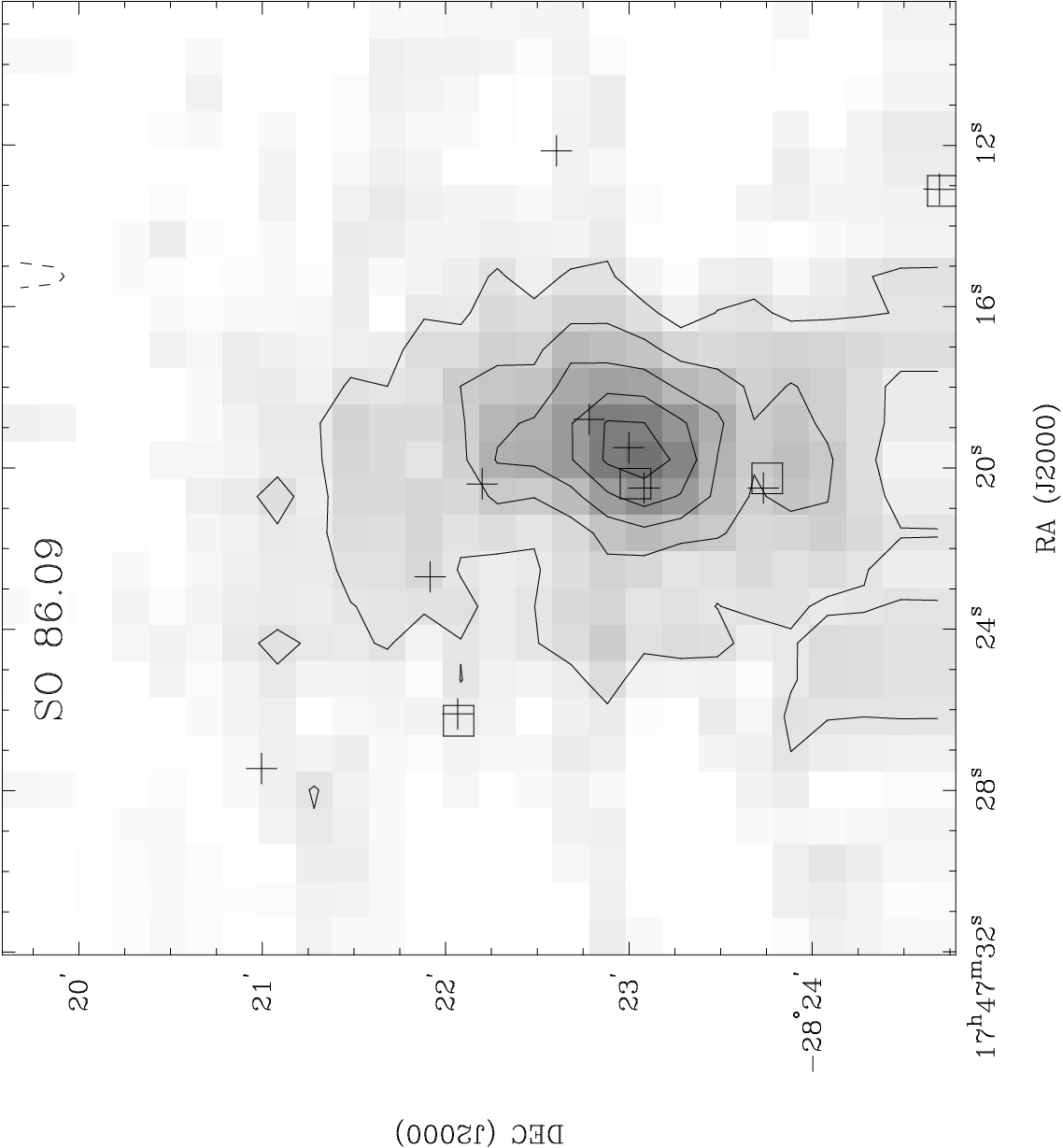}
\includegraphics[width = 5.2 cm,angle=-90]{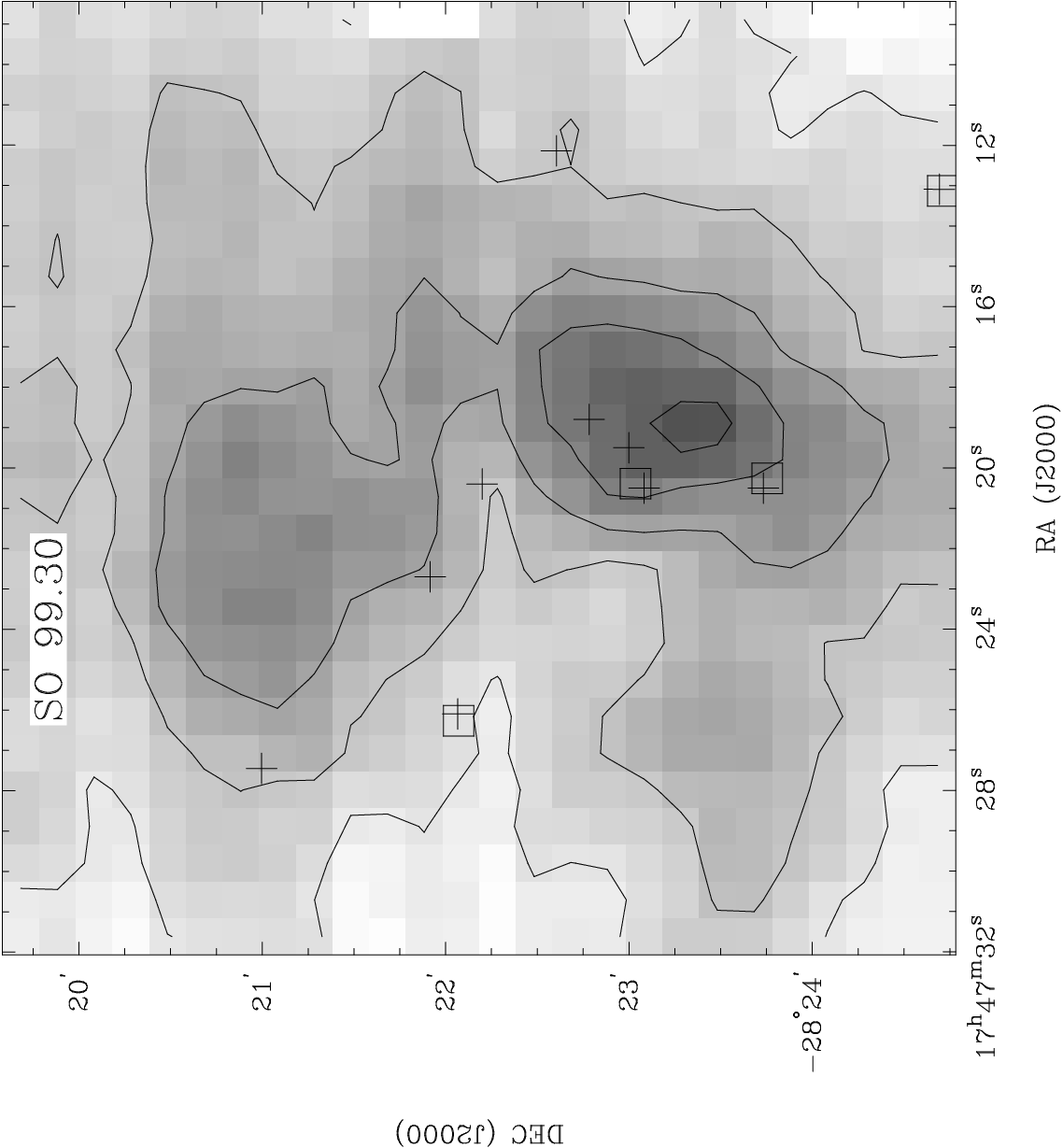}
\includegraphics[width = 5.2 cm,angle=-90]{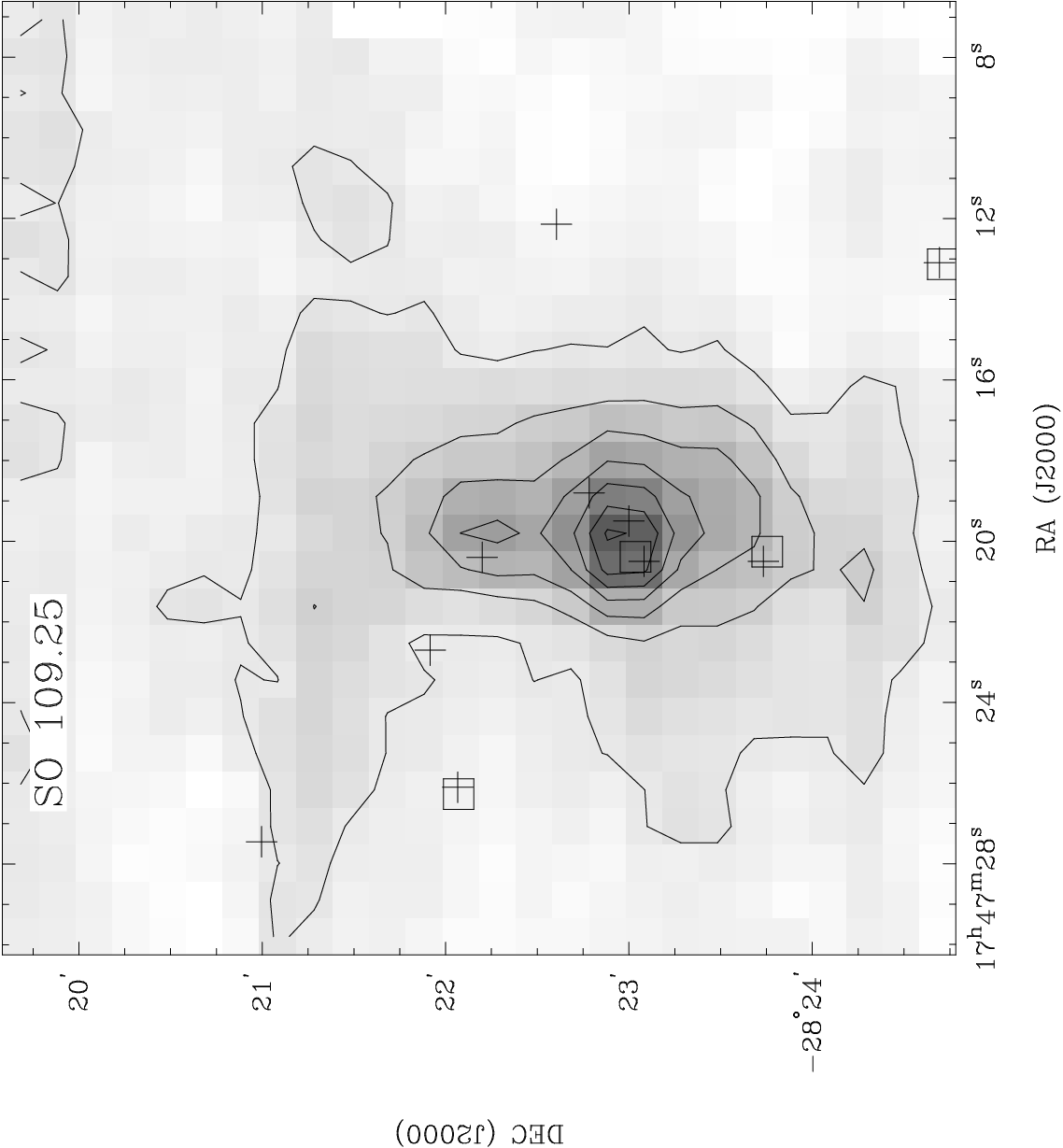}
\includegraphics[width = 5.2 cm,angle=-90]{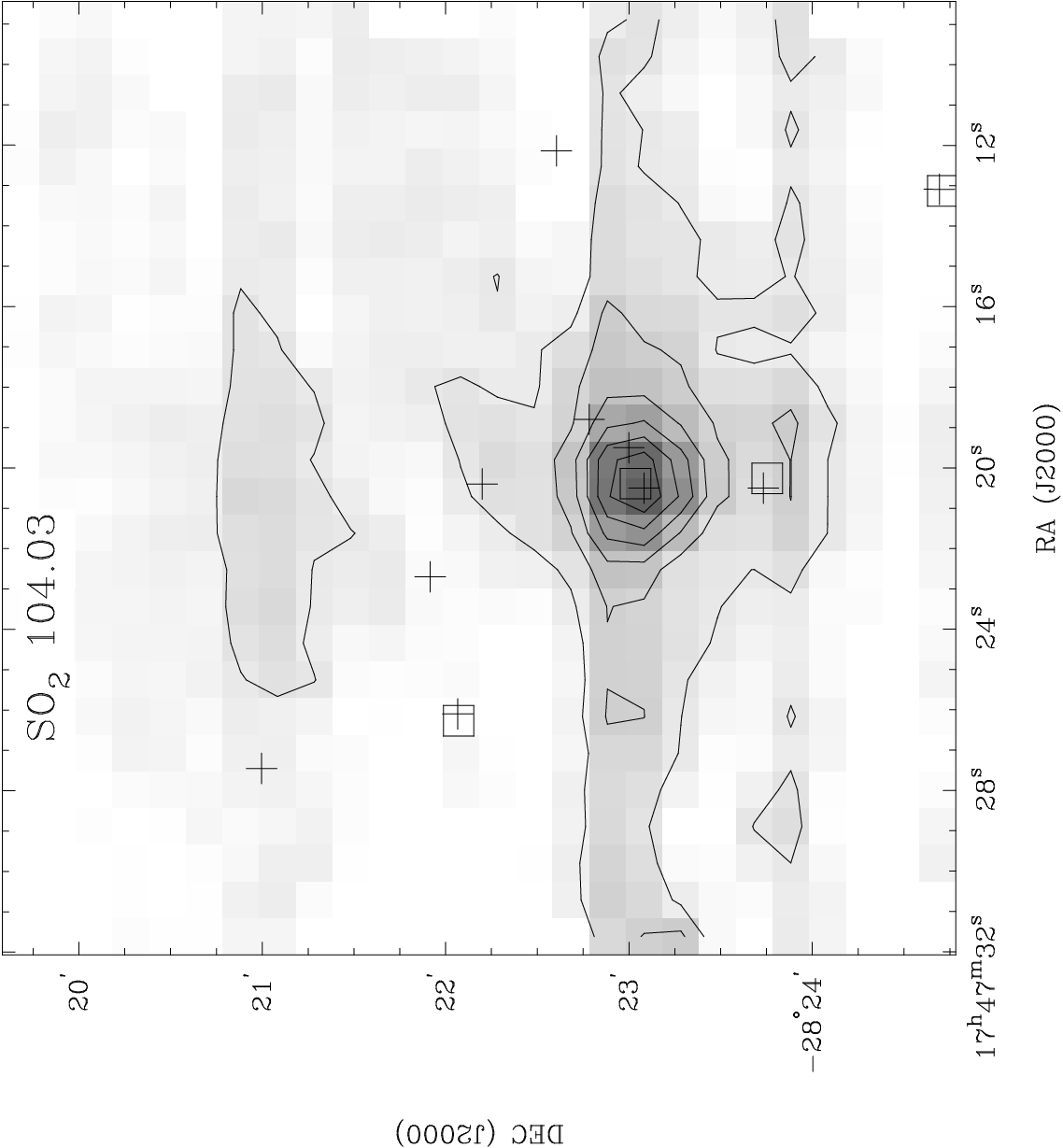}
\includegraphics[width = 5.2 cm,angle=-90]{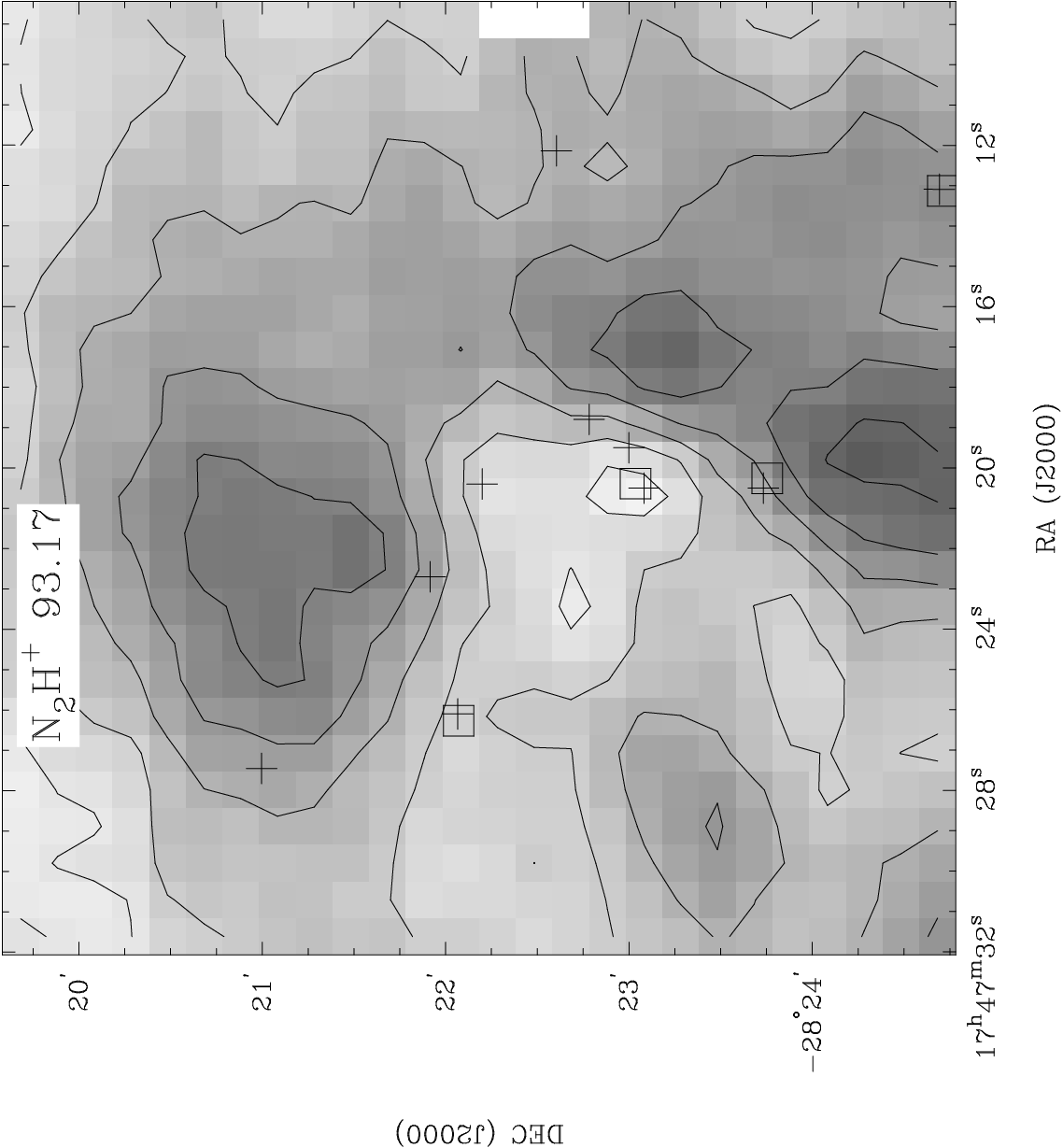}
\includegraphics[width = 5.2 cm,angle=-90]{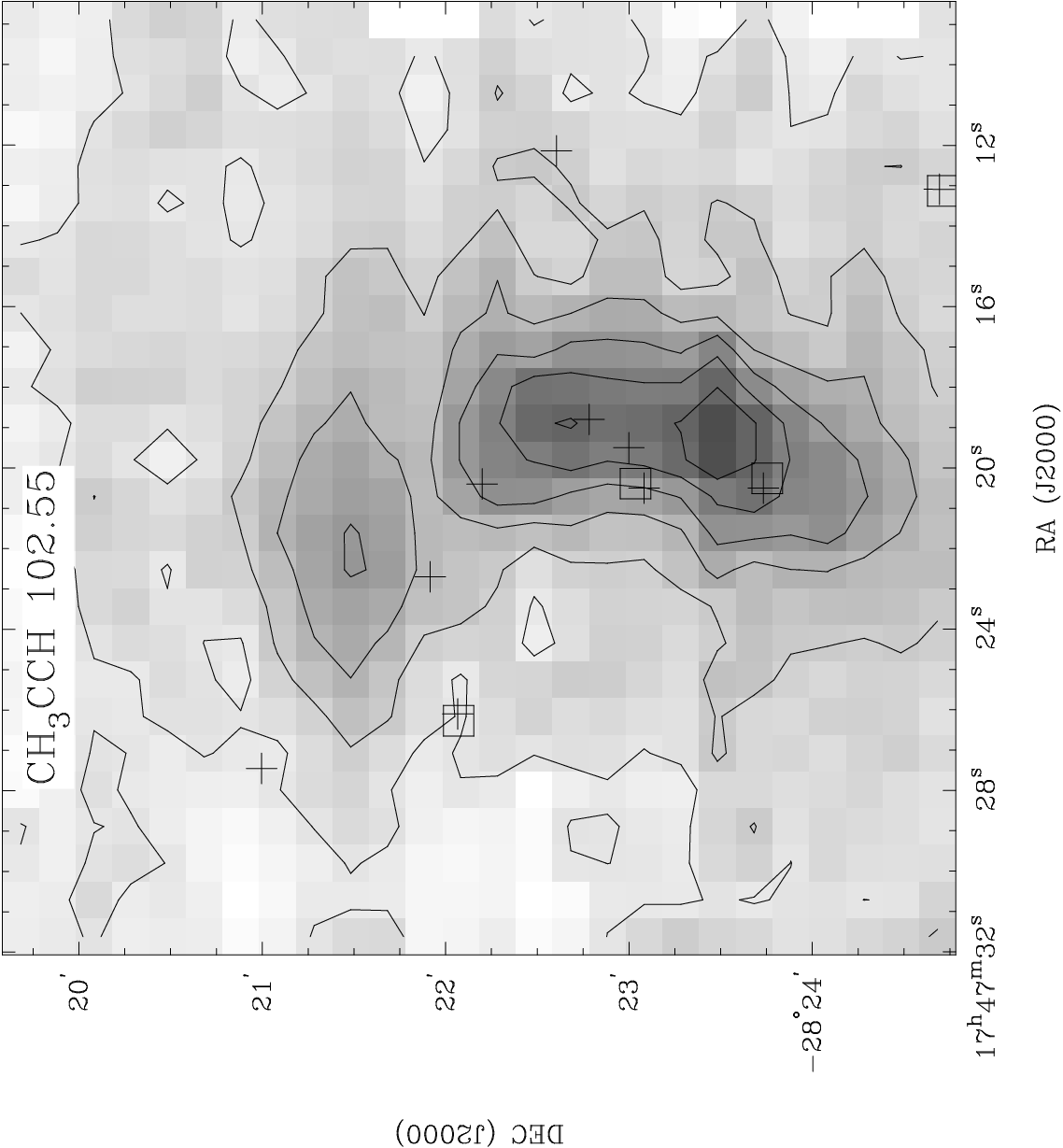}
\caption{Integrated emission for SO (86.09 GHz, contour step 2 K km~s$^{-1}$, 
peak
11.1 K km~s$^{-1}$; 99.30 GHz, step 4 K km~s$^{-1}$, peak 25 K km~s$^{-1}$; 
109.25 GHz, step
5 K km~s$^{-1}$, peak 36 K km~s$^{-1}$), SO$_{2}$ (step 5 K km~s$^{-1}$, 
peak 35 K km~s$^{-1}$),
N$_{2}$H$^{+}$ (step 5 K km~s$^{-1}$, peak 52 K km~s$^{-1}$) and CH$_{3}$CCH 
(step 5 K km~s$^{-1}$, peak 35 K km~s$^{-1}$).}
\label{so_etc_int}
\end{figure*}

The integrated emission from sulphur dioxide SO$_{2}$ 3(1,3) -- 2(0,2)
(104.03 GHz) is also shown in Fig. \ref{so_etc_int}. The peak near Sgr B2(M)
dominates, but there is also weak emission seen from the north cloud.
The peak near Sgr B2(M) is at 17 47 20.4, -28 23 04, velocity 52 km~s$^{-1}$, 
width
26 km~s$^{-1}$, which is lower velocity here than as seen in other lines.
The north cloud has velocity 68 km~s$^{-1}$, width 21 km~s$^{-1}$, and near 
Sgr B2(N)
velocity 61 km~s$^{-1}$, width 29 km~s$^{-1}$. The low level east-west 
extension is an
artifact of the east-west scanning and baseline variations.

We detect eight more lines of SO$_{2}$ at 83.69, 91.55, 97.70, 100.88, 104.24,
107.06, 107.84 and 109.75 GHz (Table \ref{extra_lines}) concentrated
at Sgr~B2(N) and Sgr~B2(M).

\subsection{N$_{2}$H$^{+}$}

\begin{figure*}
\includegraphics[width = 5.2 cm,angle=-90]{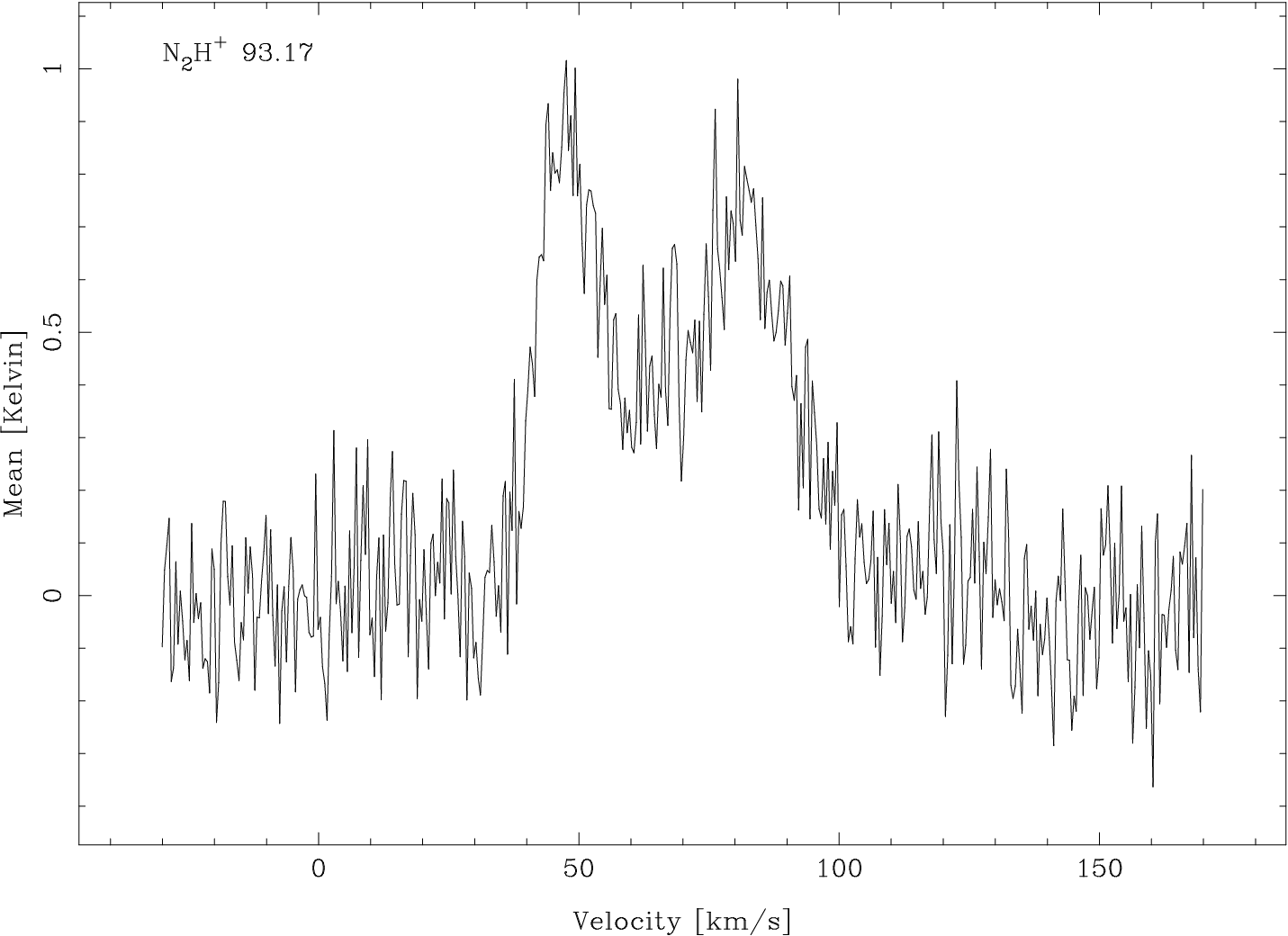}
\includegraphics[width = 5.2 cm,angle=-90]{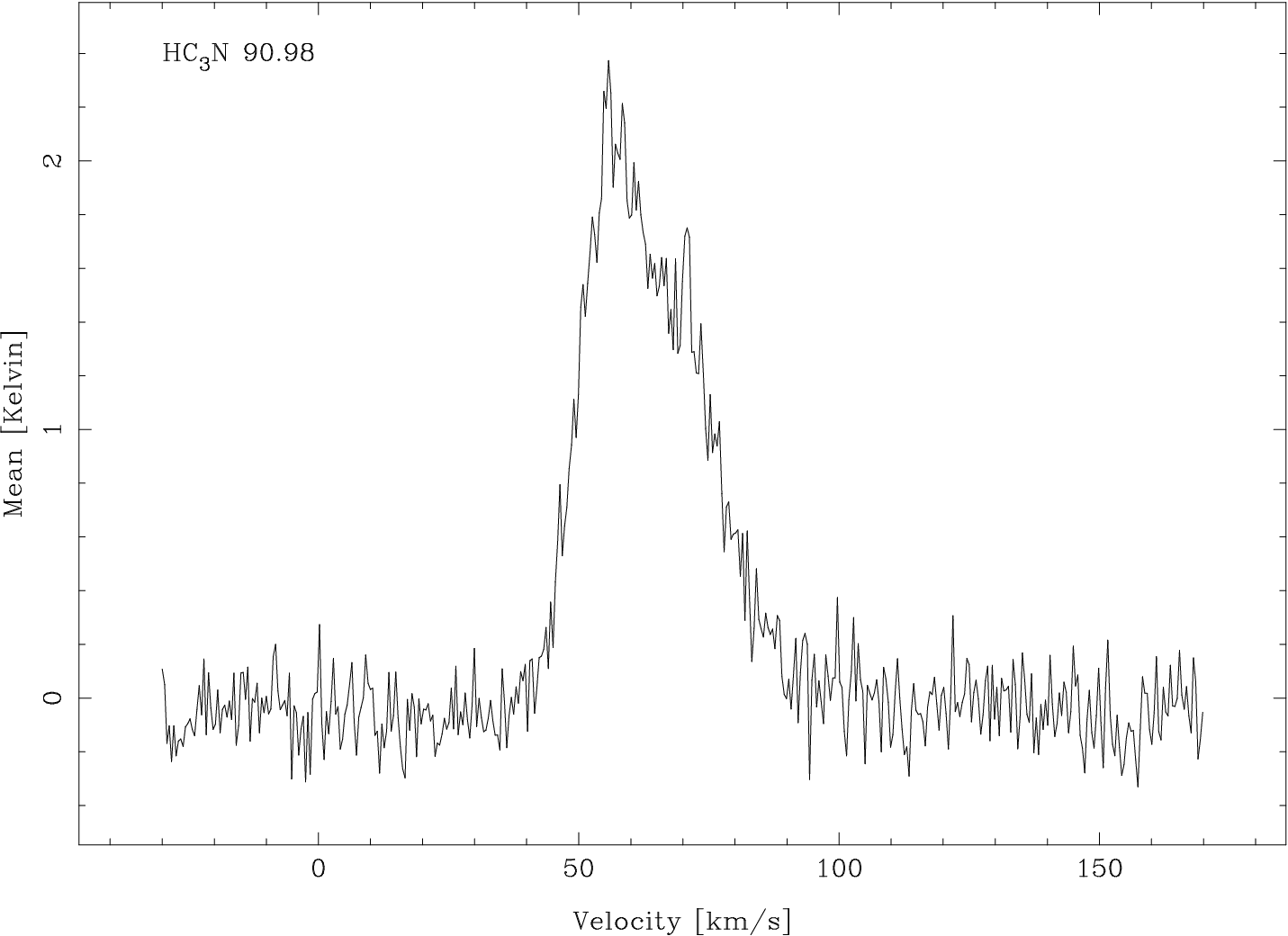}
\caption{Spectra of N$_{2}$H$^{+}$ and HC$_{3}$N near Sgr~B2(M) illustrating
how some spectral lines show absorption at around 60 km~s$^{-1}$, from gas 
seen in emission at this velocity in other lines. There are multiple 
components to the N$_{2}$H$^{+}$ line,
which contributes to broadening the line, but the frequency range
is too small to explain the double profile.}
\label{sample_spectra}
\end{figure*}

The integrated emission from diazenylium N$_{2}$H$^{+}$ 1~--~0 (93.17 GHz)
is shown in Fig. \ref{so_etc_int}. The data cube shows complicated structure,
with deep absorption at Sgr B2(N) and Sgr B2(M) at around 66 km~s$^{-1}$, and
double-peaked spectra over most of the area (Fig. \ref{sample_spectra}), 
which we attribute to widespread
absorption at a similar velocity. There are multiple components to the 1~--~0
line,
which contributes to broadening the fitted line width, but the frequency range
is too small to explain the double profiles. The major features are fitted as:
(a) the north cloud at 17 47 21.4, -28 21 23, velocity 51 and 81 km~s$^{-1}$;
(b) peak to the west of Sgr B2(M) at 17 47 17.2, -28 23 06, velocity 47 and
79 km~s$^{-1}$;
(c) peak south of Sgr B2(S) at 17 47 20.1, -28 24 09, velocity 46 and
71 km~s$^{-1}$;
(d) west ridge at 17 47 15.1, -28 22 38, velocity 120 km~s$^{-1}$, width 22 
km~s$^{-1}$;
and (e) south-east peak at 17 47 27.2, -28 23 22, velocity 43 km~s$^{-1}$, 
width 29 km~s$^{-1}$.

\subsection{CH$_{3}$CCH}

The integrated emission from propyne or methyl acetylene CH$_{3}$CCH 6~--~5
(102.55 GHz) is shown in Fig. \ref{so_etc_int}. We also have data for the
CH$_{3}$CCH 5~--~4 (85.46 GHz) transition, not shown here as it is
qualitatively similar, but weaker and noisier. The distribution shows the
north cloud, and ridge-line west of the radio continuum peaks. The fitted
features are: (a) the north cloud at 17 47 21.5, -28 21 23, velocity 73 
km~s$^{-1}$,
width 23 km~s$^{-1}$;
(b) peak near Sgr B2(N) and Sgr B2(M) at 17 47 18.9, -28 22 33, velocity 70 
km~s$^{-1}$,
width 24 km~s$^{-1}$;
(c) peak near Sgr B2(M) and Sgr B2(S) at 17 47 19.5, -28 23 22, velocity 65 
km~s$^{-1}$,
width 25 km~s$^{-1}$;
and (d) peak south of Sgr B2(S) at 17 47 20.4, -28 24 04, velocity 61 km~s$^{-1}$,
width 23 km~s$^{-1}$. There are multiple blended line components, so the 
spectra are
a fit to the line blend with the velocity calculated using the rest frequency
of one of the components.
The velocity therefore is offset, but the gradient is shown,
similar to that found in CH$_{3}$CCH by
\citet{chuhol83} with lower resolution, but over a larger area.

\subsection{NH$_{2}$CHO and H$_{2}$COH$^{+}$}

\begin{figure*}
\includegraphics[width = 5.2 cm,angle=-90]{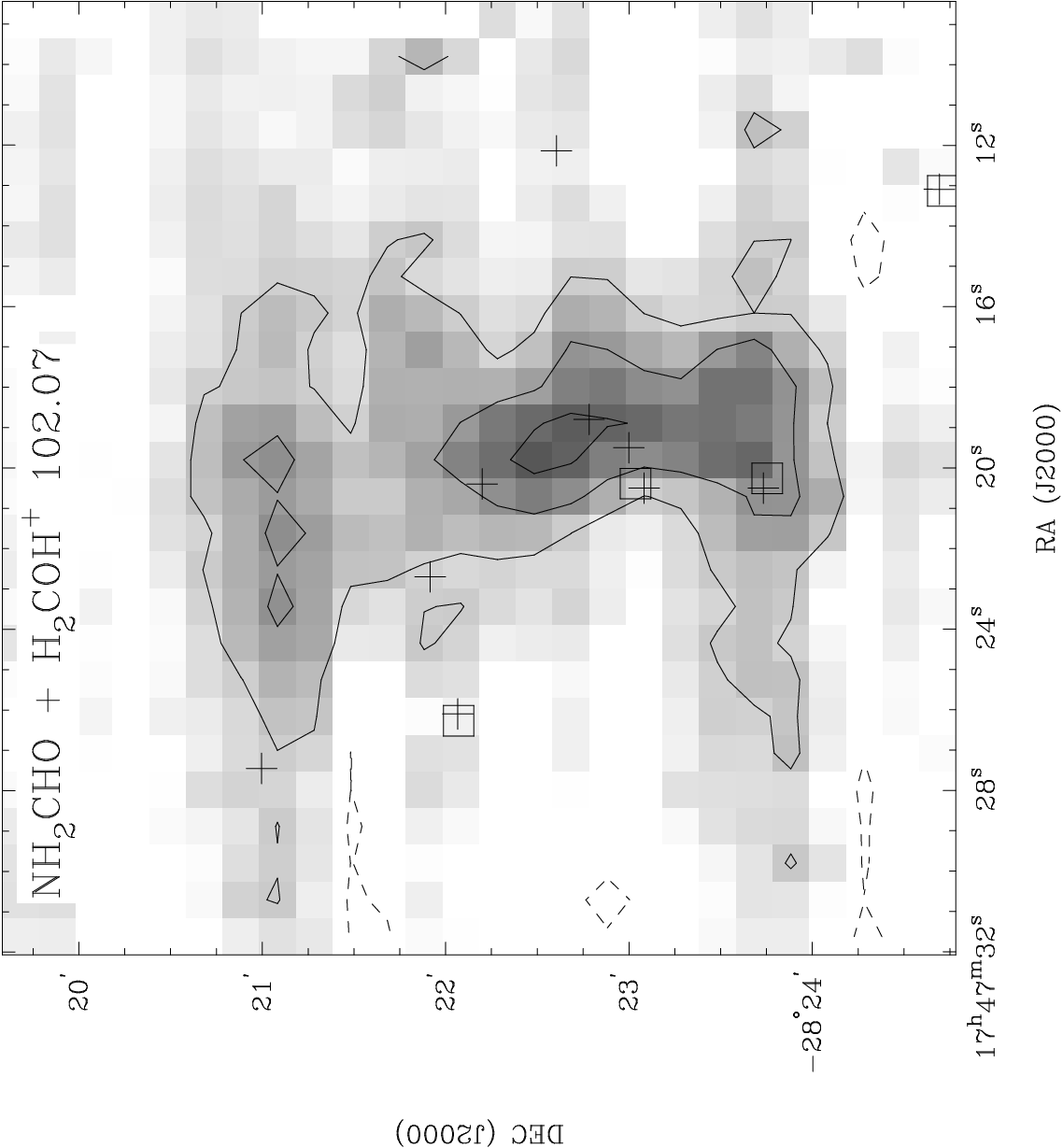}
\includegraphics[width = 5.2 cm,angle=-90]{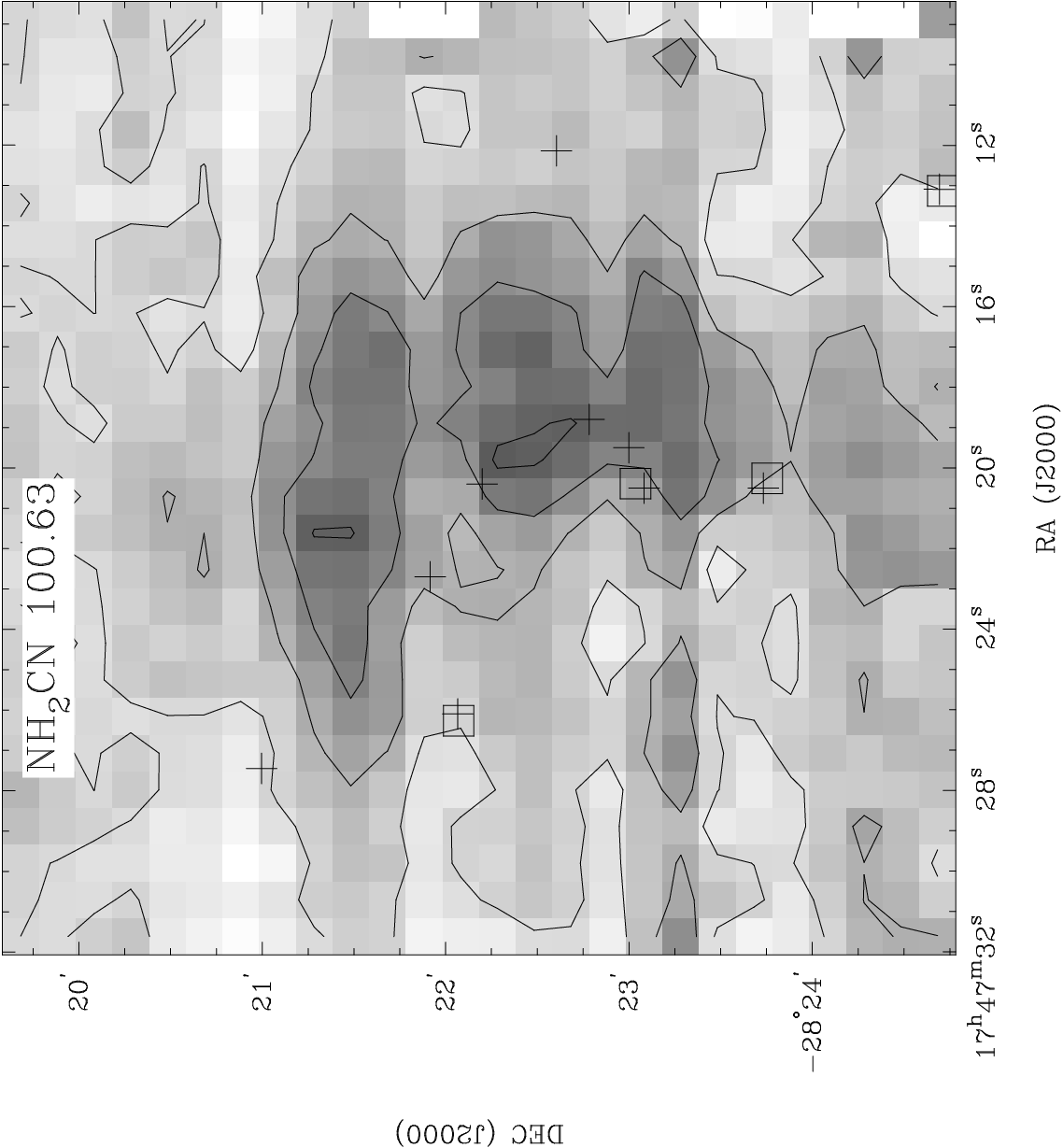}
\includegraphics[width = 5.2 cm,angle=-90]{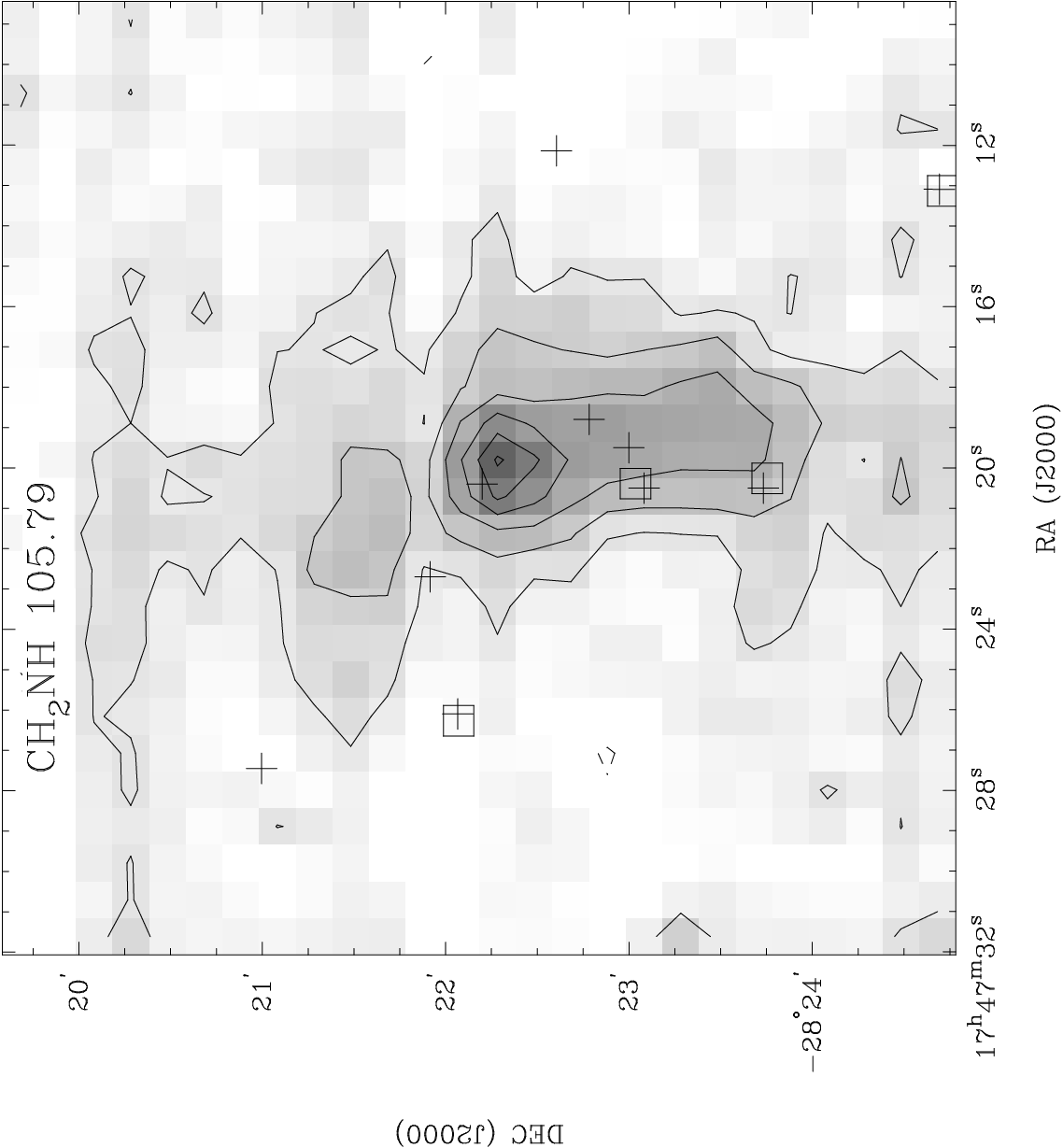}
\includegraphics[width = 5.2 cm,angle=-90]{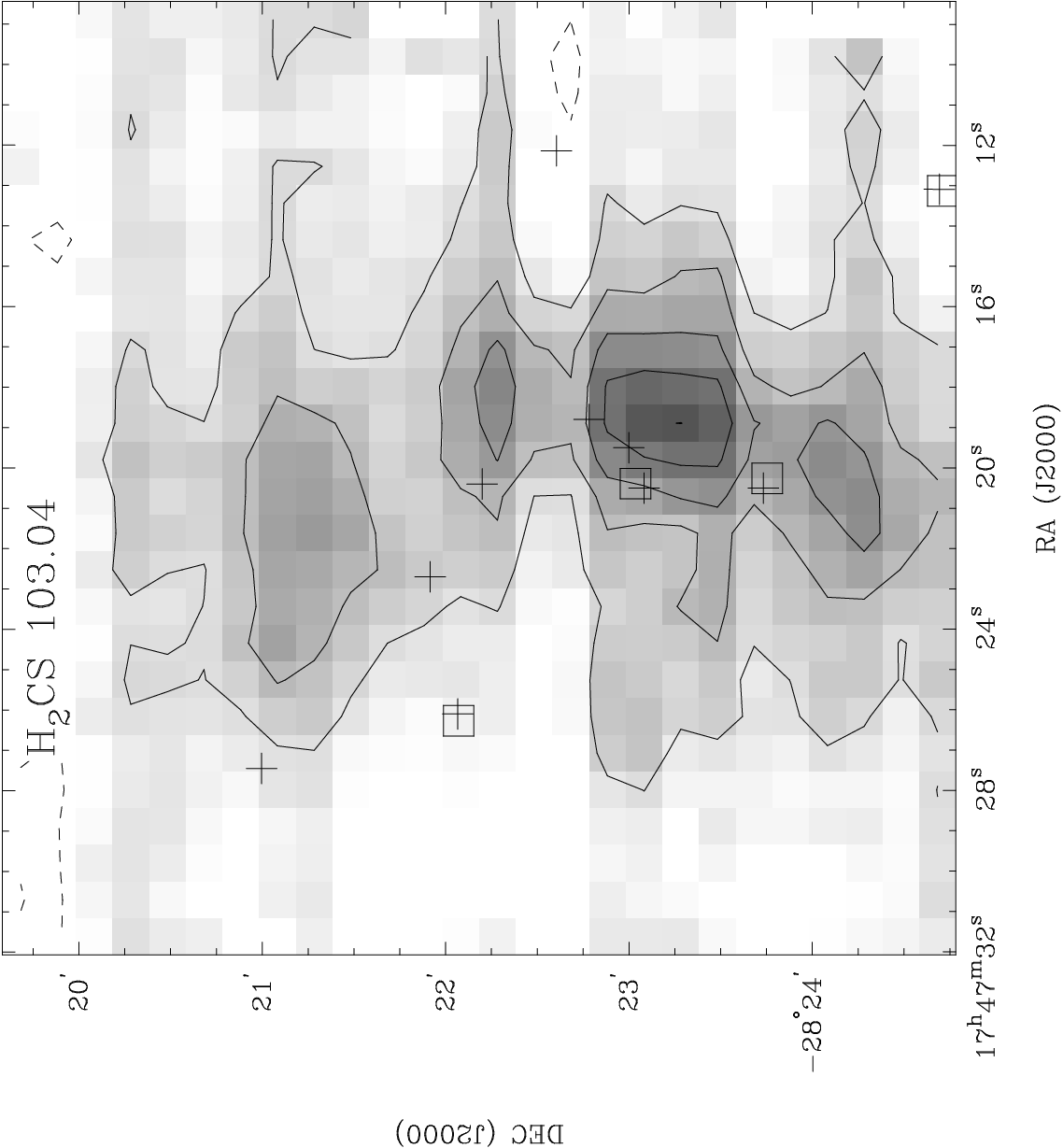}
\includegraphics[width = 5.2 cm,angle=-90]{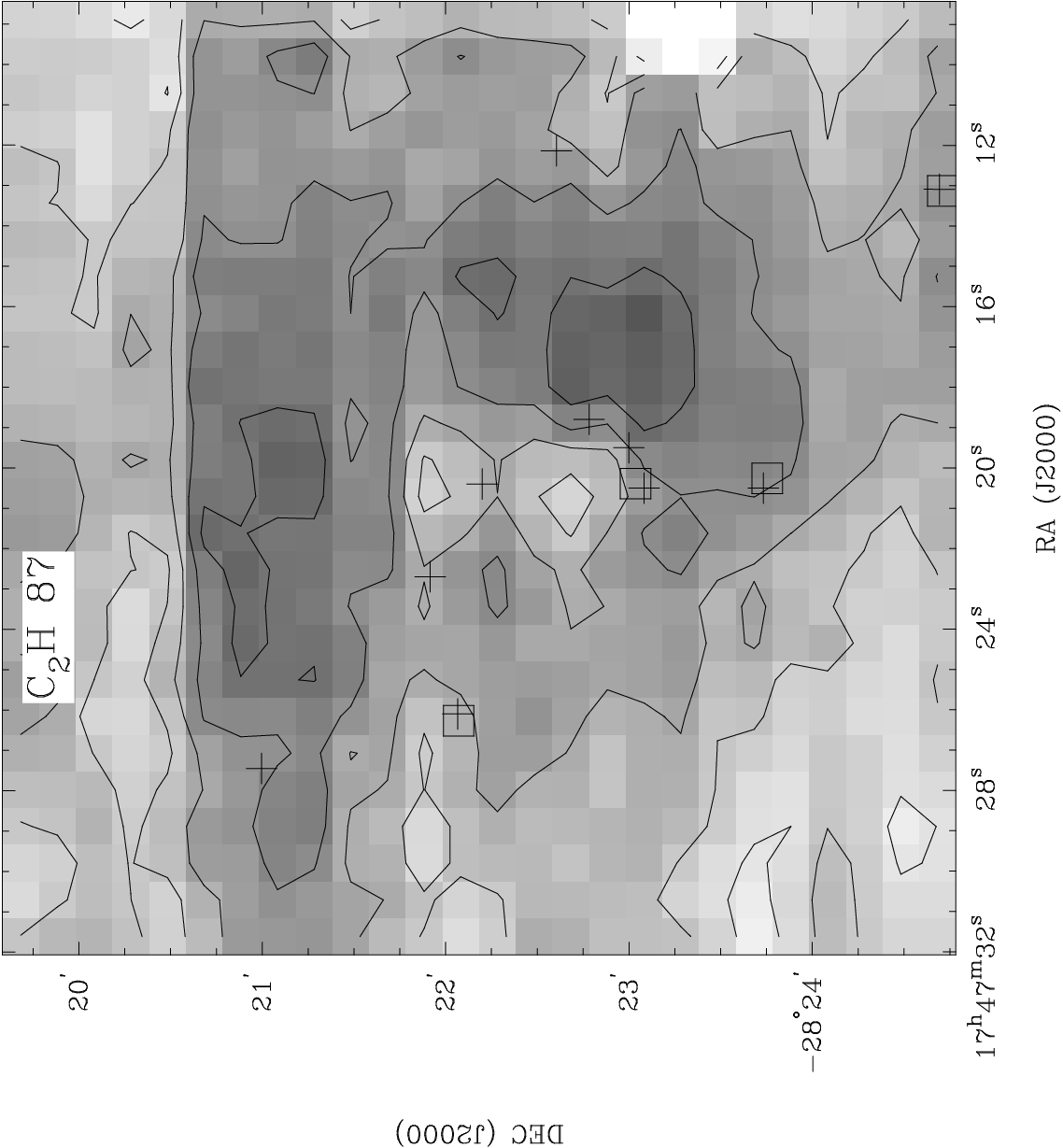}
\includegraphics[width = 5.2 cm,angle=-90]{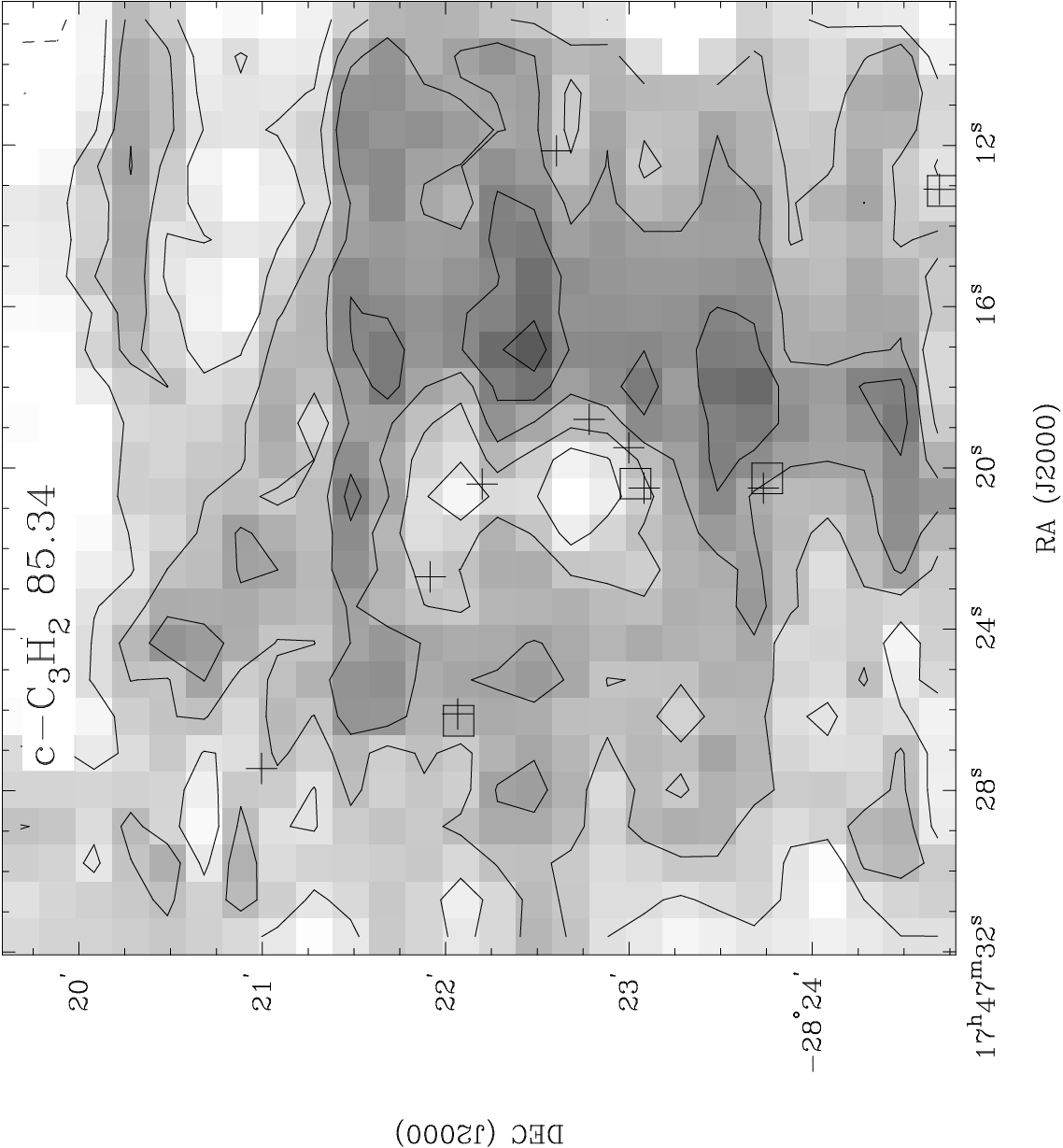}
\caption{Integrated emission for
NH$_{2}$CHO blended with H$_{2}$COH$^{+}$ (contour step 2 K km~s$^{-1}$, 
peak 6.6
K km~s$^{-1}$), NH$_{2}$CN (step 2 K km~s$^{-1}$, peak 8.4 K km~s$^{-1}$), 
CH$_{2}$NH (step 2
K km~s$^{-1}$, peak 12.3 K km~s$^{-1}$), H$_{2}$CS (103.04 GHz, step 2 
K km~s$^{-1}$, peak 10.0
K km~s$^{-1}$), C$_{2}$H (step 5 K km~s$^{-1}$, peak 40 K km~s$^{-1}$) and 
c-C$_{3}$H$_{2}$ (step
2 K km~s$^{-1}$, peak 11.1 K km~s$^{-1}$).}
\label{nh2cho_etc_int}
\end{figure*}

The integrated emission from the line at 102.07 GHz
is shown in Fig. \ref{nh2cho_etc_int}. We identify this as a blend of
formamide NH$_{2}$CHO 5(1,5) -- 4(1,4) and protonated formaldehyde
H$_{2}$COH$^{+}$ 4(0,4) -- 3(1,3), as noted by \citet{ohi+96}.
We identify four peaks, as for other lines like CH$_{3}$CCH 6~--~5
(102.55 GHz) above, but because the line is weak, we do not get good
positional fits for all of them. The features are:
(a) the north cloud with velocity 65 km~s$^{-1}$, width 24 km~s$^{-1}$;
(b) peak near Sgr B2(N) and Sgr B2(M) at 17 47 20.1, -28 22 27, velocity 64 
km~s$^{-1}$,
width 13 km~s$^{-1}$;
(c) peak near Sgr B2(M) and Sgr B2(S) at 17 47 18.7, -28 23 31, velocity 58 
km~s$^{-1}$,
width 12 km~s$^{-1}$;
and (d) peak south of Sgr B2(S) with velocity 53 km~s$^{-1}$,
width 12 km~s$^{-1}$. The velocities are calculated using the rest frequency of
NH$_{2}$CHO 5(1,5) -- 4(1,4), so again will be shifted due to the blending.

We detect eight more weak lines of NH$_{2}$CHO at 85.09, 87.85, 93.87, 105.46,
105.97, 106.13, 106.54 and 109.75 GHz (Table \ref{extra_lines}). The
spatial distribution shows excitation differences, with some of these
concentrated at Sgr~B2(N), and others extended in the north-south ridge line
a bit to the west. This is consistent with higher upper energy lines being
excited in the hot core,
although complicated by several of the lines being blended with other
species.

\subsection{NH$_{2}$CN}

The integrated emission from cyanamide NH$_{2}$CN 5(1,4) -- 4(1,3) (100.63 GHz)
is shown in Fig. \ref{nh2cho_etc_int}. The line is rather weak, and the data
affected by scanning ripples, so we do not fit the positions, but we do
see the four peaks in the data cube and fit velocities:
(a) the north cloud with velocity 59 km~s$^{-1}$, width 27 km~s$^{-1}$;
(b) peak near Sgr B2(N) and Sgr B2(M) with velocity 60 km~s$^{-1}$, width 
35 km~s$^{-1}$;
(c) peak near Sgr B2(M) and Sgr B2(S) with velocity 55 km~s$^{-1}$, width 
26 km~s$^{-1}$;
and (d) peak south of Sgr B2(S) with velocity 55 km~s$^{-1}$, width 17 
km~s$^{-1}$.

\subsection{CH$_{2}$NH}

The integrated emission from methylenimine CH$_{2}$NH 4(0,4) -- 3(1,3)
(105.79 GHz) is shown in Fig. \ref{nh2cho_etc_int}. The peak
close to Sgr B2(N) is strong, and we detect the north cloud and the
ridge line, with fits: (a) the north cloud with velocity 66 km~s$^{-1}$,
width 19 km~s$^{-1}$;
(b) peak near Sgr B2(N) at 17 47 20.0, -28 22 21, velocity 61 km~s$^{-1}$,
width 27 km~s$^{-1}$;
and (c) peak near Sgr B2(M) and Sgr B2(S) with velocity 59 km~s$^{-1}$, width 
18 km~s$^{-1}$.
This line is probably blended with HC$^{13}$CCN 12~--~11 at rest frequency
105.799093 GHz.

\subsection{H$_{2}$CS}

The integrated emission from thioformaldehyde H$_{2}$CS 3(0,3) -- 2(0,2)
(103.04 GHz) is shown in Fig. \ref{nh2cho_etc_int}. We also have data (not
plotted here as the images are similar, but noisy) for H$_{2}$CS
3(1,3) -- 2(1,2) (101.48 GHz) and 3(1,2) -- 2(1,1) (104.62 GHz). We fit four
peaks as: (a) the north cloud at 17 47 21.4, -28 21 25, velocity 68 km~s$^{-1}$,
width 21 km~s$^{-1}$;
(b) peak near Sgr B2(N) at 17 47 19.1, -28 22 23, velocity 67 km~s$^{-1}$,
width 20 km~s$^{-1}$;
(c) peak near Sgr B2(M) and Sgr B2(S) at 17 47 19.2, -28 23 21, velocity 59
km~s$^{-1}$, width 19 km~s$^{-1}$;
and (d) peak south of Sgr B2(S) at 17 47 20.2, -28 24 05, velocity 57 km~s$^{-1}$,
width 17 km~s$^{-1}$.

\subsection{C$_{2}$H}

The integrated emission from ethynyl C$_{2}$H 1~--~0  J=1/2--1/2 (87.32 GHz) and
1~--~0 J=3/2--1/2 (87.40 GHz), is shown in Fig. \ref{nh2cho_etc_int}.
As for CN, above, each set consists of components
and the integrated emission of two sets of lines are very similar, but weak,
so the sum of the two sets is plotted here. The integrated emission image
in Fig. \ref{nh2cho_etc_int} shows widespread emission peaked at the north ridge
and west of Sgr~B2(M), and a deficit of emission at the continuum peaks
Sgr~B2(N) and Sgr~B2(M). The fitting of velocity components in the data cubes
is complicated by the blended components, and the weakness of the
emission, but the emission is peaked around 60 -- 65 km~s$^{-1}$. The deficit
of integrated emission near Sgr~B2(N) and Sgr~B2(M) could be explained
by a real deficit of the molecule in this area, but is more likely
simply be due to
absorption, as \citet{grenym96} show absorption features due to intervening
clouds along the line of sight to Sgr~B2. The offset between the absorption 
and the radio continuum peaks is not considered significant, but rather due
to the baseline stripes in the east-west scanning direction causing north-south
shifts in centres of the weak features.

\subsection{c-C$_{3}$H$_{2}$}

The integrated emission from the cyclic molecule cyclopropenylidene
c-C$_{3}$H$_{2}$ 2(1,2) -- 1(0,1) (85.34 GHz)
is shown in Fig. \ref{nh2cho_etc_int}. This shows similar features to C$_{2}$H,
that is widespread emission with a deficit at continuum peaks
Sgr~B2(N) and Sgr~B2(M). The emission is weak, however, so the integrated
emission does show some spurious striping due to the RA scanning.
\citet{vrt+87} find rotation temperature $T_{rot} = 11 \pm 2$~K for
c-C$_{3}$H$_{2}$ in Sgr~B2, so absorption against the continuum
is quite plausible. As for C$_{2}$H, above, the position offset between the 
absorption and continuum peaks is not considered significant.


\subsection{Other molecules}

We list in Table \ref{extra_lines} nine more molecules, and dozens more
lines, than we have plotted and discussed above, as well as weaker transitions
of the molecules already discussed.
Most of these lines are confined to Sgr~B2(N) or Sgr~B2(M). Some of the
weaker transitions are higher energy states, some vibrationally excited,
of molecules already discussed, which trace these hot cores.

Since the main aim of this paper is the wider scale spatial distribution, we
do not concentrate here on quantitative analysis of the weaker lines. Our
Mopra OTF mapping sacrifices sensitivity on a single position to get the spatial
coverage. Therefore our data on the spectra at the Sgr~B2(N) and Sgr~B2(M)
positions
are not particularly sensitive compared to previous \citep{tur89} and current
\citep{bel+05,hie+05,bel+07} dedicated
spectral line surveys of these well studied sources.
However, by mapping it is useful to determine whether a particular line
is confined to Sgr~B2(N), Sgr~B2(M), or both, or whether it is distributed more
widely. Of the weaker lines (Table \ref{extra_lines}) a substantial fraction
are identified with blends of different species, complicating the analysis.

Four molecules in Table \ref{extra_lines} have extended spatial distribution:
acetaldehyde CH$_{3}$CHO (93.60, 95.95, 95.96 and 98.90 GHz),
dicarbon monosulphide CCS (93.87 and 106.35 GHz), methanethiol
CH$_{3}$SH (101.03 and 101.14 GHz) and ketene CH$_{2}$CHO (101.03 GHz).
These distributions appear similar to that of some other molecules, such as
HC$_{3}$N,
with the north cloud and ridge line to the west of the radio continuum
peaks, but with much lower signal to noise.

The other five molecules are confined to Sgr~B2(N), as this region is
known to be particularly rich in large molecules \citep{snykuamia94, mia+95}.
These molecules are:
ethyl cyanide or propionitrile CH$_{3}$CH$_{2}$CN (22 lines),
acrylonitrile CH$_{2}$CHCN (92.43, 94.28, 94.91, 95.33, 103.57, 104.21 and
106.64 GHz),
methyl formate CH$_{3}$OCHO (89.32, 98.18 and 107.54 GHz),
dimethyl ether CH$_{3}$OCH$_{3}$ (82.46, 100.46 and 105.77 GHz) and
formaldehyde H$_{2}$CO (101.33 GHz).

More sensitive observations of Sgr~B2(N) and Sgr~B2(M) with the IRAM 30-m
\citep{bel+05, bel+07} have been modelled with the XCLASS software
\citep{com+05}, which simultaneously fits multiple lines with the LTE 
approximation and handles line blends well. We note here that for Sgr~B2(N),
in particular, this allows us to identify some extra lines that may confuse
the Mopra images. These lines are:
CH$_{2}$CH$_{3}$CN 10(1,10) -- 9(1,9) at 86.819848 GHz for SiO;
CH$_{2}$CHCN multiplet around 85.5329236 GHz for HOCO$^{+}$;
HC$_{3}$N 12 -- 11 $\nu_{5}=1 l=1f$ at 109.244339 GHz for SO;
CH$_{2}$CHCN 9(1,8) -- 8(1,7) at 87.312827 GHz for C$_{2}$H.
However, the effect of this line confusion does not appear to be significant.

\section{Discussion}
\label{sec:discuss}

\begin{figure*}
\includegraphics[width = 7.0 cm,angle=-90]{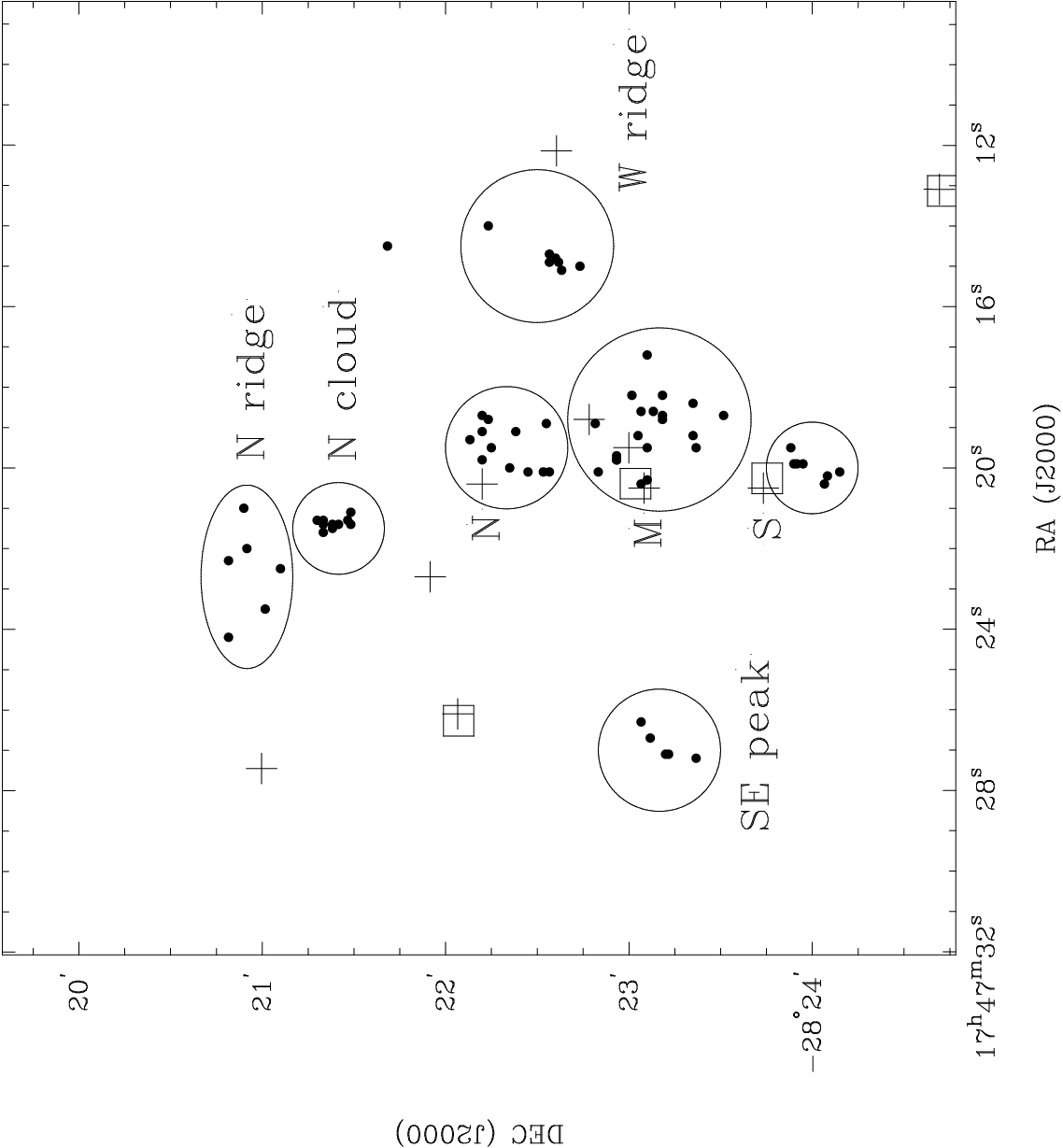}
\includegraphics[width = 7.0 cm,angle=-90]{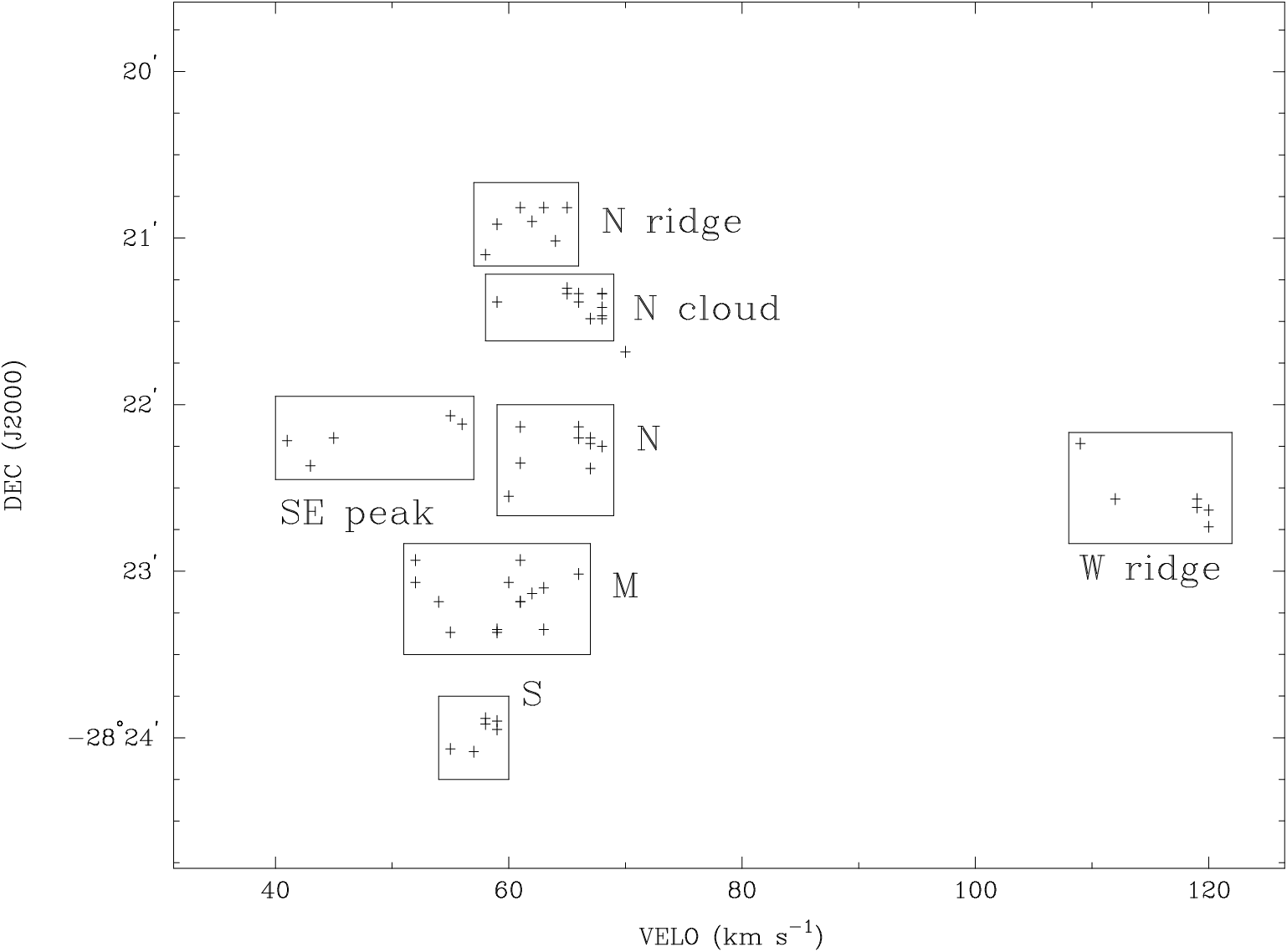}
\caption{The position (left) of the peaks fitted for the 3-mm molecular lines,
and (right) the velocity as a function of declination. Note that in the
velocity-declination plot, the points for the SE peak have been shifted
1~arcmin north for clarity, to avoid overlapping the points near Sgr~B2(M).
In the left diagram, the points are the 3-mm molecular peaks, the crosses radio
sources and the open boxes mid-IR sources.}
\label{mol_peaks}
\end{figure*}

We now consider the comparison of spatial and velocity structure in the Sgr B2
complex, as traced by the different 3-mm lines. Figure \ref{mol_peaks} shows
the positions of the molecular peaks listed in Section \ref{sec:results}, and
Table \ref{peaks_list} lists these fitted peaks.

The strongest lines, such as $^{13}$CO, C$^{18}$O, CS, HCN, HCO$^{+}$, HNC, SiO,
N$_{2}$H$^{+}$ and CH$_{3}$OH (96.74 GHz) show up three features which we have
called here the north ridge, the west ridge and the south-east peak (Tables
\ref{peaks_list} and
\ref{peak_summary}).  These features are detected in the strongest
transitions, which are also the lines which are optically thick in the densest
regions of the complex (near Sgr B2(N) and (M)), 
so the relative prominence of these three features
(Figure \ref{13co_planes})
is partly due to this optical depth effect. However, they do trace the weaker
surrounding structure of the complex. We have not imaged a large enough
area to show the `hole' around 40 km~s$^{-1}$ \citep{sat+00, has+07} well,
so we do not consider the wider surrounding structure.

We note that the south-east peak
is much more obvious in the CS, than in say $^{13}$CO or C$^{18}$O, as noted by
\citet{yus+96}, indicating that there is a chemical difference from the 
main sources.

The west ridge and
south-east peak are offset both
spatially and in velocity (at around 117 and 48 km~s$^{-1}$ respectively) from 
the main north-south axis of the Sgr B2 complex.
The other features (Table
\ref{peak_summary}), that we have called the north ridge, the north cloud,
and the three groups of peaks near Sgr B2(N), Sgr B2(M) and Sgr B2(S) are
in a north-south line, with a velocity gradient, as shown on
the right of Figure \ref{mol_peaks} and in Table \ref{peak_summary}.

The north
ridge is (as noted) seen only in the strongest lines, while the other four
features are best traced by weaker, optically thin lines. We find a spatial
and velocity difference between the north ridge, and nearby chemically enriched
\citep{min+98} north cloud. The north ridge is elongated east-west, so there
is a scatter of the peak positions along this axis, but the north cloud
has a surprisingly tight distribution of peaks fitted from the different lines.

The feature near Sgr B2(S) also has a fairly tight distribution of fitted
peak positions, given the 36 to 39 arcsec beamsize of the observations.
However, there is a significant difference in the peak positions, for the groups
of fitted peaks near Sgr B2(N) and Sgr B2(M). This is attributed to a real
difference in the positions of the peaks in different lines, where some more
excited lines
are associated with the compact hot cores Sgr B2(N) and Sgr B2(M), or
particularly for Sgr B2(N) some molecules are concentrated there. Other
lower excitation lines peak in the ridge further to the west of Sgr B2(N) and
Sgr B2(M) and avoid the hot core positions as the molecules are destroyed there.
The excitation effect can be seen clearly in the SO lines
(Figure \ref{so_etc_int}) where the 86.09 and 109.25 GHz lines are concentrated
at Sgr B2(M), while the 99.30 GHz line traces the ridge-line more to the west.

From some of the stronger lines in Table \ref{extra_lines} which are
concentrated at Sgr B2(N)
and Sgr B2(M) we fit the hot core positions and velocities as: Sgr B2(N)
17 47 19.9, -28 22 11, velocity 63 km~s$^{-1}$, width 24 km~s$^{-1}$;
and Sgr B2(M) 17 47 20.3, -28 22 58, velocity 59 km~s$^{-1}$, width 22 
km~s$^{-1}$. From
these lines (mostly CH$_{3}$OH and CH$_{3}$CH$_{2}$CN for Sgr B2(N) and SO
and SO$_{2}$ for Sgr B2(M)) we find that the hot cores are unresolved
relative to the 36 to 39 arcsec Mopra beam.

\begin{figure}
\includegraphics[width = 5.2 cm,angle=-90]{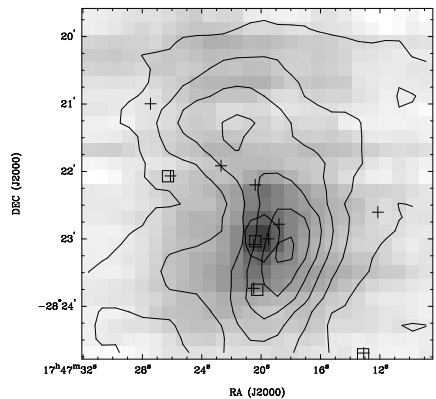}
\caption{The C$^{18}$O integrated emission as grey-scale, with the 90.98 GHz
HC$_{3}$N integrated emission as contours, showing how molecules such as
HC$_{3}$N peak in the ridge-line to the west of the hot cores.}
\label{c18o+hc3n}
\end{figure}

The distribution of optically thin C$^{18}$O, which should be a good
tracer of CO column density, and hence H$_{2}$ column density, peaks at the
Sgr B2(N) and Sgr B2(M) cores, whereas there are several molecules, such as
HC$_{3}$N, CH$_{3}$CN, CH$_{3}$OH and OCS, which peak in the ridge-line to the
west of the cores. This is shown in Figure \ref{c18o+hc3n}, and in
the integrated emission images, by
the alignment of the distributions relative to the reference crosses (radio
peaks) and squares (mid-IR peaks).

\begin{table}
\caption{Compilation of fitted peaks of the molecular features in the Sgr B2
complex. We are mostly considering here the spatial and velocity structure,
but include in this table, for completeness, the intensity of the
fitted peaks in the T$_{A}^{*}$ scale. For some molecules with multiple 
transitions, where we have used the mean spatial position and velocity for
higher signal to noise, we list the intensities for the different transitions
in consecutive lines, in order of frequency, as given in Table \ref{lines}.}
\begin{tabular}{lccccc}
\hline
Feature /             &  R.A.      & Dec.     & Vel. & Width & T$_{A}^{*}$ \\
Molecule              &  (J2000)   & (J2000)  & km~s$^{-1}$ & km~s$^{-1}$ & K \\
\hline
{\bf N ridge}         &            &           &     &    &      \\
$^{13}$CO             & 17 47 24.2 & -28 20 49 &  65 & 42 & 3.03 \\
C$^{18}$O             &            &           &  63 & 36 & 0.56 \\
CS                    & 17 47 22.3 & -28 20 49 &  61 & 57 & 1.01 \\
HNC                   & 17 47 22.0 & -28 20 55 &  59 & 54 & 0.96 \\
SiO                   & 17 47 22.5 & -28 21 06 &  58 & 41 & 0.32 \\
HC$_{3}$N             & 17 47 21.0 & -28 20 54 &  62 & 27 & 1.15 \\
                      &            &           &     &    & 1.48 \\
                      &            &           &     &    & 1.30 \\
                      &            &           &     &    & 1.33 \\
CH$_{3}$CN            & 17 47 23.5 & -28 21 01 &  64 & 41 & 0.37 \\
                      &            &           &     &    & 0.54 \\
                      &            &           &     &    &      \\
{\bf N cloud}         &            &           &     &    &      \\
HC$_{3}$N             & 17 47 21.4 & -28 21 29 &  68 & 23 & 1.48 \\
                      &            &           &     &    & 1.98 \\
                      &            &           &     &    & 1.79 \\
                      &            &           &     &    & 1.88 \\
CH$_{3}$CN            & 17 47 21.3 & -28 21 28 &  68 & 35 & 0.51 \\
                      &            &           &     &    & 0.70 \\
CH$_{3}$OH            & 17 47 21.4 & -28 21 20 &  68 & 25 & 3.53 \\
                      &            &           &     &    & 0.40 \\
                      &            &           &     &    & 0.69 \\
                      &            &           &     &    & 2.17 \\
                      &            &           &     &    & 0.81 \\
                      &            &           &     &    & 0.53 \\
$^{13}$CH$_{3}$OH     &            &           &     &    & 0.52 \\
HNCO                  & 17 47 21.6 & -28 21 20 &  65 & 25 & 2.94 \\
                      &            &           &     &    & 4.03 \\
HOCO$^{+}$            & 17 47 21.1 & -28 21 29 &  67 & 23 & 0.54 \\
                      &            &           &     &    & 0.60 \\
OCS                   & 17 47 21.3 & -28 21 18 &  65 & 23 & 0.54 \\
                      &            &           &     &    & 0.68 \\
                      &            &           &     &    & 0.74 \\
SO                    & 17 47 21.3 & -28 21 20 &  66 & 25 & 0.10 \\
                      &            &           &     &    & 0.56 \\
                      &            &           &     &    & 0.33 \\
SO$_{2}$              &            &           &  68 & 21 & 0.30 \\
N$_{2}$H$^{+}$        & 17 47 21.4 & -28 21 23 &     &    & 1.39 \\
CH$_{3}$CCH           & 17 47 21.5 & -28 21 23 & (73) & (23) & 0.30 \\
                      &            &           &     &    & 0.46 \\
NH$_{2}$CHO           &            &           & (65) & (24) & 0.16 \\
/H$_{2}$COH$^{+}$     &            &           &     &    &      \\
NH$_{2}$CN            &            &           &  59 & 27 & 0.30 \\
CH$_{2}$NH            &            &           &  66 & 19 & 0.22 \\
H$_{2}$CS             & 17 47 21.4 & -28 21 25 &  68 & 21 & 0.66 \\
                      &            &           &     &    & 0.29 \\
                      &            &           &     &    & 0.47 \\
                      &            &           &     &    &      \\
{\bf near}            &            &           &     &    &      \\
{\bf Sgr B2(N)}       &            &           &     &    &      \\
C$^{18}$O             & 17 47 19.5 & -28 22 15 &  68 & 22 & 1.60 \\
HC$_{3}$N             & 17 47 18.7 & -28 22 12 &  67 & 23 & 1.66 \\
                      &            &           &     &    & 2.29 \\
                      &            &           &     &    & 2.16 \\
                      &            &           &     &    & 2.69 \\
\hline
\end{tabular}
\label{peaks_list}
\end{table}

\begin{table}
\begin{tabular}{lccccc}
\multicolumn{2} {l} {{\bf Table \ref{peaks_list} } continued.} &  &  &  & \\
\hline
Feature /             &  R.A.      & Dec.     & Vel. & Width & T$_{A}^{*}$ \\
Molecule              &  (J2000)   & (J2000)  & km~s$^{-1}$ & km~s$^{-1}$ & K \\
\hline
CH$_{3}$CN            & 17 47 19.1 & -28 22 12 &  66 & 30 & 0.72 \\
                      &            &           &     &    & 1.09 \\
CH$_{3}$OH            & 17 47 18.8 & -28 22 14 &  67 & 19 & 2.69 \\
                      &            &           &     &    & 1.05 \\
                      &            &           &     &    & 1.18 \\
                      &            &           &     &    & 1.77 \\
                      &            &           &     &    & 1.19 \\
                      &            &           &     &    & 0.32 \\
$^{13}$CH$_{3}$OH     &            &           &     &    & 0.43 \\
OCS                   & 17 47 19.8 & -28 22 12 &  66 & 21 & 0.62 \\
                      &            &           &     &    & 1.11 \\
                      &            &           &     &    & 1.28 \\
SO                    & 17 47 19.3 & -28 22 08 &  66 & 27 & 0.35 \\
                      &            &           &     &    & 0.36 \\
                      &            &           &     &    & 0.87 \\
SO$_{2}$              &            &           &  61 & 29 & 0.23 \\
CH$_{3}$CCH           & 17 47 18.9 & -28 22 33 & (70) & (24) & 0.36 \\
                      &            &           &     &    & 0.61 \\
NH$_{2}$CHO           & 17 47 20.1 & -28 22 27 & (64) & (13) & 0.42 \\
/H$_{2}$COH$^{+}$     &            &           &     &    &      \\
NH$_{2}$CN            &            &           &  60 & 35 & 0.23 \\
CH$_{2}$NH            & 17 47 20.0 & -28 22 21 &  61 & 27 & 0.46 \\
H$_{2}$CS             & 17 47 19.1 & -28 22 23 &  67 & 20 & 0.67 \\
                      &            &           &     &    & 0.35 \\
                      &            &           &     &    & 0.39 \\
                      &            &           &     &    &      \\
{\bf near}            &            &           &     &    &      \\
{\bf Sgr B2(M)}       &            &           &     &    &      \\
C$^{18}$O             & 17 47 20.3 & -28 23 06 &  63 & 21 & 1.87 \\
CS                    & 17 47 19.2 & -28 23 03 &     &    & 2.70 \\
$^{13}$CS             & 17 47 18.7 & -28 23 11 &  54 & 15 & 0.35 \\
C$^{34}$S             &            &           &     &    & 0.64 \\
HCO$^{+}$             & 17 47 20.1 & -28 22 34 &     &    & 1.44 \\
HCN                   & 17 47 20.1 & -28 22 32 &     &    & 1.66 \\
HNC                   & 17 47 19.8 & -28 22 56 &      &    & 1.37 \\
H$^{13}$CO$^{+}$      &            &           & (50) & (8) & 0.48 \\
H$^{13}$CN            &            &           & (47) & (12) & 0.41 \\
HN$^{13}$C            &            &           &  52 &  16 & 0.32 \\
SiO                   & 17 47 18.9 & -28 22 49 &     &    & 0.54 \\
CN                    & 17 47 20.1 & -28 22 50 &     &    & 0.74 \\
HC$_{3}$N             & 17 47 18.6 & -28 23 04 &  60 & 22 & 2.02 \\
                      &            &           &     &    & 2.95 \\
                      &            &           &     &    & 2.83 \\
                      &            &           &     &    & 3.72 \\
CH$_{3}$CN            & 17 47 18.8 & -28 23 11 &  61 & 33 & 0.74 \\
                      &            &           &     &    & 1.08 \\
CH$_{3}$OH            & 17 47 18.2 & -28 23 11 &  61 & 22 & 3.01 \\
                      &            &           &     &    & 0.99 \\
                      &            &           &     &    & 1.18 \\
                      &            &           &     &    & 1.91 \\
                      &            &           &     &    & 1.19 \\
                      &            &           &     &    & 0.65 \\
$^{13}$CH$_{3}$OH     &            &           &     &    & 0.41 \\
HNCO                  & 17 47 18.2 & -28 23 01 &  66 & 29 & 1.83 \\
                      &            &           &     &    & 2.97 \\
HOCO$^{+}$            & 17 47 18.4 & -28 23 21 &  63 & 22 & 0.23 \\
                      &            &           &     &    & 0.35 \\
OCS                   & 17 47 18.6 & -28 23 08 &  62 & 21 & 0.54 \\
                      &            &           &     &    & 0.81 \\
                      &            &           &     &    & 0.95 \\
SO                    & 17 47 19.8 & -28 22 56 &  61 & 20 & 0.59 \\
                      &            &           &     &    & 0.96 \\
                      &            &           &     &    & 1.66 \\
SO$_{2}$              & 17 47 20.4 & -28 23 04 &  52 & 26 & 1.00 \\
\hline
\end{tabular}
\end{table}

\begin{table}
\begin{tabular}{lccccc}
\multicolumn{2} {l} {{\bf Table \ref{peaks_list} } continued.} &  &  &  & \\
\hline
Feature /             &  R.A.      & Dec.     & Vel. & Width & T$_{A}^{*}$ \\
Molecule              &  (J2000)   & (J2000)  & km~s$^{-1}$ & km~s$^{-1}$ & K \\
\hline
N$_{2}$H$^{+}$        & 17 47 17.4 & -28 23 06 &     &    & 0.92 \\
CH$_{3}$CCH           & 17 47 19.5 & -28 23 22 & (65) & (25) & 0.39 \\
                      &            &           &     &    & 0.79 \\
NH$_{2}$CHO           & 17 47 18.7 & -28 23 31 & (58) & (12) & 0.40 \\
/H$_{2}$COH$^{+}$     &            &           &     &    &      \\
NH$_{2}$CN            &            &           &  55 & 26 & 0.21 \\
CH$_{2}$NH            &            &           &  59 & 18 & 0.35 \\
H$_{2}$CS             & 17 47 19.2 & -28 23 21 &  59 & 19 & 0.90 \\
                      &            &           &     &    & 0.51 \\
                      &            &           &     &    & 0.67 \\
AlF                   & 17 47 19.7 & -28 22 56 &     &    & 0.29 \\
                      &            &           &     &    &      \\
{\bf near}            &            &           &     &    &      \\
{\bf Sgr B2(S)}       &            &           &     &    &      \\
HC$_{3}$N             & 17 47 19.9 & -28 23 55 &  58 & 20 & 1.53 \\
                      &            &           &     &    & 2.63 \\
                      &            &           &     &    & 2.45 \\
                      &            &           &     &    & 2.52 \\
CH$_{3}$CN            & 17 47 19.9 & -28 23 54 &  59 & 32 & 0.73 \\
                      &            &           &     &    & 0.94 \\
CH$_{3}$OH            & 17 47 19.9 & -28 23 57 &  59 & 20 & 2.00 \\
                      &            &           &     &    & 0.75 \\
                      &            &           &     &    & 1.10 \\
                      &            &           &     &    & 2.04 \\
                      &            &           &     &    & 1.07 \\
                      &            &           &     &    & 0.66 \\
$^{13}$CH$_{3}$OH     &            &           &     &    & 0.42 \\
OCS                   & 17 47 19.5 & -28 23 53 &  58 & 19 & 0.42 \\
                      &            &           &     &    & 0.58 \\
                      &            &           &     &    & 0.67 \\
N$_{2}$H$^{+}$        & 17 47 20.1 & -28 24 09 &     &    & 1.41 \\
CH$_{3}$CCH           & 17 47 20.4 & -28 24 04 & (61) & (23) & 0.20 \\
                      &            &           &     &    & 0.46 \\
NH$_{2}$CHO           &            &           & (53) & (12) & 0.32 \\
/H$_{2}$COH$^{+}$     &            &           &     &    &      \\
NH$_{2}$CN            &            &           &  55 & 17 & 0.25 \\
H$_{2}$CS             & 17 47 20.2 & -28 24 05 &  57 & 17 & 0.66 \\
                      &            &           &     &    & 0.43 \\
                      &            &           &     &    & 0.48 \\
                      &            &           &     &    &      \\
{\bf W ridge}         &            &           &     &    &      \\
$^{13}$CO             & 17 47 14.0 & -28 22 14 & 109 & 32 & 1.68 \\
CS                    & 17 47 14.9 & -28 22 37 & 119 & 14 & 0.42 \\
HCO$^{+}$             & 17 47 14.8 & -28 22 36 &     &    &      \\
HCN                   & 17 47 14.7 & -28 22 34 & 119 & 24 & 0.63 \\
HNC                   & 17 47 14.9 & -28 22 34 & 112 & 21 & 0.27 \\
CH$_{3}$OH            & 17 47 15.0 & -28 22 44 & 120 & 21 & 0.60 \\
N$_{2}$H$^{+}$        & 17 47 15.1 & -28 22 38 & 120 & 22 & 0.35 \\
                      &            &           &     &    &      \\
{\bf SE peak}         &            &           &     &    &      \\
CS                    & 17 47 27.1 & -28 23 13 &  41 & 20 & 1.53 \\
SiO                   & 17 47 27.1 & -28 23 12 &  45 & 29 & 0.36 \\
HC$_{3}$N             & 17 47 26.3 & -28 23 04 &  55 & 23 & 0.70 \\
                      &            &           &     &    & 0.98 \\
                      &            &           &     &    & 0.94 \\
                      &            &           &     &    & 1.06 \\
CH$_{3}$OH            & 17 47 26.7 & -28 23 07 &  56 & 34 & 1.72 \\
N$_{2}$H$^{+}$        & 17 47 27.2 & -28 23 22 &  43 & 29 & 0.89 \\
\hline
\end{tabular}
\end{table}

\begin{table*}
\caption{Summary of molecular features in the Sgr B2 complex, from the Mopra
3-mm peaks. We give the mean and standard deviation of position, velocity and 
velocity width, from the
fits to different lines, and include positions in galactic coordinates for
reference.}
\begin{tabular}{lcccccccccc}
\hline
Feature & R.A. (J2000)  & Dec. (J2000) &  $\sigma$(R.A.) & $\sigma$(Dec.) & 
lat. & long. & Velocity & $\sigma$(Vel.) & Width & $\sigma$(Width) \\
                &               &              & arcsec   & arcsec  & degree &
degree & km~s$^{-1}$   & km~s$^{-1}$     & km~s$^{-1}$  & km~s$^{-1}$ \\
\hline
north ridge     & 17 47 22.6 & -28 20 56 & 15 & ~7 & 0.702 & -0.024 &
~62 & ~3 & 43 & 10 \\
north cloud      & 17 47 21.4 & -28 21 24 & ~2 & ~4 & 0.693 & -0.024 &
~66 & ~3 & 24 & ~4 \\
near Sgr B2(N)  & 17 47 19.3 & -28 22 18 & ~7 & ~8 & 0.676 & -0.026 &
~65 & ~3 & 25 & ~5 \\
near Sgr B2(M)  & 17 47 19.2 & -28 23 04 & 11 & 15 & 0.665 & -0.032 &
~59 & ~4 & 22 & ~5 \\
near Sgr B2(S)  & 17 47 20.0 & -28 24 00 & ~4 & ~6 & 0.653 & -0.043 &
~58 & ~2 & 21 & ~6 \\
west ridge      & 17 47 14.8 & -28 22 34 & ~5 & ~9 & 0.664 & -0.014 &
117 & ~5 & 22 & ~6 \\
south-east peak & 17 47 26.9 & -28 23 12 & ~5 & ~7 & 0.678 & -0.057 &
~48 & ~7 & 27 & ~6 \\
\hline
\end{tabular}
\label{peak_summary}
\end{table*}

\begin{figure*}
\includegraphics[width = 5.2 cm,angle=-90]{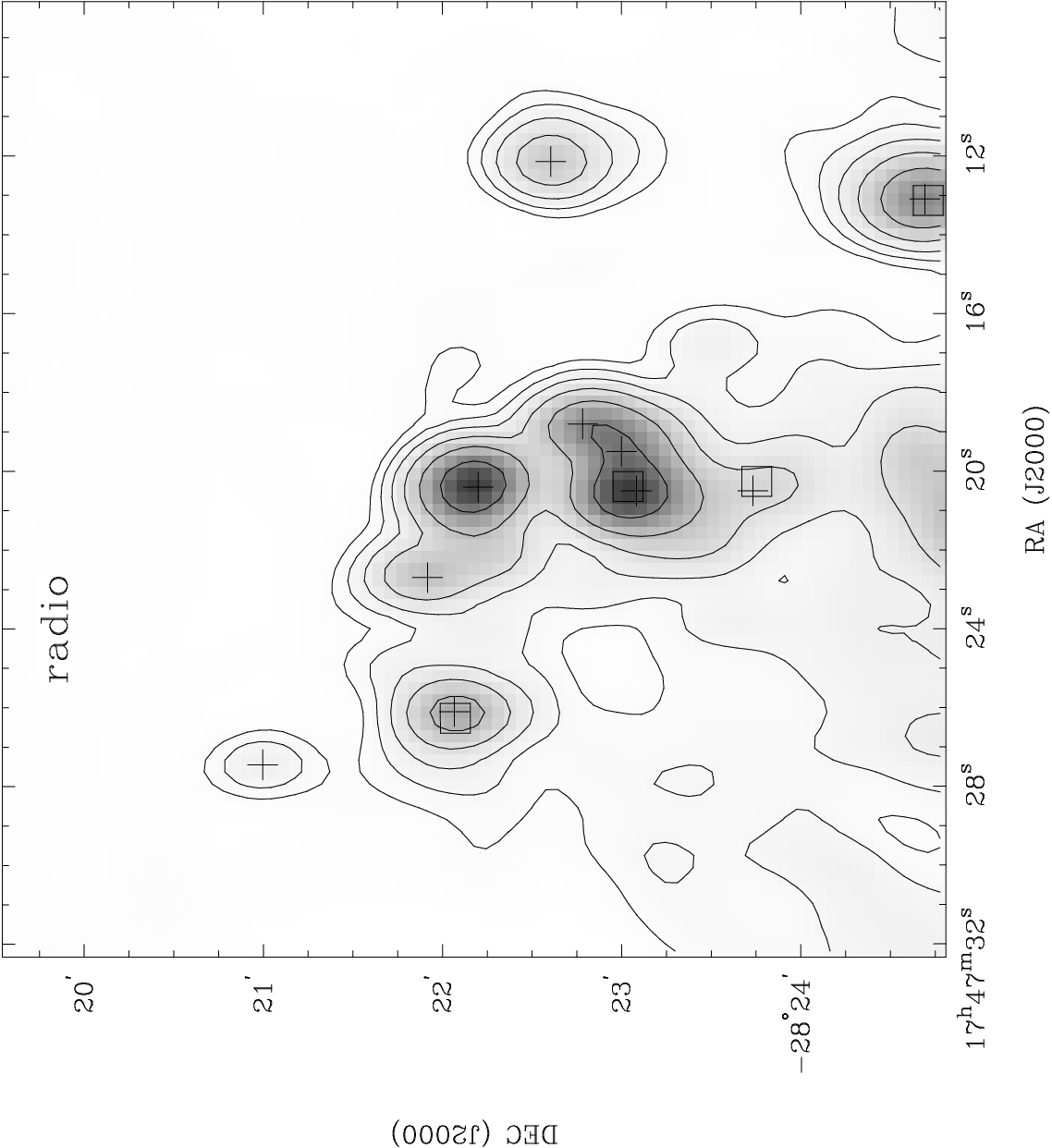}
\includegraphics[width = 5.2 cm,angle=-90]{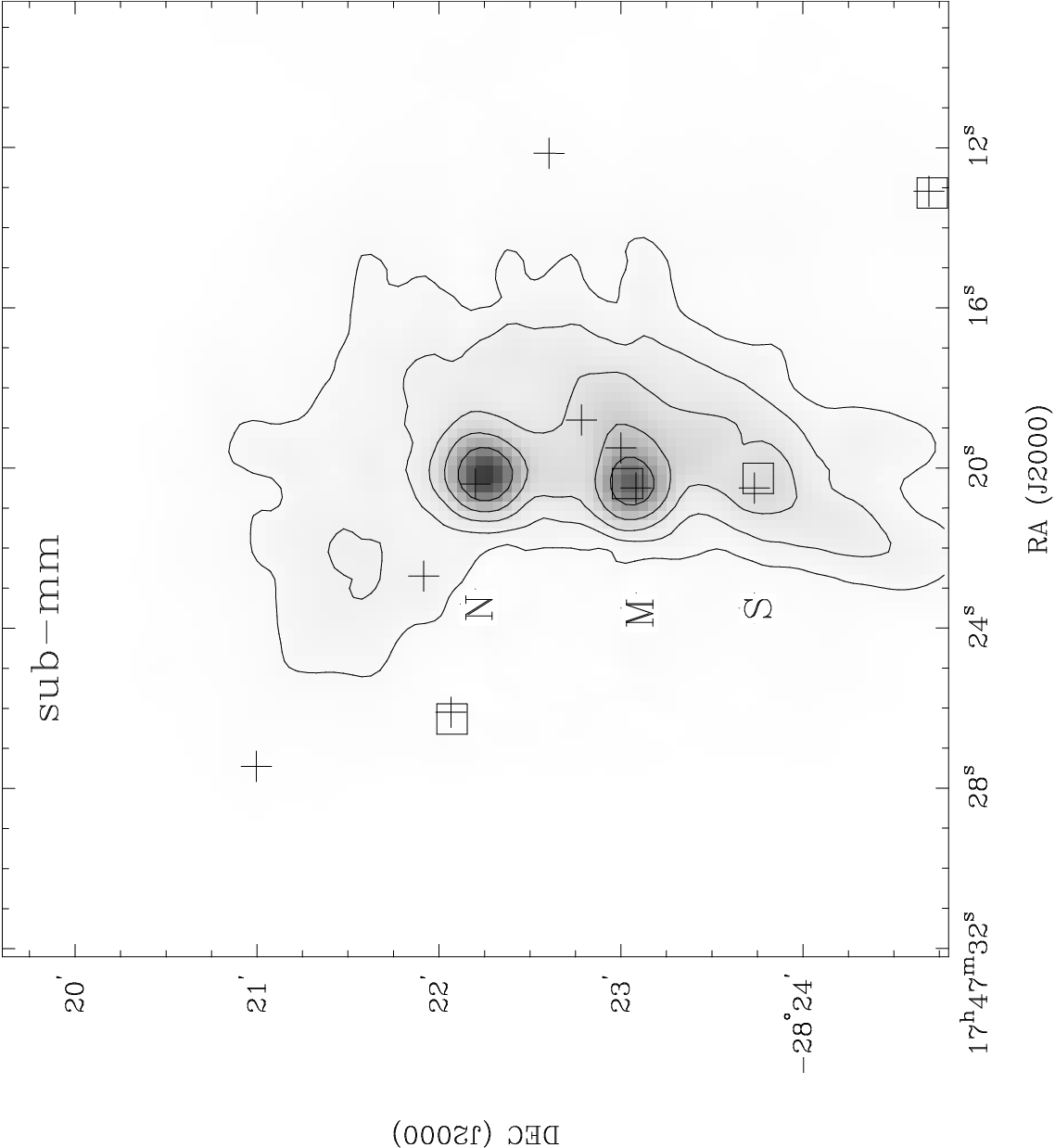}
\includegraphics[width = 5.2 cm,angle=-90]{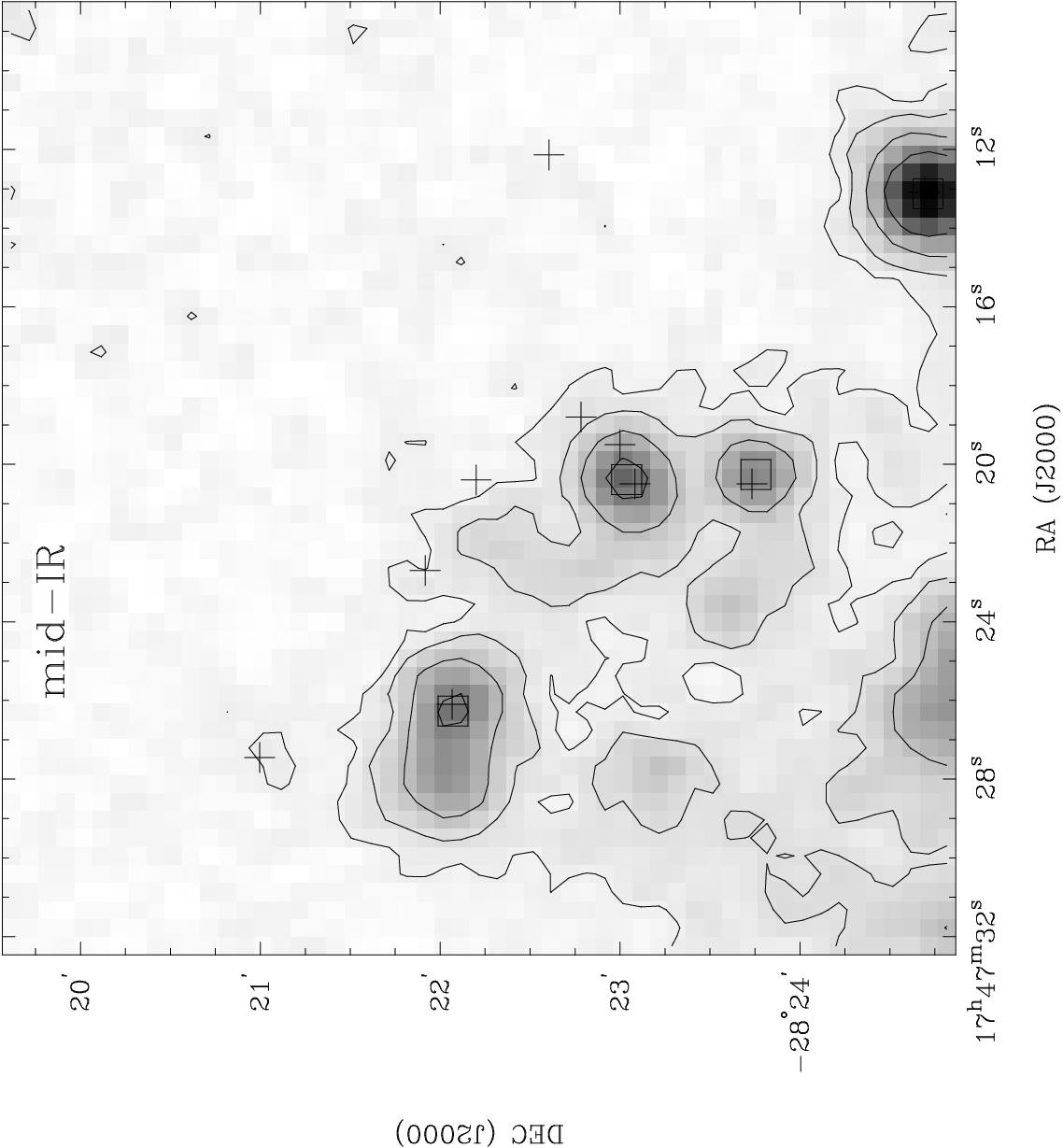}
\caption{The continuum emission in radio (left) from the VLA at 20-cm, sub-mm
(middle) from SCUBA at 850~$\micron$ and mid-IR
(right) from MSX at 21~$\micron$. The overlaid crosses are for radio sources
and open squares for mid-IR sources. Note how the sub-mm traces diffuse
cool dust to the west and north of the Sgr B2(N), (M) and (S) cores, as well
as compact emission from the cores. The radio and mid-IR trace star formation
in the cores and to the south-east.}
\label{radio_IR}
\end{figure*}

We also show in Figure \ref{radio_IR}
the 20-cm radio, from the VLA 
\footnote{http://imagelib.ncsa.uiuc.edu/imagelib.html},
the 850-$\micron$ sub-mm from SCUBA
\footnote{http://www3.cadc-ccda.hia-iha.nrc-cnrc.gc.ca/jcmt/}, and
21-$\micron$ mid-IR,
from MSX \footnote{http://irsa.ipac.caltech.edu/applications/MSX/}.

We point out that,
for the many molecules here that peak in the
ridge-line to the west of the Sgr B2(N) and Sgr B2(M) cores, this
distribution of molecular emission `wraps around' the north and west side,
where there is little radio and mid-IR emission tracing recent star formation,
and avoids the south-east area where there is recent star formation.
The north cloud, in particular, is quite isolated from the recent star formation activity. In contrast, the SCUBA sub-mm emission, tracing cooler dust than the
mid-IR, shows extended emission around the Sgr B2(N) and Sgr B2(M) cores,
to the north and west which matches well the north cloud and the
molecular ridge-line.

Each line that we have imaged here has its own particular distribution,
corresponding to the spatial distribution of the molecule, and the effect
of the excitation of the different levels. There is also a complicated velocity
structure in the region. However, we can make some generalisations and comments
here to bring some order to the overall results.

The CO ($^{13}$CO, C$^{18}$O and C$^{17}$O) shows that the densest region is 
around
Sgr~B2(M) at velocity 63 km~s$^{-1}$. The $^{13}$CO is optically thick at this 
core,
so the density there is better traced by the C$^{18}$O. The column density,
would be further concentrated at Sgr~B2(N) and Sgr~B2(M), than the C$^{18}$O
1~--~0 integrated line emission shown here. The higher temperature at the 
cores ($\sim$ 200 K,
compared to $\sim$~20 K for the surrounding gas) leads to an extra factor 
there, when converting, with the standard LTE analysis, from integrated
line emission
to total CO column density, and hence total H$_{2}$ column density.

The HCO$^{+}$, HCN and HNC are strong and widespread with absorption at the
Sgr~B2(M) and Sgr~B2(N) cores. The column density is likely to be peaked
at these cores, but the integrated emission is strongly affected by this
absorption, leading to local minima in the emission intensity
at the cores. There are differences
in the detailed distribution of these three lines, as expected: HCN should be
a good tracer of high gas density ($\ge 10^{4}$ cm$^{-3}$), the isomer HNC
should trace cool quiescent gas, and the ion HCO$^{+}$ should trace ionisation
due to cosmic rays. We have further Mopra data of these lines over a larger
area, from broad-band observations over the 85.3 to 93.3 GHz range, which show
the differences more clearly, so we postpone further discussion for a later
paper.

The CS and SiO distributions are also affected by absorption at the 
Sgr~B2(M) and Sgr~B2(N)
cores, so the column density distribution is hard to determine
from the integrated emission images. CS is expected to be, like HCN,
a good tracer of high density gas, and SiO is expected to trace shocks, but
is quite widespread here.

The CN emission is quite uniform over the $5 \times 5$~arcmin area observed 
here, except
for the absorption at the Sgr~B2(M) and Sgr~B2(N) cores. It is expected to be
associated with an enhanced ultraviolet (UV) field, so this would indicate
a widespread UV field in the Sgr B2 complex. The weak lines of C$_{2}$H
and c-C$_{3}$H$_{2}$ also have absorption at the Sgr~B2(M) and Sgr~B2(N) cores
and widespread emission, but some excess emission
on the ridge-line west of the cores.

Most of the lines imaged here trace the ridge-line west of the Sgr~B2(M),
Sgr~B2(N) and Sgr~B2(S) cores, and north-east to what we are calling the
north cloud. These lines include HC$_{3}$N, CH$_{3}$CN, CH$_{3}$OH, HNCO, OCS,
N$_{2}$H$^{+}$, CH$_{3}$CCH, NH$_{2}$CHO/H$_{2}$COH$^{+}$, NH$_{2}$CN,
CH$_{2}$NH and H$_{2}$CS. These more complex molecules, as noted above, trace
the cooler dust seen at sub-mm wavelengths, and avoid the areas with the 
warmer dust
(mid-IR) and radio continuum associated with the active star formation.
N$_{2}$H$^{+}$ and the 96.74 GHz transition of CH$_{3}$OH are strong, and
also show some absorption at the Sgr~B2(M) and Sgr~B2(N) cores.

The relative prominence of the peaks in the ridge-line west of the
Sgr~B2(M), Sgr~B2(N) and Sgr~B2(S) cores in these different molecules,
and between different transitions of the same molecule (e.g.CH$_{3}$OH),
indicate differences in chemistry and excitation conditions.

The lines of HNCO and HOCO$^{+}$ highlight the north cloud, and are tracers of
shock chemistry.

The lines of SO and SO$_{2}$ are also tracers of shocks, and are concentrated
at Sgr~B2(M), although the lower excitation 99.30 GHz SO line also traces
the more extended gas in the north cloud and ridge to the west.

\section{Summary}
\label{sec:summ}

We have undertaken a 3-mm spectral-line imaging survey of the Sgr~B2 area, of
5 arcmin square, with
the Mopra telescope, at resolution $\sim$ 36~arcsec.
We covered almost the
complete spectral the range 81.7 to 113.5 GHz, with 2.2 MHz or 
$\sim 6$~km~s$^{-1}$ spectral channels, and have observed 24 lines, 
with 0.033 MHz, or $\sim 0.1$~km~s$^{-1}$ channels.  
We have discussed the distribution of around 50
lines, and presented integrated emission images for 38 of the lines. In
addition, we have detected around 120 more lines, mostly concentrated
at Sgr~B2(N).

By fitting the peak position and velocity of the emission in the various 
lines, we find that there are seven distinct molecular features in the region, 
which show distinct differences in both molecular abundances and excitation 
conditions.

\section*{Acknowledgments}

The Mopra telescope is funded by the Commonwealth of Australia as a National
Facility managed by CSIRO as part of the Australia Telescope. The UNSW MOPS
digital filterbank was provided with funding from the Australian Research
Council, University of New South Wales, Sydney University, Macquarie University
and the CSIRO ATNF.
PAJ thanks the Max-Planck-Institut f\"{u}r Radioastronomie, Bonn, for a Visiting Fellowship in 2006, and the anonymous referee for comments that improved
the presentation of the paper.


\bsp

\label{lastpage}

\clearpage
\end{document}